%% file: ojcoms.tex
\documentclass{IEEEtran}
\usepackage{cite}
\usepackage{amsmath,amssymb,amsfonts}
\usepackage{color}
\usepackage{textcomp}
\usepackage[pdftex]{graphicx} 
\usepackage{algorithmic} 
\usepackage{array} 
\usepackage{url} 
\usepackage{hyperref} 
\usepackage[nolist,nohyperlinks]{acronym}
\usepackage[normalem]{ulem}
\usepackage{tikz}
\usepackage{tabularray}
\usepackage{multirow}

\usepackage[pscoord]{eso-pic}
\usetikzlibrary{decorations.markings}
\usepackage{pgfplots}
\pgfplotsset{compat=newest}

\definecolor{lbluew}{RGB}{222,242,252}
\definecolor{dbluew}{RGB}{186,229,250}
\definecolor{lbluef}{RGB}{233,247,254}
\definecolor{dbluef}{RGB}{12,114,184}

\definecolor{dgreen}{RGB}{208,232,196}
\definecolor{dred}{RGB}{239,178,184}
\definecolor{dorange}{RGB}{250,211,187}
\newcommand{\thickhline}{\noalign {\hrule height 1pt}}

\def\BibTeX{{\rm B\kern-.05em{\sc i\kern-.025em b}\kern-.08em
    T\kern-.1667em\lower.7ex\hbox{E}\kern-.125emX}}
\AtBeginDocument{\definecolor{ojcolor}{cmyk}{0.93,0.59,0.15,0.02}}

\usepackage{fancyhdr}
\fancypagestyle{firststyle}{\fancyhf{}
	\fancyhead[L]{T. S. Rappaport, M. Ying, N. Piovesan, A. De Domenico, and D. Shakya, ``Waste Factor and Waste Figure: A Unified Theory for Modeling and Analyzing Wasted Power in Radio Access Networks for Improved Sustainability", in \textit{IEEE Open Journal of the Communications Society (OJ-COMS)}, July 2024, pp. 1--28 (Final submission).}
}

\usepackage{times}
\begin{document}

\title{Waste Factor and Waste Figure: A Unified Theory for Modeling and Analyzing Wasted Power in Radio Access Networks for Improved Sustainability}


\author{
    \IEEEauthorblockN{Theodore S. Rappaport,\IEEEauthorrefmark{1}~\IEEEmembership{Fellow,~IEEE}, Mingjun Ying,\IEEEauthorrefmark{1}~\IEEEmembership{Student Member,~IEEE}, Nicola Piovesan,\IEEEauthorrefmark{2}~\IEEEmembership{Member,~IEEE}, Antonio De Domenico,\IEEEauthorrefmark{2}~\IEEEmembership{Member,~IEEE}, Dipankar Shakya,\IEEEauthorrefmark{1}~\IEEEmembership{Student Member,~IEEE}}
    
    \IEEEauthorblockA{\IEEEauthorrefmark{1}NYU WIRELESS, New York University, Tandon School of Engineering, Brooklyn, NY 11201 USA}
    
    \IEEEauthorblockA{\IEEEauthorrefmark{2}Huawei Technologies, Paris Research Center, Boulogne Billancourt 92100 France}
    \thanks{This research is supported by the New York University (NYU) WIRELESS Industrial Affiliates Program and the NYU Tandon School of Engineering graduate fellowship.}
}


\maketitle

\thispagestyle{firststyle}

\begin{abstract}
This paper introduces Waste Factor (\textit{W}), also denoted as Waste Figure (\textit{WF}) in dB, a promising new metric for quantifying energy efficiency in a wide range of circuits and systems applications, including data centers and Radio Access Networks (RANs). Creating and managing 5G and future 6G networks that are energy-efficient is of paramount importance as the wireless industry evolves to become a major consumer of energy. Also, the networks used to connect data centers and artificial intelligence (AI) computing engines with users for machine learning (ML) applications must become more power efficient. This paper illustrates the limitations of existing energy efficiency metrics that inadequately capture the intricate energy dynamics of RAN components. We show here that \textit{W} provides a generalized analysis of power utilization and energy waste at both the component and system levels for any source-to-sink communication system. We delineate the methodology for applying \textit{W} across various network configurations, including multiple-input single-output (MISO), single-input and multiple-output (SIMO), and multiple-input multiple-output (MIMO) systems, and demonstrate the effectiveness of \textit{W} in identifying energy optimization opportunities. Our findings reveal that \textit{W} not only offers nuanced insights into the energy consumption of RANs but also facilitates informed decision-making for network design and operational efficiency. Furthermore, we show how \textit{W} can be integrated with other key performance indicators (KPIs) and key value indicators (KVIs) to guide the development of optimal strategies for enhancing network energy efficiency under different operational conditions. Additionally, we present simulation results for a distributed multi-user MIMO (MU-MIMO) system at 3.5, 17, and 28 GHz, demonstrating overall network power efficiency on a per square kilometer basis, and show how overall \textit{W} decreases (e.g., energy efficiency increases over the entire network) with an increasing number of base stations and increasing carrier frequency. This paper shows that adopting \textit{W} as a figure of merit (FoM) can enable the design of more sustainable next-generation wireless communication networks, paving the way for greener and more sustainable, energy-efficient 5G and 6G technologies.
\end{abstract}


\section{INTRODUCTION}
\label{sec:intro}

\IEEEPARstart{I}{nformation} and communication technology (ICT) infrastructures, such as data centers, mobile communication networks, and consumer devices, are essential in driving digital advancements and supporting various applications such as Internet of Things (IoT) and Artificial Intelligence (AI). However, the ICT industry is a great consumer of energy, which incurs significant environmental costs. As of 2020, ICT infrastructures were responsible for consuming between 5-8\% of global electricity production \cite{WorldBank2023, itu2020greenhouse} and were estimated to contribute between 1.5-4\% of global greenhouse gas emissions \cite{silver2019mobile, bieser2023review} and this was well before the ramp of AI. 

With two-thirds of the global population now online \cite{ayers2023measuring}, and despite advances in energy efficiency and renewable energy adoption, the environmental impact of the ICT sector is poised to rise, potentially consuming over 20\% of global power due to increased data traffic \cite{ying2023waste}.

The introduction of 5G and the anticipation of 6G, both of which use bandwidths and deliver end-user data rates far beyond anything in history \cite{Rappaport2013Itwillwork, Rappaport2019Beyond6G}, underline the urgency to address energy efficiency. 
Although 5G networks are about four times more energy-efficient per unit of data than 4G networks, their total energy consumption is considerably higher \cite{Huawei2020}. This increase is primarily due to the need for additional power-intensive processing to support wider bandwidths, more radio chains per base station, and increased cell site density to support higher frequency bands like C-band, mid-band, and millimeter wave \cite{Rappaport2013Itwillwork, Shakya2024Indoor, Kang2024FR3, Shakya2024Pen}.

\begin{figure}[!t]
	\centering
 \includegraphics[width=3.0in]{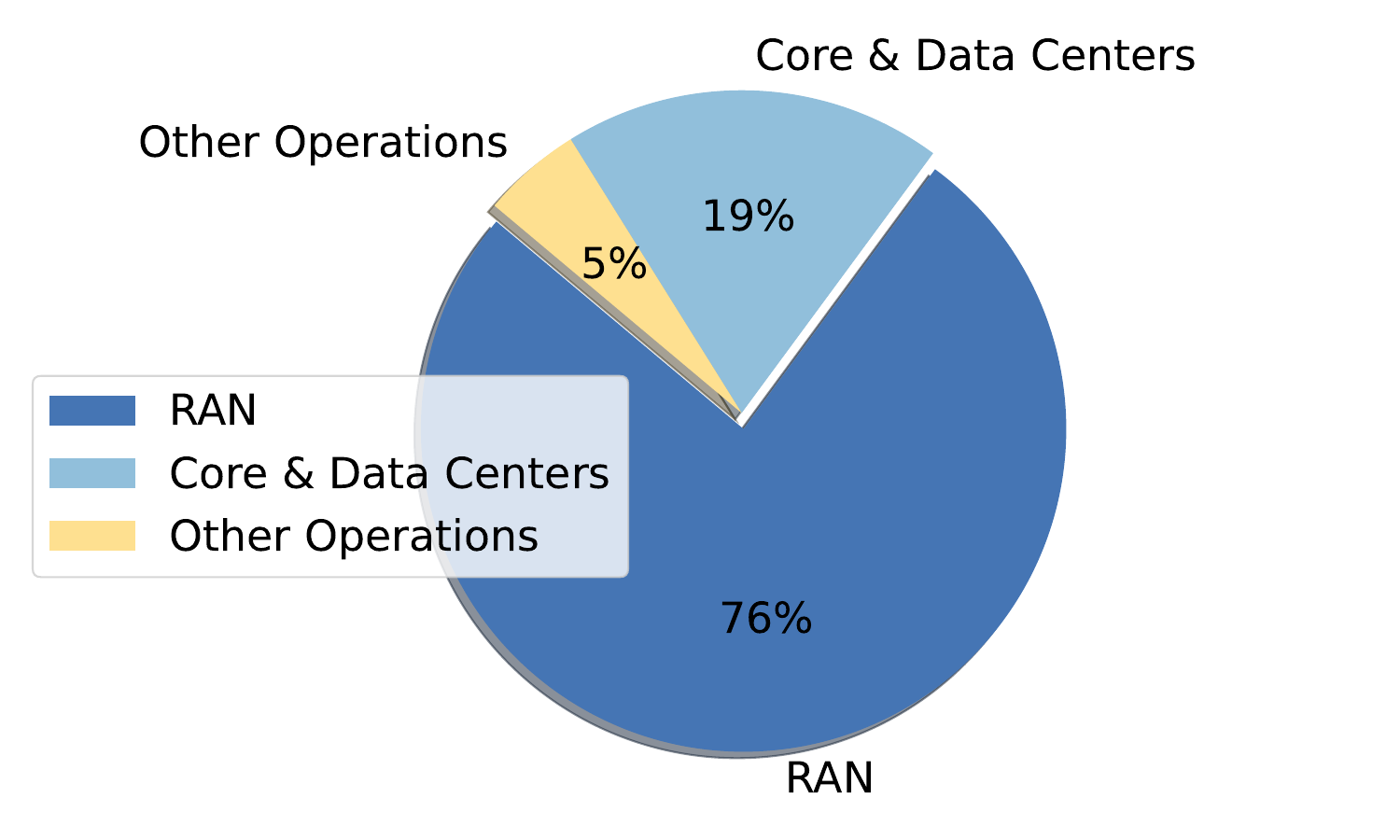}
	\caption{Energy consumption distribution in telecommunication operators \cite{gsma2024going}.}
	\label{fig:ecdis}
\end{figure}

In the typical cellular network, the \ac{RAN} is the largest power consumer, accounting for 76\% of the total energy usage. The core network and data centers consume 19\%, while other operations account for the remaining 5\% of energy use, as depicted in Figure \ref{fig:ecdis} \cite{NGMN2021NetworkEnergy, gsma2024going}. 
The financial implications of increased energy consumption are significant, particularly for mobile network operators. Network costs, which on average constitute about 25\% of the \ac{OPEX} of a mobile network operator, are heavily impacted by energy expenses, with approximately 90\% of network costs dedicated to energy bills\cite{GSMA20205Genergy}. 
A nationwide carrier with 100 million subscribers pays about \$2 billion USD/year for electricity, and when multiplied over many hundreds of network operators (also called carriers) and 4 billion subscribers across the globe, the electricity bill for the cellular industry is approximately \$1 trillion USD/year.
The challenges associated with reducing network energy consumption are exacerbated by a projected threefold increase in mobile data traffic from 2023 to 2029~\cite{ericsson2023}. Addressing this demand requires the deployment of additional infrastructure and innovative strategies to optimize energy usage.

In response to the rising energy demands, the \ac{3GPP} has initiated efforts to mitigate \ac{RAN} energy consumption. The 3GPP \ac{NR} Release 15 has introduced intra-NR network energy-saving solutions, including autonomous cell switch-off and re-activation features for capacity-boosting cells through Xn/X2 interfaces. Furthermore, \ac{3GPP} \ac{NR} Release 17 has specified inter-system network energy-saving solutions and recognized network energy saving as a potential application for AI.

The trajectory of the network energy consumption growth underscores the urgent need for sustainable network expansion strategies and accurate, standardized energy efficiency metrics.

Many researchers have attempted to quantify and unify the analysis of energy consumption in mobile communications. In \cite{chen2010energy}, the authors overview various energy efficiency metrics used in wireless systems at the component, equipment, and system/network levels. They note that energy efficiency metrics at the component and equipment levels are well-developed, while those for system/network levels need more attention. Authors in \cite{chen2010energy} note existing studies predominantly focus on macro-level energy efficiency metrics, often overlooking the intricate energy dynamics within RAN components. Such approaches do not fully capture the granular energy consumption patterns essential for devising targeted energy-saving strategies \cite{boyle2017energy, hossfeld2022analytical}. To achieve long-term energy efficiency goals for the sustainable growth of the wireless industry, a unified metric is necessary to fill the gaps in previous energy efficiency metrics. In \cite{Mccune2019globecom}, McCune highlights that energy efficiency in wireless links varies by more than eleven orders of magnitude and stresses the need for a uniform approach to energy efficiency across different systems. In \cite{murdock2013consumption}, the consumption factor evaluates the power consumption (e.g. the energy efficiency) of a relay network and cascaded circuits. In \cite{kanhere2022power}, the consumption efficiency factor (CEF) provides a quantitative metric for the trade-off between the data rate and the energy and power consumed by a communication system and provides insights for network energy efficiency with different cell sizes. In the \ac{ETSI}, \ac{NGMN}, \ac{ITU}, and \ac{3GPP} standard bodies, diverse energy efficiency (EE) metrics have been defined \cite{etsiES203228, etsiTS103786, NGMN2021NetworkEnergy, ituL13502016, ituL13102020, 3gpp28813}.

This paper introduces Waste Factor ($W$), or Waste Figure (\textit{WF}) in dB, as a novel, powerful, and extensible metric to quantify energy efficiency within a cascade and for paralleled systems such as multiple-chain signal processing and antenna systems within RANs. Unlike traditional metrics, \textit{W} provides a detailed perspective on how energy is utilized, wasted, or converted into useful output, thereby offering a comprehensive tool for energy analysis that can be applied far beyond just simple circuits or communication networks~\cite{ying2023waste, Rappaport2024Microwave, kanhere2022power, ying2024MIMOWF}. The waste factor is capable of quantifying the energy efficiency of \textit{any} source-to-sink communication path, including computers, data centers, or other systems over which communications are carried out~\cite{Rappaport2024Microwave}. The Waste Factor theory is the first approach that we are aware of that breaks down the power consumption of any device or cascade into different components via superposition in order to characterize the work done for source-to-sink transport while considering the additive wasted power or squandered energy of such “on-path” components, and considering separately the work done by other components that are not even on the signal path of information flow. When such a dichotomy is made, it becomes an easy task to quantify energy usage and efficiency using the invariable fact that energy is neither created nor destroyed. That which is not delivered is wasted, and by using such quantification, as is done in the Waste Factor theory, it becomes possible to measure, model, and improve energy efficiency. Said another way, this theory lays bare the wasted and transported power and allows its modeling and optimization in a systematic way for any linear, cascaded, source-to-sink system.

This paper comprehensively examines energy efficiency within RANs, emphasizing how \textit{W} can be used to compare and contrast energy efficiency within RAN systems and sub-systems. Our analysis compares the use of \textit{W} with conventional energy efficiency metrics used in the standards bodies of today's mobile industry such as \ac{ITU} and \ac{ETSI}, and demonstrates how \textit{W} provides valuable and nuanced insights into the energy dynamics of RAN components. We use \textit{W} to evaluate the energy performance of various RAN elements and highlight its utility in pinpointing areas where energy optimization can yield significant benefits.

The paper also explores the application of \textit{W} for quantifying energy efficiency in different network configurations, including Multiple-Input Single-Output (MISO), Single-Input Multiple-Output (SIMO), and Multiple-Input Multiple-Output (MIMO) systems, showcasing the versatility of \textit{W} in assessing energy efficiency across a spectrum of network architectures.

Dynamic network conditions, such as fluctuating traffic demands, varying channel conditions, and diverse network topologies, benefit from a \textit{W}-centered analysis to guide targeted enhancements, such as deploying energy-efficient hardware or implementing adaptive resource allocation algorithms, as well as long-term planning decisions, like network expansion or technology upgrade, to maintain high energy efficiency as the network evolves. 

The remainder of the paper is organized as follows. Section~II defines \textit{W} from theoretical underpinnings and highlights its advantages compared to conventional energy efficiency and power efficiency metrics. Section~III extends the Waste Factor theory to paralleled systems, such as MISO and MIMO. Section~IV applies \textit{W} to a 5G RAN case study and explains how \textit{W} is used to determine power efficiency for both passive and active components. Section~V emphasizes the utility of \textit{W} in providing a comprehensive and nuanced perspective on energy efficiency in RANs, and illustrates the potential of \textit{W} to inform decisions that can lead to more energy-efficient network operations than is currently found when using ETSI, NGMN, ITU or 3GPP energy efficiency metrics. Finally, Section~VI concludes the paper by highlighting how the Waste Factor provides detailed insights into power utilization and energy efficiency optimization. We provide a realistic simulation comparing carrier frequency, antenna gain, and bandwidth considerations to demonstrate the practical use of \textit{W} for determining power consumption for an operator, thereby illustrating how Waste Factor may be used to develop more sustainable wireless networks.

\section{Waste Factor for a Cascaded System}
\label{sec:WFcascadedsystem}

In this section, we derive the Waste Factor theory using a model of additive wasted power along a cascade, and show how this framework for studying energy efficiency was inspired by the model of additive noise used by Harald Friis in his 1944 derivation of Noise Factor ($F$) \cite{a12}.

\subsection{Noise Factor - A Brief Review}
Noise Factor, denoted as $F$, quantifies the additive noise contributed by each component in a signal cascade. This metric is essential for analyzing the degradation of the signal-to-noise ratio (SNR) throughout the cascade. Specifically, $F$ is defined as the ratio of the input SNR ($\text{SNR}_{i}$) to the output SNR ($\text{SNR}_{o}$), expressed as $F = {\text{SNR}_{i}}/{\text{SNR}_{o}}$. In decibel units, this factor is known as the Noise Figure (\textit{NF}), where a value of 0 dB indicates no added noise and thus no degradation in SNR across a component or cascade within a receiver \cite{a12}.

\subsubsection{Significance and Impact}

The concept of the Noise Factor is fundamental in understanding how each component, such as amplifiers, mixers, and filters, adds to the overall noise within a communication receiver, and plays a critical role in designing systems to minimize noise accumulation, especially in applications that are ultra sensitive to SNR, such as satellite communications, noise-limited cellular networks, and fixed wireless access links with small link budget margins. The key to effective noise management lies in quantifying and controlling the additive noise at each stage of the receiver chain to ensure maximum output SNR ($\text{SNR}_{o}$) at the detector \cite{SNROptTAP, LowNoiseAmp, Rappaport2024Microwave}. 

\subsubsection{Friis's Formula for Noise factor}
Friis's formula for a cascaded receiver system calculates the total $F$ of cascaded devices, each device having its own individual $F$ and power gain $G$, on a linear scale. The formula for the overall noise factor $F$ of a cascade is given by \cite{a12}
\begin{equation}\label{b1}
F = F_{1} + \frac{(F_{2}-1)}{G_1} + \frac{(F_{3}-1)}{G_1 G_{2}} + \ldots + \frac{(F_N-1)}{\prod_{i=1}^{N-1} G_i}.
\end{equation}

Here, $F_{i}$ is the noise factor and $G_{i}$ is the power gain of the $i^{th}$ device in the cascade where the first device is closest to the source, and the $N^{th}$ device is closest to the sink. Eq.~\eqref{b1} is essential for understanding how each component affects the overall noise performance of the system. It is readily apparent from \eqref{b1} that the noise factor of a cascade is primarily influenced by the input components with high gain, indicating that the initial component in the cascade, originating from the source side, exerts the most significant impact on the noise factor. The additional noise power contributed by the $i^{th}$ device in the cascade (independent of the input noise power) is defined by
\begin{equation}
\label{addnoise}
    P_{\text{additive-noise,i}} = (F_{i} - 1)G_{i}N_{i},
\end{equation}
where $N_i$ is the noise power of \(i^{th}\) device input \cite{Rappaport2024Microwave, a12}.

\subsubsection{Real-world Applications of Noise Factor}
The practical importance of minimizing $F$ is critical for any communication system since improved SNR leads to better communication performance (e.g., lower bit error rate). Maximizing SNR is especially critical for extending the link budget in low-power scenarios or for enabling extremely sophisticated multi-level digital modulations that are ultra-sensitive to SNR. $F$ facilitates informed decision-making regarding component selection and system design to achieve minimal noise impact and maximum detector performance. For example, in deep-space communication, even small increases in noise can lead to significant data loss \cite{hemmati2006deep}. In proposed 5G and 6G systems, Orthogonal frequency-division multiplexing (OFDM) with 4096-level quadrature amplitude modulation (QAM), which encodes 12 bits per symbol, requires SNR levels to exceed 22 dB when a coding rate of 1/3 is used under the extended pedestrian A (EPA) fading mode to achieve a $10^{-2}$ bit error rate (BER)~\cite{tian2018ber}.

In modern communication systems, both signal quality and energy consumption are paramount but usually offer conflicting design choices. As originally proposed in~\cite{a8,x1}, the transition to wider bandwidth communications was inevitable over time. It was recognized that analogous to how the Noise Figure quantifies additive noise, a unifying metric was necessary to quantify the additive power waste in a cascade.
This reasoning, first inspired by consideration of why it is impossible to build perpetual motion machines, was the genesis of Waste Factor $W$. While \textit{F} assesses additive noise power, \textit{W} focuses on additive wasted power along the source-to-sink path, in order to provide a unified method to model and analyze power efficiency and sustainability in any communication network.

With the move to wide bandwidth wireless channels (going from 20 MHz in 4G to 200 MHz or more in 5G), there is a need to allocate mobile services at higher carrier frequencies where spectrum is more plentiful \cite{Shakya2024Pen, Shakya2024Indoor, Rappaport2013Itwillwork, Rappaport2019Beyond6G}. At such higher frequencies, the use of directional antennas on both ends of the link (e.g. dual directional channels) makes the channels more noise-limited and less interference-limited (the early cellular generations below 6 GHz were interference-limited \cite{Rappaport2024wireless}). As carrier frequencies and the commensurate channel bandwidths increase, the channels become more noise-limited or power-limited \cite{Rappaport2013Itwillwork, Rappaport2019Beyond6G, Rappaport2024wireless}. This paradigm shift to wider RF channel bandwidths brings about much greater power consumption due to the raw increase in consumed bandwidth and underscores the importance of optimizing power efficiency while also simultaneously minimizing noise for system sustainability. As shown here, \textit{W} may become an essential metric in next-generation wireless communications for evaluating the power wasted in a cascade, which as shown here, is useful well beyond just receiver analysis, but may be used for understanding the power efficiency of any device, entire transmitter and receiver chains, and even the entire cascade between transmitters, receivers, and the propagation path loss of one or multiple channels.

\begin{figure}[!t]
	\centering
	\includegraphics[width=0.9\columnwidth]{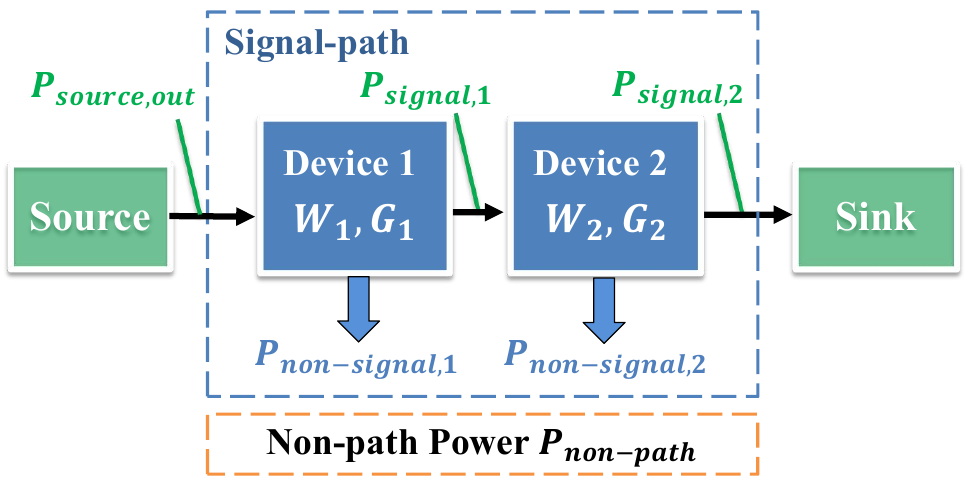}
	\caption{A general cascaded communication system composed of two devices, and components not on the signal path are deemed to consume $P_{non-path}$ power.}
	\label{figa1}
\end{figure}

\subsection{Waste Factor}
$W$ characterizes the power efficiency of a cascaded system by considering the power wasted by components along a cascade. 
We note that power, $P$, is the rate of work done while energy is being consumed, whereas energy, $E$, is the ability to do work, and may be consumed or stored. It is well understood that over a unit of time $T$, $E \text{ (Joule)} = P \text{ (Watt)}\times T \text{ (s)}$. In this context, one may consider power efficiency and energy efficiency synonymously as long as a device or system is in a particular operating condition or steady state.
Akin to $F$ which studied additive noise, the power wasted by a device/cascade can be modeled by using \textit{W} to relate the additive power waste, based on the output power at each successive stage, as the signal propagates from source to sink along the cascade. Such formulation provides an intuitive way to understand power waste at each cascade stage, whereby $W$ becomes a key metric by which to compare the power efficiency of different devices/systems through their wasted power. A key concept in the development of $W$ is that wasted power is defined as the amount of power consumed by a component that carries the signal which is not found in the signal power output sent to the subsequent component in the cascade. In other words, the consumed energy that is not transmitted along the signal path cascade is classified as wasted energy. To derive $W$, the power consumption of an entire communication system between a source and a sink is represented by $P_\text{consumed,total}$. Then, it is noted that $P_\text{consumed,total}$ is equal to the superposition of the power $P_\text{consumed,path}$, which denotes the power consumed by all components along the signal path cascade, and for which $W$ is a useful parameter for computing, in addition to all of the power consumed by all components that are not on the path of the source-to-sink (this off-path power is denoted as $P_\text{non-path}$). To illustrate how W is used to compute the signal-path waste, which is only a portion of the total power consumed in any system, we break down the total consumed power into finer details.   We observe first that $P_\text{consumed,total}$ may be split into four principal power components by superposition \cite{ murdock2013consumption, kanhere2022power, x1, Rappaport2024Microwave}: 
\begin{itemize}
        \item Input power from the source ($P_\text{source, out}$): the power applied to the device or the cascaded system that comes from the signal source or input side.
        \item Output power ($P_\text{signal}$): the signal power delivered to the device or cascade output.
        \item Non-signal path power ($P_\text{non-signal}$): the power consumed by devices within the cascade (e.g., along the signal path) to facilitate the signal transfer, but which is not part of the delivered signal power, and is viewed as wasted power which includes wasted power consumed by active mixers, amplifiers, or power lost from heat dissipation and other losses in components along the signal path.
        \item Non-path power ($P_\text{non-path}$) refers to the power consumed by components in the system that do not contribute to the signal power.\footnote{Oscillators in electronic circuits are designed primarily for frequency synthesis, generating low-power periodic waveforms that serve as reference or control signals, and are not on the direct signal path of a cascade, so their power consumption would be accounted for as non-path power. For amplifiers that are on the signal path, when they are turned off or not in use, the quiescent or standby power is shown in Section~\ref{sec:RAN_intro} to be properly categorized as non-path power even though amplifiers are along the signal path during normal operation \cite{Rappaport2024Microwave}.} This includes the quiescent power of amplifiers that are not carrying signal, and the power used by oscillators, power dissipation units, displays, and other similar components not directly involved in the source-to-sink signal path. 
\end{itemize}

\textit{W} is defined in \eqref{defW} as the ratio of the power consumed on the signal path along a cascade $P_\text{consumed,path}$, where $P_\text{consumed,path} = P_\text{signal} + P_\text{non-signal}$, to the signal output power $P_\text{signal}$, \cite{Rappaport2024Microwave,kanhere2022power,ying2023waste}, and $\eta_{\text{w}}$ is defined as the Waste Factor Efficiency, $\eta_{\text{w}}$, which is the reciprocal of \textit{W}.
\begin{equation}
W = \frac{1}{\eta_{\text{w}}}=\frac{P_\text{consumed,path}}{P_\text{signal}} = \frac{P_\text{signal} + P_\text{non-signal} }{P_\text{signal}}.
\label{defW}
\end{equation}

The total power consumption of any system $P_\text{consumed,total}$ is the sum of power consumed on the signal path $P_\text{consumed,path}$ together with the power consumed by non-path components $P_\text{non-path}$, where $W$ is used to characterize the wasted power of each component and the entire cascade on the signal path between source and sink (this is similar to $F$ which represents the noise of a device or a cascade) \cite{Rappaport2024Microwave,ying2023waste,a8,murdock2013consumption}.
\begin{align}
\label{powercontotal}
\nonumber
    P_\text{consumed,total} &= P_\text{source,out} + P_\text{system,added} + P_\text{non-signal} + P_\text{non-path}\\ \nonumber 
    &= \underbrace{ P_\text{signal} + P_\text{non-signal} }_{\text{signal-path power}} + P_\text{non-path} \\ \nonumber
    &= P_\text{consumed,path} + P_\text{non-path}  \\
    &= W P_\text{signal} + P_\text{non-path},
\end{align}
where $P_\text{system,added} = P_\text{signal} -  P_\text{source,out}$ is the power added (e.g., contributed) to the signal output power solely by the device or cascade with the applied signal source power removed from consideration under the assumption of linearity.

$W$ is determined based on the ratio of power consumed along the signal path to useful output power, as defined in \eqref{defW}. 
This leads to the practical application of \textit{W} in determining the power wasted across a device or cascaded system from \eqref{defW}
\begin{equation}
\label{powerwasted}
P_\text{wasted} = P_\text{non-signal} = (W-1) P_\text{signal}.
\end{equation}

Furthermore, the wasted power for a cascade can also be computed based on the input power from the source and the gain of each stage
\begin{equation}
P_\text{wasted} = (W-1) P_\text{source,out} \prod_{i = 1}^{N}G_{i},
\end{equation}
where $P_\text{signal}=P_\text{source,out} \prod_{i = 1}^{N}G_{i}$  and $G_{i}$ denotes the gain of the \(i^{th}\) stage in the cascade, with $i = 1$ representing the stage closest to the source and $N$ being the total number of cascaded components in practice, $P_\text{signal}$ is used for measurement reference. 

Note (\ref{defW}), (\ref{powercontotal}), and (\ref{powerwasted}) offer new ways to analyze, design, and compare the power efficiency and energy efficiency of circuits and cascaded systems, since $W = 1$ denotes no wasted power exists (all consumed power on the signal path is delivered as output signal power,) and $W = \infty$ denotes all power is wasted and no power is delivered to the output. Just like $F$, $W$ is generally defined regardless of the presence or level of signal input power applied by a signal source may be measured in-site, and is only a function of the gain and the Waste Factor of the device or system.\footnote{As discussed in Section~\ref{sec:RAN_intro}-\ref{sec:active_components}, certain devices that transform between the analog and digital domains may use power levels to define $W$ as in \eqref{defW}-\eqref{powerwasted},  and we show in Section~\ref{sec:RAN_intro}-\ref{sec:WFmeasurements} that measurements of actual power consumption and power delivered may always be used to define \textit{W} for a device or system using \eqref{defW}-\eqref{powerwasted}.}

\subsection{Waste Factor for a cascaded system}

\textit{W}  for a system consisting of two cascaded devices is explained using Fig. \ref{figa1}. Here, we concentrate on the signal path, ignoring any non-path power in the system. Using \eqref{powercontotal}, we see 
\begin{equation}\label{d1}
	\begin{aligned}
		P_\text{consumed,path}&=W \times P_\text{signal,2}.\\
	\end{aligned}
\end{equation}

Now, we can define the signal-path power consumed before the terminal of device 1 as
\begin{equation}\label{d2}
	P_\text{consumed,path1}=W_{1}  P_\text{signal,1},
\end{equation}
where $P_\text{consumed,path1}$ in eq. \eqref{d2} includes the input power from the source output and also comprises both the useful signal power transmitted to the subsequent device and the power wasted by device 1 itself. When $P_\text{source,out}$ is subtracted, we arrive at the standalone signal-path power consumption of device 1 ($D1$).
\begin{equation}\label{d3}
	P_\text{consumed,path}^{D1}=W_{1}  P_\text{signal,1}-P_\text{source,out},
\end{equation}
and it also follows that the signal-path power consumption of device 2 ($D2$) alone is
\begin{equation}\label{d4}
	P_\text{consumed,path}^{D2}=W_{2} P_\text{signal,2}-P_\text{signal,1}.
\end{equation}

 The total power consumption on the signal path can be expressed by \eqref{d1_1}, which is the sum of power consumed by each device and the signal power input (e.g., signal source output power) of the system.

\begin{equation}\label{d1_1}
		P_\text{consumed,path}  =P_\text{consumed,path}^{D1}+P_\text{consumed,path}^{D2} +P_\text{source,out}.
\end{equation}
Further, we know that the output power of the system is
\begin{equation}\label{d5}
P_\text{signal,2}=G_{2}P_\text{signal,1}.
\end{equation}
Hence, it follows from (\ref{d1})-(\ref{d5})
\begin{equation}\label{d6}
	\begin{aligned}
		P_\text{consumed,path}&=W_{2}P_\text{signal,2}+(W_{1}-1)P_\text{signal,1}\\
		&=\left(W_{2}+\frac{W_{1}-1}{G_{2}}\right)P_\text{signal,2}.
	\end{aligned}
\end{equation}

Since (\ref{d6}) is equal to (\ref{d1}), $W$ for the cascaded system is given by 
\begin{equation}\label{d7}
	W=\left(W_{2}+\frac{W_{1}-1}{G_{2}}\right),
\end{equation}
where \( W \) is the Waste Factor for the cascade, \( W_1 \) and \( W_2 \) are the Waste Factors of the individual devices, and \( G_2 \) is the gain of the second device.

Now consider a three-device cascade. If we model the cascade of device 1 and device 2 as a single entity, and then cascade this entity with device 3, we see the Waste Factor for the first two devices can be expressed in the same form as \eqref{d7}
\begin{equation}\label{proof1}
    W_{1,2} = W_2 + \frac{W_1 - 1}{G_2}.
\end{equation}

Taking device 1 and device 2 as a whole, and cascading this combined entity with a third stage closer to the sink, device 3, we have
\begin{equation}\label{proof2}
    W_{1,2,3} = W_3 + \frac{W_{1,2} - 1}{G_3}.
\end{equation}

Substituting \eqref{proof1} into \eqref{proof2}, we obtain $W$ for the three-device cascade
\begin{align}\label{proof3}
    W_{1,2,3} &= W_3 + \frac{\left( W_2 + \frac{W_1 - 1}{G_2} - 1 \right)}{G_3} \nonumber \\
              &= W_3 + \frac{W_2 - 1}{G_3} + \frac{W_1 - 1}{G_2 G_3}.
\end{align}

Based on \eqref{d7} and \eqref{proof3}, we generalize $W$ for a cascaded system consistsing of \(N\) devices
\begin{align}\label{Wcas}
    W_{1,2,...,N} &= W_N + \frac{\left( W_{N-1} + \cdots + \frac{W_1 - 1}{\prod_{i=2}^{N-1} G_i} - 1 \right)}{G_N} \nonumber \\
                  &= W_N + \frac{W_{N-1}-1}{G_N} + \cdots + \frac{W_1-1}{\prod_{i=2}^N G_i},
\end{align}
where the final expression of \eqref{Wcas} is similar to the cascaded $F$ in \eqref{b1} \cite{a12} except \textit{W} is referred to the output of the cascade, where the signal power out of the cascade and applied to the sink may be readily measured \cite{Rappaport2024Microwave}.

Despite the similarity of the math between $F$ and $W$, \cite{Rappaport2024Microwave, ying2023waste}, it is worth noting that $F$ concerns additive noise, which is of primary consequence in the design of a receiver, whereas $W$ concerns the additive wasted power which, as shown in this paper, has broad application in many realms wherever there is a transmission of information from a source to a sink. For example, it is shown in \cite{Rappaport2024Microwave} and in Section~\ref{sec:RAN_intro} that the Waste Factor of an attenuator is simply its loss, and the Waste Factor of any channel is simply the channel loss. These principles are used subsequently.

\section{Waste Factor Analysis for Parallel Communication Systems}
\label{sec:WFparallelsystem}

MISO, SIMO, and MIMO are prevalent architectures in wireless networks, known for enhancing system performance, capacity, and reliability. This section introduces the concept of Waste Factor for paralleled systems and outlines a comprehensive methodology for evaluating the power efficiency of MISO, SIMO, and MIMO configurations. By analyzing the efficiencies and gains of individual components within these paralleled chains, we establish a generalized theory to compute the overall \(W\) of these systems. The analysis provided herein expands the utility of Waste Factor for analyzing power efficiency in complex multi-antenna and multi-user system and network architectures.

\subsection{W for Multiple-Input Single-Output (MISO) System}
\label{sec:MISO}

To investigate the MISO structure, we first consider a configuration consisting of two transmitters (TXs) and one receiver (RX). In this setup, each TX communicates with the RX via uncorrelated channels. These channels are assumed to be zero-mean additive white Gaussian noise (AWGN).

\subsubsection{\textit{W} for non-coherent combining MISO}

\begin{figure}[!t]
    \centering
    \includegraphics[width=0.9\columnwidth]{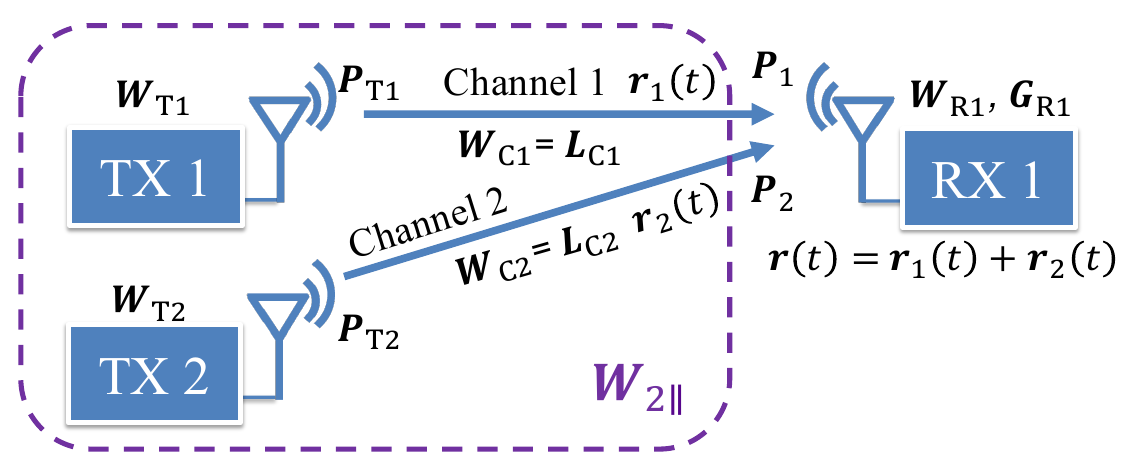}
    \caption{A two TX and one RX MISO communication system.}
    \label{figp1}
\end{figure}

In Fig.~\ref{figp1}, TX1 is characterized by a Waste Factor \(W_\text{T1}\), operating over a channel with Waste Factor \(W_{\text{C1}} = L_\text{C1} = 1/G_\text{C1}\), (see \cite{Rappaport2024Microwave} and Section~\ref{sec:RAN_intro}-G)) and delivering received power \(P_1\) into RX1, which has its own Waste Factor \(W_\text{R1}\) and receiver gain \(G_\text{R1}\). Similarly, TX2 has Waste Factor $W_\text{T2}$ and transmits to RX1 with channel loss of $L_\text{C2}= 1/G_\text{C2}$ which yields $W_\text{C2}$ and received power \(P_2\) to RX1. According to the results of a lossy channel study in Section~\ref{sec:RAN_intro} and \cite{Rappaport2024Microwave}, the Waste Factor \(W\) for each of these two parallel systems applied to RX1 can be considered. When assuming non-coherent combining at the input of RX1, as discussed in \cite{Shu2014Beamcomb}, the received powers \(P_1\) and \(P_2\) are combined in the input to the last stage of the MISO system (shown as a dashed box in Fig. \ref{figp1}), where the Waste Factor of the system just before the last stage of the MISO system (e.g., just before the receiver in Fig. \ref{figp1}) is defined by a single Figure of Merit \(W_{\text{2}\parallel}\). For two received signals \(r_1(t)\) and \(r_2(t)\) with powers \(P_1\) and \(P_2\) respectively, non-coherent combining involves no phase information and assumes that the phases of incoming signals from each transmitter are uniformly and identically distributed (i.i.d.). Therefore, the combined receiver power average in non-coherent combining at the RX antenna is simply the sum of individual powers: \(P_\text{signal,noncoh} = P_1 + P_2\) \cite{Shu2014Beamcomb}. Then, using eqs. \eqref{defW}, \eqref{d6}, and \eqref{Wcas}, the Waste Factor for a MISO system is given by
\begin{align}
    \label{p1}
    W_{\text{2}\mathbin{\parallel}}^\text{noncoh} 
    &= \frac{P_\text{consumed,path}}{P_\text{signal,noncoh}}  \nonumber\\
    &= \frac{{P_{1} \left(W_\text{C1} +\frac{W_\text{T1}-1}{G_\text{C1}} \right)}+ {P_{2} \left(W_\text{C2} +\frac{W_\text{T2}-1}{G_\text{C2}} \right)}}{P_{1} + P_{2}}.
\end{align}

Assuming the power received from each transmitter (TX) is proportional to a certain power \(P\), i.e., \(P=P_1\gamma_1\) and \(P=P_2\gamma_2\), the received powers \(P_1\) and \(P_2\) can be expressed as \( [P_1, P_2] = P[\gamma_1, \gamma_2] \). Consequently, eq. \eqref{p1} can be simplified without  dependence on particular power levels, requiring only the ratios of received power levels of the two channels where the largest $\gamma$, say $\gamma_1$, may be arbitrarily set to unity and the smaller $\gamma$ may be normalized to $\gamma_2/\gamma_1$
\begin{equation}
    W_{\text{2}\mathbin{\parallel}}^\text{noncoh} = \frac{{\gamma_1 \left(W_\text{C1} +\frac{W_\text{T1}-1}{G_\text{C1}} \right)}+ {\gamma_2 \left(W_\text{C2} +\frac{W_\text{T2}-1}{G_\text{C2}} \right)}}{\gamma_1 + \gamma_2}.
    \label{W2par}
\end{equation}

\begin{figure}[!t]
    \centering
    \includegraphics[width=0.9\columnwidth]{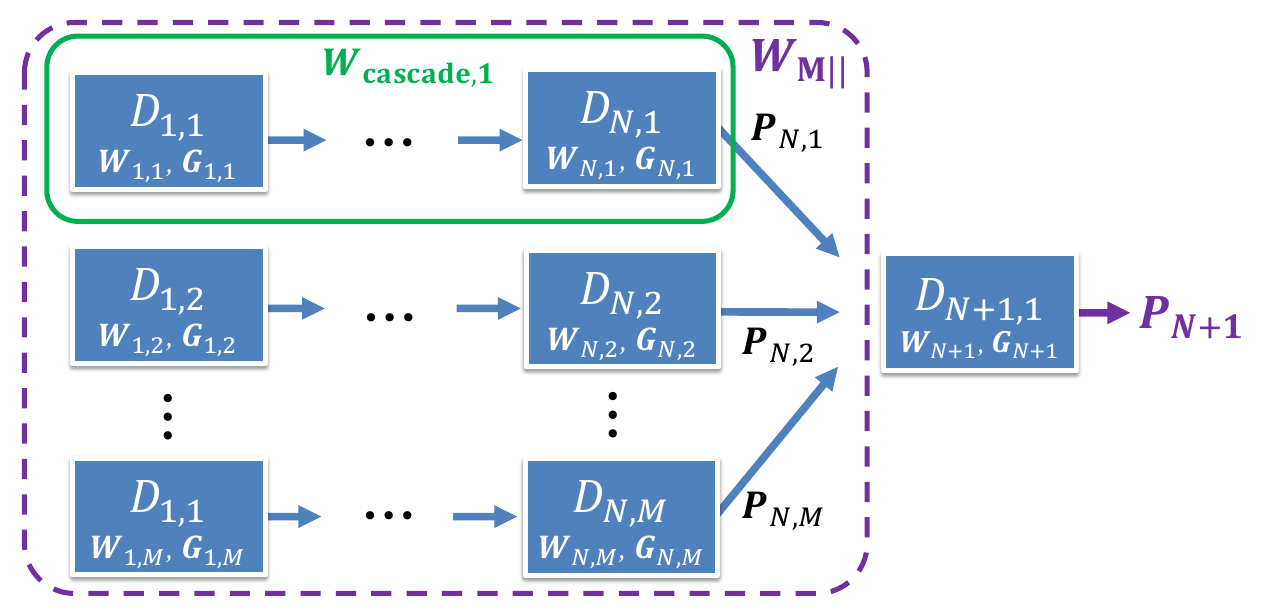}
    \caption{A MISO system with M paralleled input cascade.}
    \label{fig:MISO_M}
\end{figure}

In Fig.~\ref{fig:MISO_M}, we generalize the previous analysis and we consider a system consisting of \(M\) parallel cascades, where each cascade includes \(N\) devices. The output signal power from these devices is combined at device \(N+1\) using non-coherent combining methods. To determine Waste Factor for the entire system with a single output node, we initially compute the \(W\) for each individual cascade \(m\) using eq.~\eqref{Wcas}
\begin{equation}\label{p5}
    W_{\text{cascade},m} = W_{N,m} + \frac{(W_{N-1,m}-1)}{G_{N,m}} + \cdots + \frac{(W_{1,m}-1)}{\prod_{i=2}^N G_{i,m}}.
\end{equation}

Subsequently, Waste Factor (\(W\)) for the entire paralleled system, when using non-coherent combining at the \((N+1)^\text{th}\) device, is given by
\begin{equation}
\label{W_par_inco}
    W_{\text{M}\mathbin{\parallel}}^\text{noncoh} = \frac{\sum_{i=1}^{\text{M}} \left({P_{N,i} \times W_{\text{cascade},i}}\right)}{\sum_{i=1}^{\text{M}} {P_{N,i}}}. 
\end{equation}

As done in \eqref{W2par}, here in \eqref{W_par_inco}, we assume that the power output from each cascade is represented by \([P_{N,1}, P_{N,2}, \ldots, P_{N,M}] = P[\gamma_1, \gamma_2, \ldots, \gamma_M]\), where \(P\) is a constant and \(\gamma_i\) are scaling factors illustrating the relative contribution of each cascade to some constant (e.g., the total received power). It is sensible to normalize all of the received powers to the maximum power level in the strongest path, such that $\gamma_{i}$ will be less or equal to 1 for all $i$ and not dependent on actual received power levels, thus we have \([P_{N,1}, P_{N,2}, \ldots, P_{N,M}] = P_{\max}[\gamma_1, \gamma_2, \ldots, \gamma_M]\). The equation for the Waste Factor for non-coherent combining at the \((N+1)^\text{th}\) device at the terminus of all parallel paths in Fig. \ref{fig:MISO_M} (e.g. the input into the final MISO stage) follows from \eqref{W2par} and is given by \eqref{W_par_inco} 
\begin{equation}
\label{W_par_inco_gamma}
    W_{\text{M}\mathbin{\parallel}}^\text{noncoh}  = \frac{\sum_{i=1}^{\text{M}} {\gamma_{i} W_{\text{cascade},i}}}{ \sum_{i=1}^{\text{M}} {\gamma_{i}} }.
\end{equation}

\subsubsection{\textit{W} for coherent combining MISO}

For coherent combining, the receiver must be phase-synchronized with the signals \( r_1(t) \) and \( r_2(t) \), each signal combining with identical phases for maximum power combining. If \( r_1(t) \) and \( r_2(t) \) are phase-aligned and their respective powers to the input of a cascade are \( P_1 \) and \( P_2 \) as shown in Fig. \ref{figp1}, the combined signal power average at the RX antenna, denoted as \( P_\text{signal,coh} \), is calculated by the square of the magnitude of the vector sum of the two signals \cite{Shu2014Beamcomb}
\begin{equation}
     P_\text{signal,coh} = \left|\sqrt{P_1} + \sqrt{P_2}\right|^2.
     \label{p3}
\end{equation}

Considering coherent combining in a $M$ TX single RX MISO system, we define the power output of the \(i^\text{th}\) TX as \(P_{\text{T}i}\), and the received power at the input of the RX from TX \(i\) before combining as \(P_i\). After coherent combining at the RX antenna, the resulting average power is denoted as \(P_\text{signal,coh}\). As in the previous derivations, we assume \([P_1, P_2, \ldots, P_M] = P_{\max}[\gamma_1, \gamma_2, \ldots, \gamma_M]\), i.e., that the received powers from each TX are proportional to the maximum power level (\(P_{\max}\)) in the strongest path.

Using eq. \eqref{defW} and \eqref{p3}, and following an approach similar to eq. \eqref{W2par}, \textit{W} for this coherent combining MISO (example in Fig. \ref{figp1}) can be expressed as
\begin{equation}
\label{p4}
    W_{\text{M}\mathbin{\parallel}}^\text{coh} = \frac{\sum_{i=1}^{M} \left[ {\gamma_{i} \left(W_{\text{C}i} +\frac{W_{\text{T}i}-1}{G_{\text{C}i}} \right)} \right]}{ \left|\sum_{i=1}^{M} {\sqrt{\gamma_i} } \right|^2 }.
\end{equation}

Based on \eqref{p4}, consider a general coherently combined MISO system in Fig.~\ref{fig:MISO_M}. It is readily seen that \textit{W} for the entire paralleled system using coherent combining is
\begin{equation}
\label{W_par_co}
    W_{\text{M}\mathbin{\parallel}}^\text{coh} = \frac{\sum_{i=1}^{M} \left({P_{N,i} \times W_{\text{cascade},i}}\right)}{\left|\sum_{i=1}^{M} {\sqrt{P_{N,i}} } \right|^2 } 
    = \frac{\sum_{i=1}^{M} \left( {\gamma_{i} W_{\text{cascade},i}} \right)}{ \left|\sum_{i=1}^{M} {\sqrt{\gamma_i} } \right|^2 }.
\end{equation}

 We see for MISO in general, using $W_{\text{M}\mathbin{\parallel}}$ for either non-coherent \eqref{W_par_inco_gamma} or coherent \eqref{W_par_co} combining, we can calculate \textit{W} of a MISO system shown in Fig.~\ref{fig:MISO_M} as
\begin{equation}
\label{WMISO}
    W_{\text{MISO}} =  W_{N+1}  + \frac{(W_{\text{M}\mathbin{\parallel}} - 1)}{G_{N+1}},
\end{equation}
where $W_{N+1}$ and $G_{N+1}$ represents the Waste Factor and gain of the ${(N+1)}^\text{th}$ single output device which is the stage following the terminus of parallel input cascades in Fig.~\ref{fig:MISO_M}.
It should be clear that MISO systems using equal gain combining may be computed following the approach detailed above.

\subsection{W for Single-Input Multiple-Output (SIMO) System}
\label{sec:SIMO}

\begin{figure}[!t]
    \centering
    \includegraphics[width=0.9\columnwidth]{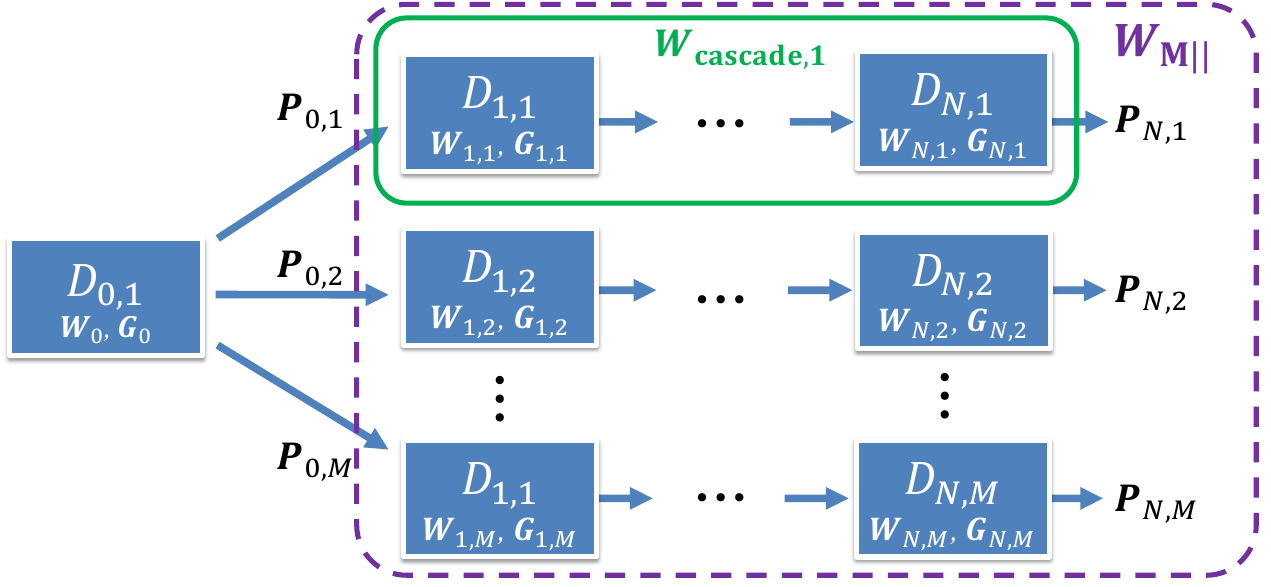}
    \caption{A SIMO system with M paralleled output cascade.}
    \label{fig:SIMO}
\end{figure}

Consider a SIMO system as depicted in Fig.~\ref{fig:SIMO}.
In this setup, the Waste Factors of each parallel cascade are individually computed using eq.~\eqref{Wcas} and are not a function of the applied power levels as in the previous section. Since \textit{W} is a function of the output, we need to consider the \textit{W} of each parallel cascade. If the parallel paths are to be combined at a subsequent stage (to the right of the parallel structures) beyond the depicted configuration in Fig.~\ref{fig:SIMO}, one can apply the MISO combining techniques outlined in Section III-\ref{sec:MISO}. 

Additionally, this analysis exploits the fundamental Waste Factor theory given in Section ~\ref{sec:WFcascadedsystem}, and quantifies the cumulative impact of Waste Factors when cascades are combined using MISO or SIMO structures. \textit{W} for any device remains constant if each component operates within its linear range, regardless of the transceiver implementation. For instance, in OFDM systems, linearity is ensured through backoff to prevent distortion. The linearity of each component is also required for the Waste Factor theory to be applied to single carrier and carrier aggregation scenarios, which involve specific hardware components in the cascade to determine overall system energy efficiency.

\subsection{W for Multiple-Input Multiple-Output (MIMO) System}
\label{sec:MIMO}

\subsubsection{\textit{W} for non-coherent combining MIMO}

\begin{figure}[!t]
    \centering
    \includegraphics[width=0.9\columnwidth]{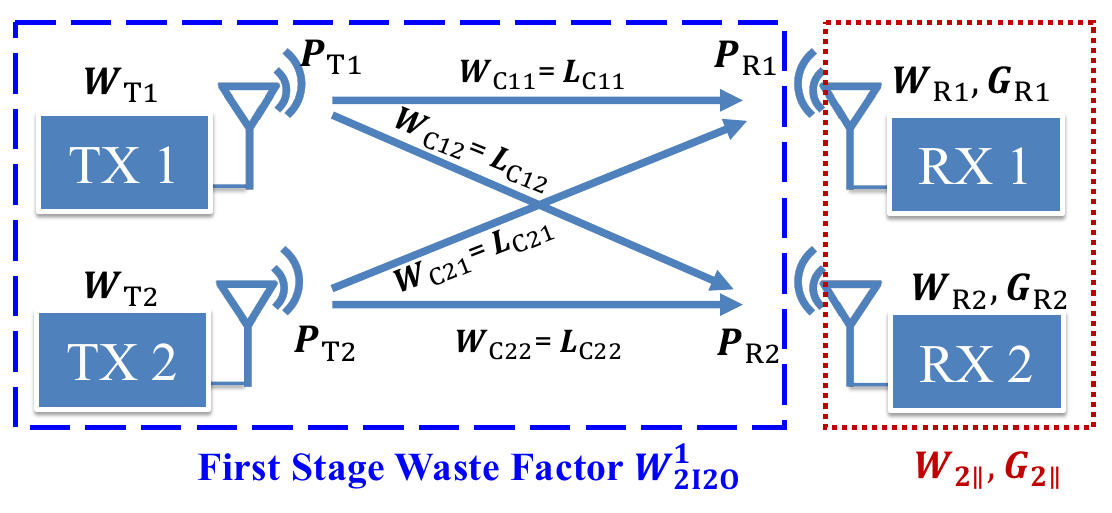}
    \caption{A two-input and two-output MIMO system.}
    \label{fig:MIMO}
\end{figure}

For a 2-Input 2-Output (2I2O) MIMO system in Fig.~\ref{fig:MIMO}, the received average power of each RX using non-coherent combining is
\begin{equation}
\label{PRX_noncoh}
\begin{bmatrix}
P_\text{R1}^\text{noncoh} & P_\text{R2}^\text{noncoh}
\end{bmatrix}
= \begin{bmatrix}
P_\text{T1} & P_\text{T2}
\end{bmatrix}
\begin{bmatrix}
W_\text{C11}^{-1} & W_\text{C12}^{-1} \\
W_\text{C21}^{-1} & W_\text{C22}^{-1}
\end{bmatrix},
\end{equation}
where \(P_{\text{R}i}^\text{noncoh}\) represents the power received at the antenna of the $i^\text{th}$ RX when signals are combined non-coherently. \(P_{\text{T}i}\) denotes the power transmitted by the $i^\text{th}$ TX. In \eqref{PRX_noncoh}, we compute the received powers of each RX cascade as if they are two separate MISO systems. \(W_{\text{C}ij}\) is the Waste Factor of the channel between the $i^\text{th}$ TX and the $j^\text{th}$ RX. The total signal-path power consumption of the system before the receiver (e.g., the first stage, which considers only the TXs and the channels) using non-coherent combining from eq.~\eqref{defW} and combining the consumed power of the two MISO systems is
\begin{equation}
P_\text{consumed,path}^\text{noncoh} = \sum_{i=1}^{2} { \left({P}_{\text{R}i}^\text{noncoh} W_{2\parallel}^\text{noncoh} \right) },
\end{equation}
where \(W_{2\parallel}^{\text{noncoh}}\) represents Waste Factor for a 2-TX parallel MISO system together with the channel, as defined in \eqref{W2par} in Section III-\ref{sec:MISO}.

Prior to the power entering the receivers, we define the first stage Waste Factor of the 2I2O system using non-coherent combining in the left-most (first stage) dashed box of Fig.~\ref{fig:MIMO} by applying the basic definition of \textit{W} as given in \eqref{W_par_inco}
\begin{equation}
\label{W2I2O_1st}
W_{\text{2I2O}}^\text{1,noncoh} = \frac{\sum_{i=1}^{2} { \left({P}_{\text{R}i}^\text{noncoh} W_{2\parallel}^\text{noncoh} \right) }}{\sum_{i=1}^{2} { {P}_{\text{R}i}^\text{noncoh} }}.
\end{equation}

Letting $[P_\text{R1}^\text{noncoh},P_\text{R2}^\text{noncoh}]=P^\text{noncoh}[\gamma_1,\gamma_2]$, eq.~\eqref{W2I2O_1st} can be written as
\begin{equation}
\label{2I2O_W_nonco_gamma}
W_{\text{2I2O}}^{1,\text{noncoh}} = \frac{\sum_{i=1}^{2} { \left({\gamma}_{i} W_{2\parallel}^\text{noncoh} \right) }}{\sum_{i=1}^{2} { \left({\gamma}_{i} \right) }}.
\end{equation}

To capture the entire power efficiency of the system, including the receivers, the complete Waste Factor for a 2I2O system using non-coherent combining at the input of the receivers (\(W_{\text{2I2O}}^\text{noncoh}\)) is calculated by cascading the first-stage \(W_{\text{2I2O}}^{1,\text{noncoh}}\) and the multiple receiver stage, which is the right-most (second stage) of the system in Fig. \ref{fig:MIMO}. The Waste Figures for each RX are computed separately, and the outputs of the MIMO receivers (not shown) may then be combined coherently or non-coherently to a further single-path stage to the right by using MISO formulations in Section III-\ref{sec:MISO} or kept separately as parallel paths to parallel sinks, with each cascaded path having its own system Waste Factor.  
\begin{equation}
\label{W2I2O_noncoh}
W_{\text{2I2O}}^\text{noncoh} = W_{\text{2}\mathbin{\parallel}} + \frac{(W_{\text{2I2O}}^{1,\text{noncoh}} - 1)}{G_{\text{2}\mathbin{\parallel}}}.
\end{equation}

In \eqref{W2I2O_noncoh}, the Waste Factor for the parallel receivers is given by $W_{\text{2}\mathbin{\parallel}}$, where $W_{\text{2}\mathbin{\parallel}}$ is at the parallel receiver side, based on \eqref{W_par_inco_gamma}, and the gain of the parallel receiver is given by $G_{\text{2}\mathbin{\parallel}}=\frac{ \sum_{i=1}^{\text{2}} P_{\text{R}i} G_{\text{R}i}}{\sum_{i=1}^{\text{2}} P_{\text{R}i}}$ if we consider non-coherent combining at the outputs of the MIMO receivers, or $G_{\text{2}\mathbin{\parallel}}=\frac{ \left|\sum_{i=1}^{\text{2}} {\sqrt{P_{\text{R}i} G_{\text{R}i}} } \right|^2 }{\sum_{i=1}^{\text{2}} P_{\text{R}i}}$ when we assume coherent combining at the outputs of the MIMO receivers.

\subsubsection{\textit{W} for coherent combining MIMO}

The average received power of each RX input using coherent combining can be described as follows
{\small
\begin{equation}
\label{PRX_coh}
\begin{bmatrix}
P_{\text{R}1}^\text{coh} & P_{\text{R}2}^\text{coh}
\end{bmatrix}
= \left(\begin{bmatrix}
\sqrt{P_{\text{T}1}} & \sqrt{P_{\text{T}2}}
\end{bmatrix}
\begin{bmatrix}
\sqrt{W_{\text{C}11}^{-1}} & \sqrt{W_{\text{C}12}^{-1}} \\
\sqrt{W_{\text{C}21}^{-1}} & \sqrt{W_{\text{C}22}^{-1}}
\end{bmatrix}\right) ^{2},
\end{equation}
}
where the square indicates an element-wise operation. 

The total power consumed by the signal path in the left-most dashed system (first stage) depicted in Fig. \ref{fig:MIMO} is calculated before applying the coherently combined received power to the MIMO RXs, and is given by
\begin{equation}
P_\text{consumed,path}^\text{coh} = \sum_{i=1}^{2} { \left({P}_{\text{R}i}^\text{coh} W_{2\parallel}^\text{coh} \right) },
\end{equation}
where \(P_{\text{R}i}^{\text{coh}}\) is the coherently combined power at the antenna of the $i^\text{th}$ RX, and \(W_{2\parallel}^{\text{coh}}\) represents the Waste Factor for a 2-TX paralleled MISO system together with the channel, which is the same as \eqref{p4} with \(M = 2\). Before the power enters the receivers, we define the first-stage Waste Factor of the 2I2O system using coherent combining
\begin{equation}
W_{\text{2I2O}}^{1,\text{coh}} = \frac{\sum_{i=1}^{2} { \left({P}_{\text{R}i}^\text{coh} W_{2\parallel}^\text{coh} \right) }}{\sum_{i=1}^{2} { {P}_{\text{R}i}^\text{coh} }}.
\label{W2I2O_1st_coh}
\end{equation}
Letting $[P_{\text{R}1}^\text{coh},P_{\text{R}2}^\text{coh}]=P^\text{coh}[\gamma_1,\gamma_2]$, where \eqref{W2I2O_1st_coh} can be written as
\begin{equation}
W_{\text{2I2O}}^{1,\text{coh}} = \frac{\sum_{i=1}^{2} { \left({\gamma}_{i} W_{2\parallel}^\text{coh} \right) }}{\sum_{i=1}^{2} { {\gamma}_{i} }}.
\label{W2I2O_1st_coh_gamma}
\end{equation}

To capture the entire power efficiency of the system, including the receivers, the complete Waste Factor for a 2I2O system using coherent combining ($W_{\text{2I2O}}^\text{coh}$) is calculated by cascading the first stage $W_{\text{2I2O}}^{1,\text{coh}}$ and the Waste Factor of the parallel receivers shown in the second (right-most) stage.  Each receiver in the rightmost stage has its own Waste Factor, and eventually, the parallel chains may be combined coherently or non-coherently to a single path in a later stage to the right or a sink (not shown) whereby the Waste Factor for the 2I2O system is given by
\begin{equation}
W_{\text{2I2O}}^\text{coh} = W_{\text{2}\mathbin{\parallel}} + \frac{(W_{\text{2I2O}}^{1,\text{coh}} - 1)}{G_{\text{2}\mathbin{\parallel}}}.
\label{W2I2O_coh}
\end{equation}

\subsubsection{\textit{W} for general MIMO system}
\begin{figure}[!t]
    \centering
    \includegraphics[width=0.9\columnwidth]{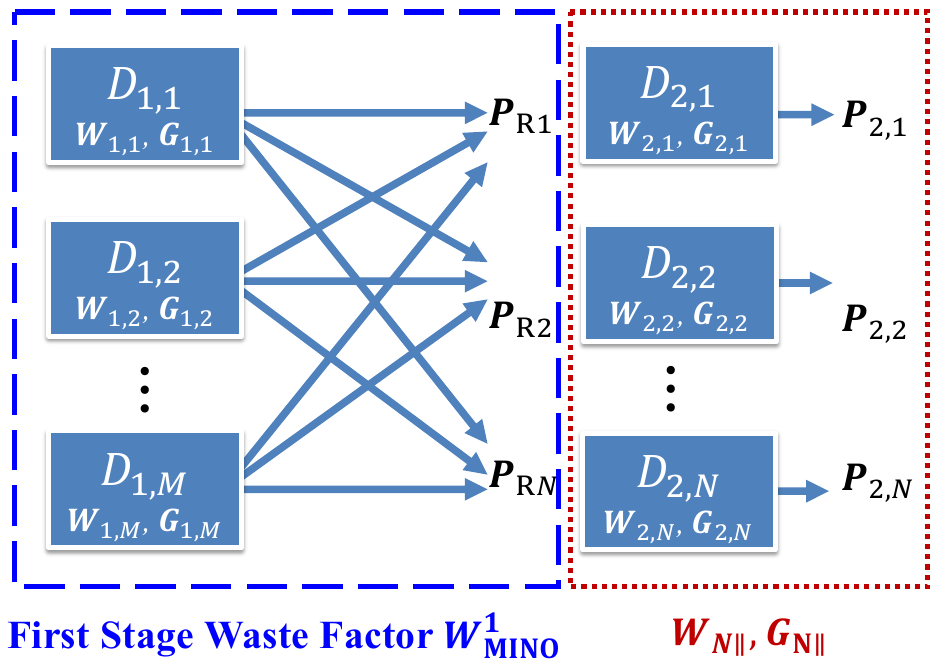}
    \caption{An M-input and N-output MIMO system.}
    \label{fig:MINO}
\end{figure}

In Fig. \ref{fig:MINO}, we show an $M$-Input $N$-Output general MIMO (e.g., MINO) system. If we assume proportional combined powers at the input of each device in the right-most dotted box, e.g., such as in \eqref{2I2O_W_nonco_gamma} or \eqref{W2I2O_1st_coh_gamma}, the input power average of each device can be denoted as
\begin{equation} 
[P_{\text{R}1},P_{\text{R}2},\ldots,P_{\text{R}M} ] = P[\gamma_1,\gamma_2,\ldots,\gamma_{M} ],
\end{equation}
then \eqref{2I2O_W_nonco_gamma} and \eqref{W2I2O_1st_coh_gamma} can be extended to
\begin{equation}
W_{\text{MINO}}^{1} = \frac{\sum_{i=1}^{N} { \left({P}_{\text{R}i} W_{\text{M}\parallel} \right) }}{\sum_{i=1}^{N} { \left({P}_{\text{R}i} \right) }} = \frac{\sum_{i=1}^{N} { \left({\gamma}_{i} W_{\text{M}\parallel} \right) }}{\sum_{i=1}^{N} { \left({\gamma}_{i} \right) }},
\label{WMIMO_1st}
\end{equation}
where $W_{\text{MINO}}^{1}$ is the first stage Waste Factor for the left-most dashed box consisting of $M$ parallel device from $D_{1,1}$ to $D_{1,M}$ and channels. 
\(W_{\text{M}\parallel}\) is used to calculate the power consumption of the left-most parallel system based on a single input of $D_{2,i}$, as shown in Fig. \ref{fig:MINO}. This calculation is based on the input power from a device located in the right-most red-dotted box. The specific formula used depends on the method of MISO signal combining and is referenced as either \eqref{W_par_co} for coherent combining or \eqref{W_par_inco_gamma} for non-coherent combining.

Based on \eqref{WMIMO_1st}, $W$ for the generalized $M$-Input $N$-Output MINO system terminating into a single path to the right of the RXs (not shown) is given by
\begin{equation}
\label{W_MIMO_general}
W_{\text{MINO}} =  W_{\text{N}\mathbin{\parallel}}  + \frac{(W_{\text{MINO}}^{1} - 1)}{G_{\text{N}\mathbin{\parallel}}},
\end{equation}
where $G_{\text{N}\mathbin{\parallel}}$ is the gain of the parallel RXs, which can be calculated based on the ratio of the total output power of parallel RXs to the input power of parallel RXs assume non-coherent or coherent combining to the right of the RXs (not shown)
\begin{equation}
\label{gain_co_nonco}
    G_{\text{N}\mathbin{\parallel}} = 
        \begin{cases}
        \frac{ \sum_{i=1}^{N} P_{\text{R}i} G_{\text{R}i}}{\sum_{i=1}^{N} P_{\text{R}i}} & \textrm{non-coherent combining}, \\
        \frac{ \left|\sum_{i=1}^{N} {\sqrt{P_{\text{R}i} G_{\text{R}i}} } \right|^2 }{\sum_{i=1}^{N} P_{\text{R}i}} & \textrm{coherent combining}.
        \end{cases}
\end{equation}

\subsection{Examples of Waste Factor in practical systems}

Several practical applications of Waste Factor are detailed in \cite{Rappaport2024Microwave}, where calculations and usage of $W$ are demonstrated for making energy-efficient decisions and to accurately determine power consumption in circuits, systems, channels, and data centers. The exploration of $W$ in MIMO structures notably broadens its applicability, particularly in advanced systems like emerging cell-free massive MIMO and ultra-dense networks. When the input and output structures of these networks are well-defined, applying the Waste Factor is straightforward by referring to the method proposed in Section~\ref{sec:WFparallelsystem}-A, B, and C. This method can be applied in general to any wireless or wired communication systems with parallel structures, as discussed in Section~\ref{sec:RAN_intro}.

In the following sections, we apply the Waste Factor theory for cascaded systems (presented in Section~\ref{sec:WFcascadedsystem}) and the Waste Factor theory for parallel systems (derived in Section~\ref{sec:WFparallelsystem}-A, B, and C) to the \ac{RAN} scenario. In particular, Section~\ref{sec:RAN_intro} introduces the \ac{RAN} architecture and derives the Waste Factor for the main cascaded components in any RAN. Section~\ref{sec:Analysis_RAN} provides analysis and simulation results of a RAN consisting of distributed MU-MIMO networks, showcasing the benefits of applying the Waste Factor theory to assess and enhance the energy efficiency of \acp{RAN} through real-world system-based power and energy consumption models.

\section{Radio Access Network (RAN) Analysis Using Waste Factor}
\label{sec:RAN_intro}

Given the rising energy demands of the RAN, it becomes imperative to develop and implement a robust metric for measuring energy efficiency. In response to the need for a new sustainability metric, this section applies \textit{W} as a novel and effective approach for analyzing energy utilization within the RAN, and shows why a new metric such as \textit{W} is needed urgently.

\subsection{RAN Architecture}
\begin{figure*} []
\centering
\includegraphics[width=16cm]{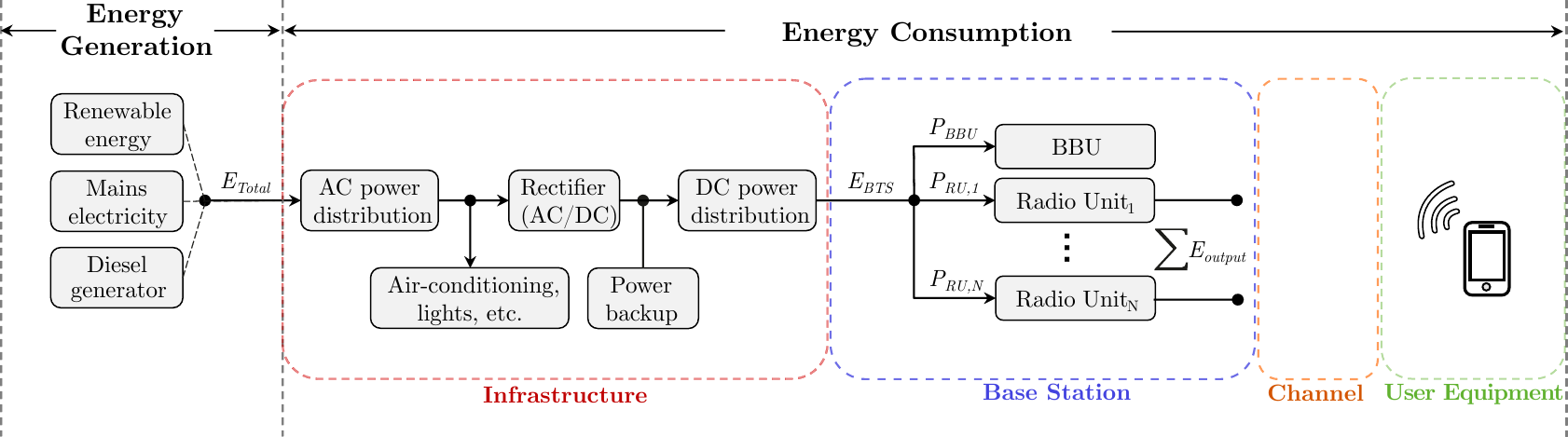}
\caption{Illustration of a typical RAN site, and energy flow from main AC input to reception at the user equipment.}
\label{fig:big_scenario}
\end{figure*}

A \ac{RAN} includes three principal types of equipment: the \textit{site infrastructure}, the \textit{\ac{BS}}, and the \textit{\acp{UE}}. A fourth component is the radio channel between each BS and UE. The site infrastructure generally includes rectifiers to convert AC to DC power, power backup systems such as batteries, and cooling mechanisms like air conditioners. The base station consists of the baseband and radio units, which can be co-located or distributed within the RAN, and it is in charge of facilitating the transmission of data to \acp{UE} through the air interface (using the propagation channel).

An illustrative breakdown of the RAN setup and the associated equipment is provided in Fig.~\ref{fig:big_scenario}, which also delineates the path of energy flow, starting from the AC power input from the electrical grid, proceeding through DC power conversion, and culminating in the delivery of energy to the main equipment for subsequent transmission over the air interface to user equipment. 

It should be noted that since infrastructure components do not directly contribute to the signal path, our analysis in this section will primarily concentrate on signal-path power, including the power consumption of \ac{RU} and the \ac{UE}.

\begin{figure*} []
\centering
\includegraphics[width=16cm]{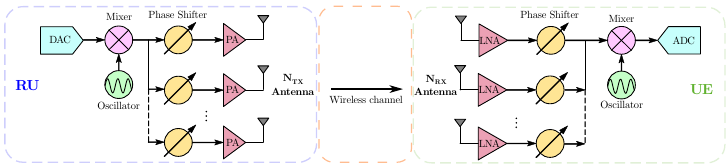}
\caption{General structure of the RU and UE.}
\label{fig:TXRXStructure}
\end{figure*}

Fig. \ref{fig:TXRXStructure} provides a focus to our reference scenario, which is composed of the transmitting \ac{RU}, the wireless channel, and the receiving \ac{UE}. The essential components responsible for signal transmission at the \ac{RU} and signal reception at the UE are also depicted. 

As can be seen in Fig. \ref{fig:TXRXStructure}, the \ac{RU} comprises a \ac{DAC}, Mixer, \ac{LO}, \ac{PS}, \ac{PA}, and Antenna, while the \ac{UE} receives the signal and inverts the transmission process of the RU using
a \ac{LNA} at the RX front end instead of a TX \ac{PA}, to amplify signals without elevating noise above the signal level. 
The transition from analog to digital signals in the receiver is executed by an \ac{ADC}. 

As described in \cite{Habibi2019RAN}, a typical RAN site includes a variety of equipment essential for mobile communications. It is important to recognize that RAN architecture can vary significantly, ranging from traditional setups to more modern configurations like Cloud RAN (C-RAN), Distributed RAN (D-RAN), Virtualized RAN (vRAN), and Open RAN (O-RAN). Each of these architectures offers unique benefits and challenges in terms of deployment and energy consumption.\\
A key observation is that all RAN architectures, and in fact, virtually all practical communication systems, require linear signal replication over huge orders of magnitude of power levels between a source to sink so that signal fidelity and integrity are preserved and so that a receiver may differentiate between digital waveforms. The Waste Factor, \textit{W}, provides a universal metric capable of assessing total power consumption across all forms of RAN architectures. By applying \textit{W}, network operators can evaluate and compare the energy efficiency of different RAN setups under a standardized framework, enabling targeted improvements and optimizations. Moreover, \textit{W} is particularly valuable in identifying and quantifying the most energy-hungry components or processes within each RAN structure, guiding the development of more sustainable network solutions.

\subsection{RAN Energy Efficiency Metrics}
\label{sec:EEmetrics}

\begin{table}[]
    \caption{Energy efficiency metrics by different standard working groups}
    \label{tab:EEmetrics}
    \begin{tblr}{
  colspec = {X[c]X[c]X[c]},
  stretch = 0,
  rowsep = 6pt,
  hlines = {black},
  vlines = {black},
}
\textbf{Standard Working Group} & \textbf{Document ID} & \textbf{Metric} \\
ETSI EE & ES 203 228 \cite{etsiES203228} & $\displaystyle\text{EE}_\text{ntw} = \frac{U_\text{output}}{E_\text{Network}}$   \\
ITU-T SG5 & L.1350 \cite{ituL13502016} & $\displaystyle\text{EE}_\text{site} = \frac{E_\text{BS}}{E_\text{site}}$ \vspace{-0.1cm} \\ 
ETSI EE & TS 103 786 \cite{etsiTS103786} & $\displaystyle\text{EE}_\text{BS} = \frac{\text{DV}_\text{BS}}{E_\text{BS}}$ \vspace{-0.1cm} \\ 
ITU-T SG5 & L.1310 \cite{ituL13102020} & $\displaystyle\text{EE}_\text{RF} = \frac{E_\text{output}}{E_\text{RF}}$\vspace{-0.1cm} \\ 
3GPP SA5 & TR 28.813 \cite{3gpp28813} & Specify target network useful output, $U_\text{output}$, for new scenarios (RAN-sharing, network slicing, etc). \\ 
\end{tblr}
\end{table}

In this section, we overview the previous work carried out by global standard bodies in defining energy efficiency metrics for the RAN both as a cascaded system and for individual RAN components along the cascade. 

To evaluate the energy-related performance of the \ac{RAN}, the mobile industry has defined several measurement methods and metrics for different network levels (network \cite{etsiES203228}, site \cite{ituL13502016}, \ac{BS} \cite{etsiTS103786}, and UE \cite{3gpp38840}) as well for different scenarios (dense urban, urban, and rural coverage) \cite{etsiES203228}, and services (enhanced mobile broadband, ultra-reliable low-latency communications, and massive machine type communications) \cite{3gpp28813}. Importantly, current standards define distinct and often decoupled approaches to compute the \ac{EE} at the component level, at the equipment level, and at the network level. The \ac{ITU} study group 5 (SG5) has defined \ac{RU} energy efficiency, $\mathrm{EE}_{\text{RU}}$, to evaluate the performance at \ac{RU} level as follows \cite{ituL13102020}
\begin{equation}
    \mathrm{EE}_{\text{RU}}= \frac{E_{\text{output}}}{E_{\text{RU}}},
    \label{eq:EE_RF}
\end{equation}
where $E_{\text{output}}$ is the \ac{RU} output energy consumption in units of Watt-hours [Wh] and $E_{\text{RU}}$ is the \ac{RU} energy consumption in units of [Wh].
The \ac{ETSI} Environmental Engineering has defined in \cite{etsiTS103786} the \ac{EE} of a \ac{BS}, $\mathrm{EE}_{\text{BS}}$, as follows
\begin{equation}
    \mathrm{EE}_{\text{BS}}= \frac{DV_{\text{BS}}}{E_{\text{BS}}},
    \label{eq:ETSI_EE_BS}
\end{equation}
where $DV_{\text{BS}}$ denotes the amount of data volume successfully transmitted by the BS in a given amount of time and $E_{\text{BS}}$ is the related energy consumption. ITU SG5 has defined the BS site EE as follows \cite{ituL13502016}
\begin{equation}
    \mathrm{EE}_{\text{site}}= \frac{E_{\text{BS}}}{E_{\text{site}}},
\end{equation}
where $E_{\text{site}}$ denotes the site overall energy consumption, including the energy \textit{losses} due to air-conditioning, rectifier, and power distribution.
Finally, \ac{ETSI} Environmental Engineering has introduced the network \ac{EE}  \cite{etsiES203228}
as follows
\begin{equation}
    \mathrm{EE}_{\text{ntw}}= \frac{U_{\text{output}}}{E_{\text{network}}},
    \label{eq:EE_ntw}
\end{equation}
where $E_{\text{network}}$ is the overall \ac{RAN} energy consumption, and $U_{\text{output}}$ denotes a target network useful output, such as one of data volume (e.g. for enhanced mobile broadband), coverage area (e.g. for massive machine type communications), latency (e.g. for ultra-reliable low-latency communications).
Furthermore, the 3GPP SA5 group has specified~\cite{3gpp28813} target metrics for use in particular scenarios, including RAN-sharing and network slicing.

Table~\ref{tab:EEmetrics} presents an overview of the diverse energy efficiency metrics as defined by various standardization bodies.

Importantly, as we demonstrate in Section~\ref{sec:RAN_intro} and \ref{sec:Analysis_RAN}, all of these widely adopted standardized energy efficiency metrics suffer a major issue that prevents them from being effectively used to assess and compare the energy efficiency of the RAN and the impact of components in transmitting or receiving data.

\subsection{Problems with the Standard RAN EE Metrics}
\label{sec:problemsEE}

Although the effort to define EE metrics highlights the importance of the sustainability issue in the industrial community, some notable problems can be identified in the current methodologies.
First, there is a lack of a standard metric that can used in different scenarios and equipment. Second, \ac{EE} is defined as the ratio of two terms often depending on each other, which makes comparison and explainability limited. 

For example, comparing two BSs operating under different traffic conditions can reveal deficiencies in the standard EE metric defined in eq.~(\ref{eq:ETSI_EE_BS}). This metric is often employed to identify the most energy-efficient BSs in a network, for instance, to facilitate user transfers to such BSs through network load balancing, thus achieving energy savings~\cite{dedomenico2023modeling}. However, an incorrect assessment of energy efficiency can lead to suboptimal optimization decisions and wasted energy.

To illustrate the problem with the ETSI standard EE metric, consider a numerical comparison between two BSs using existing metrics defined by the standards \cite{etsiTS103786}, but now using the Waste Factor approach of power superposition:

\begin{itemize}
\item BS-A has a non-path power consumption of $P_{\text{non-pathA}}=50\;\text{Wh}$ and consumes 2~Wh to transmit 1~GB of data, i.e., $P_{\text{consumed,pathA}}=2 \text{Wh/GB}$. The data volume serviced by BS-A is $DV_\text{A}=10 \;\text{GB}$.
\item BS-B has identical non-path power consumption as BS-A, $P_{\text{non-pathB}}=50\;\text{Wh}$. However, it is less efficient in transmission (e.g. an older generation of cellular technology that is less power efficient in its design for data transmission), consuming 4~Wh to transmit 1~GB of data, i.e., $P_{\text{consumed,pathB}}=4 \text{Wh/GB}$. The data volume serviced by BS-B is $DV_\text{B}=50 \;\text{GB}$.
\end{itemize}

Using the EE formula defined in eq.~(\ref{eq:ETSI_EE_BS}), the EE for BS-A is computed as
\begin{align}
    \text{EE}_\text{BS,A}&=\frac{DV_\text{A}}{P_\text{consumed,totalA}} = \frac{10 \;\text{GB}}{50\;\text{Wh} + 2\;\text{Wh/GB}\cdot 10\;\text{GB}} \nonumber \\ 
    &= 0.14\;\text{GB/Wh}.
\end{align}
Conversely, for BS-B, it is
\begin{align}
    \text{EE}_\text{BS,B}&=\frac{DV_\text{B}}{P_\text{consumed,totalB}} = \frac{10 \;\text{GB}}{50\;\text{Wh} + 4\;\text{Wh/GB}\cdot 50\;\text{GB}} \nonumber \\ 
    &= 0.2\;\text{GB/Wh}.
\end{align}

Despite being less efficient in transmitting data, BS-B is evaluated as more energy-efficient (e.g., having a higher data volume per energy) than BS-A according to the EE metric in eq.~(\ref{eq:ETSI_EE_BS}). 

This EE metric produces a misleading result because it depends on the traffic load at which it is measured rather than on the power efficiency of the equipment. For example, if BS-A were more heavily loaded, it would appear to be much more energy efficient.
In more detail, the issue arises because BS-A handles a low data volume, and under such conditions, the total consumed power is significantly influenced by the non-path power consumption, which is not related to the actual energy efficiency of source-to-sink transmission. Consequently, the true transmission efficiency is obscured by the specific power level (e.g., loading) considered in the EE metric.
In contrast, BS-B handles a large data volume, which dilutes the relative impact of its substantial per-GB data transmission energy cost.

This example reveals a critical flaw: the standard EE metric can be skewed by data volumes, making BSs appear more energy-efficient simply because they transmit more data, not because they use energy more judiciously. This can lead to incorrect optimization decisions, particularly in network load balancing scenarios, where traffic might be inappropriately offloaded to less efficient BSs under the guise of improving energy efficiency. In essence, the current EE metrics cause operators to continue to use less energy-efficient, older technology, and this lack of energy efficiency may explain the slow rollout of 5G. Indeed, 69$\%$ of the operators interviewed during the GSMA Intelligence Workshop Operator Survey 2024 \cite{gsma2024going}, indicated that, \textit{despite 5G’s theoretical energy efficiency
being well known}, carriers were not able to perceive short-term benefits from 5G \ac{NR} on overall energy efficiency,

\begin{table}
\centering
\caption{Network comparison based on standard EE and related flaws.}
\begin{tabular}[t]{lcc}
 \hline
  & Network A & Network B\\
  \hline
  Site Data Volume [GB/day] & 569 & 311 \\
  Site Energy Consumption [kW/day] & 34.8 & 34.2 \\
  Downlink Load [$\%$] & 54 & 12 \\
  Downlink UE Throughput [Mbps] & 6.1 & 21.6 \\
  EE [GB/kWh] & 16.4 & 9.1 \\
 \hline
\end{tabular}
\label{table:EEanalysis}
\end{table}

The same problems are observed at a larger scale by analyzing two BSs and two networks operating in distinct traffic conditions. An example of such inaccurate energy efficiency evaluation is reported in Table~\ref{table:EEanalysis}, where Network A is composed of products from an old generation, while Network B is built with new and more energy-efficient components.
Indeed, Network B provides much higher spectral efficiency and downlink user throughput than Network A (21.6 vs. 6.1 Mbps). However, since the time on the market of Network A is greater than Network B, the first is characterized by legacy installation and a larger data volume in the existing network deployment. Moreover, the two networks have similar energy consumption (measured in kW/day), which leads to better \ac{EE} for Network A when using the \ac{ETSI} network \ac{EE}  \cite{etsiES203228}, defined in \eqref{eq:EE_ntw}. Looking at this network EE metric from \ac{ETSI}, a mobile network operator may choose to shut down Network B to reduce its energy consumption and move the traffic from Network A to Network B. Such decision would lead to poor performance, i.e., the less-spectral-efficient Network A load would be excessively large thereby degrading the user throughput, while the operator thinks, incorrectly, it is being more energy efficient. Indeed, a better approach would be to shut down Network A, and take advantage of the better spectral efficiency of Network B and better EE (as shown by the use of \textit{W} in Section \ref{sec:Analysis_RAN}), to deliver high user throughput with a limited increase in energy consumption for much greater data throughput.

\begin{figure}
    \centering
    \includegraphics[scale=0.7]{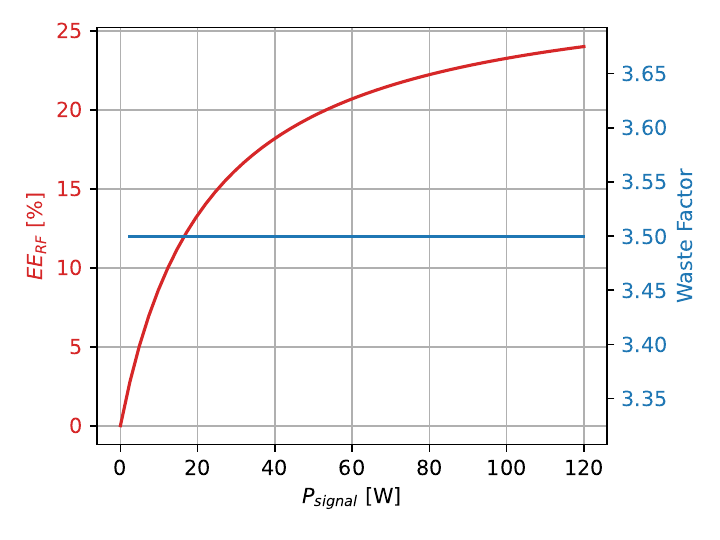}
    \caption{Comparison of energy efficiency for a \ac{RU} as defined in ITU~\cite{ituL13102020} (which varies with traffic or output power) and the corresponding Waste Factor (which is independent of output power) at different output power levels, $P_{\text{signal}}$.}
    \label{fig:EEvsW}
\end{figure}

In contrast with existing standard metrics, the Waste Factor has notable advantages. First, \textit{W} is easy to interpret as the energy waste of transmitting a signal along a source-to-sink chain, or any subset of the chain. Since \textit{W} is independent of the transmitted data load (e.g., the power needed to send a signal) and is also independent of the non-path energy consumption, it allows for fair and clear comparisons between different traffic-carrying systems operating in distinct environments and operating conditions. An example is reported in Fig.~\ref{fig:EEvsW}, which compares the \ac{ITU} SG5 EE metric defined in eq.~(\ref{eq:EE_RF}) with W in assessing the sustainability of a \ac{RU}. Notably, the \ac{ITU} SG5 energy efficiency metric varies with the operating scenario; the higher the output power, $P_\text{signal}$, the greater the apparent energy efficiency. However, \textit{W}, which focuses solely on the energy consumed in the signal path and disregards both the particular signal levels and the non-path power, remains unaffected by changes in $P_\text{signal}$ and by traffic load, thereby laying bare the true energy efficiency in source-to-sink communications without dependence on the power (e.g. traffic) operating point.

Finally, \textit{W} is proposed as a unified metric to assess the end-to-end transmission and reception from the energy perspective, including the impact of the wireless channel and of the user equipment, thereby allowing a unified, harmonious energy efficiency metric for the network, site, and RU components over widely varying operating conditions and propagation environments.
To this end, in the following subsections, we use the fundamentals of the Waste Factor theory to derive the Waste Factor for all passive and active components of the RAN, followed by the derivation of the Waste Factor for RUs and UEs.

\subsection{Waste Factor of Passive Components}
\label{sec:passive_components}

In this section, we discuss \textit{W} related to passive components found in RU and UE shown in Fig.~\ref{fig:TXRXStructure}. For any passive element, let $P_\text{in}$ denote the input power and $P_\text{signal,out}=P_\text{out}$ the output power. The loss within the passive element is thus represented by $P_\text{non-signal}=P_\text{in}-P_\text{out}$.
Accordingly, \textit{W} of such a passive element can be mathematically expressed as \cite{Rappaport2024Microwave}
\begin{align}
\label{eq:passive}
  \nonumber  W &= \frac{P_\text{non-signal}+P_\text{signal,out}}{P_\text{signal,out}} \\  &= \frac{P_\text{in}-P_\text{out}+P_\text{out}}{P_\text{out}}=\frac{P_\text{in}}{P_\text{out}}=L.
\end{align}
Eq.~(\ref{eq:passive}) underscores how $W$ is equivalent to the loss factor $L$ of the passive element. In essence, \textit{W} directly reflects the efficiency loss in passive elements, indicating that for passive components, \textit{W} and the loss are identical concepts~\cite{Rappaport2024Microwave}.

\subsubsection{\textbf{Passive RF Mixers}}
The performance efficiency of passive \ac{RF} mixers is significantly impacted by their inherent losses, predominantly characterized by conversion and insertion losses.

The conversion loss, denoted by $L_{\text{con}}$, is a critical measure that quantifies the diminution in signal power resulting from the frequency translation process, specifically from the \ac{RF} input signal to the \ac{IF} output signal~\cite{held1978conversion}. The conversion loss parameter is routinely documented within the datasheets of mixers, indicating its importance in assessing mixer performance. The conversion loss is mathematically defined as
\begin{equation}
L_{\text{con}}(\text{dB}) = 10 \log_{10}\left(\frac{P_{\text{RF}}}{P_{\text{IF}}}\right),
\end{equation}
where $P_{\text{RF}}$ represents the power of the input RF signal, and $P_{\text{IF}}$ denotes the power of the output IF signal, with the loss expressed as a positive number in decibels (dB).

Insertion loss, symbolized by $L_{\text{ins}}$ (dB), encapsulates the signal power dissipation incurred due to the integration of the mixer within the signal path, and encompasses the detriments attributable to impedance mismatches and parasitic effects inherent to the design and operation of the mixer. The aggregate loss experienced by the signal as it traverses through the passive mixer can be articulated as follows
\begin{equation}
L_{\text{Mix}}\,(\text{dB}) = L_{\text{con}}\,(\text{dB}) + L_{\text{ins}}\,(\text{dB}).
\end{equation}

Accordingly, the overall efficiency or \textit{W} of passive mixers, encompassing both conversion and insertion losses, can be represented by
\begin{equation}
W_{\text{Mix}}\, (\text{dB})= L_{\text{Mix}}\,(\text{dB}) = -G_{\text{Mix}}\,(\text{dB}).
\end{equation}

We classify the \ac{LO} as a non-path component and when calculating the total power consumption, the power utilized by the LO is considered in $P_{\text{non-path}}$ and incorporated into the overall consumed power ($P_{\text{consumed,total}}$) using \eqref{powercontotal}, ensuring a comprehensive assessment of energy use within the system.

\subsubsection{\textbf{Phase shifter}} The Phase shifter is another crucial component along the signal path, as it must provide precise phase control without introducing significant losses. Reflection loss is one such parameter that demands attention; it is quantitatively assessed by the \ac{VSWR}, which indicates impedance mismatch. Defined as the ratio of the maximum to the minimum voltage of the standing wave along the transmission line, \ac{VSWR} facilitates the calculation of the reflection coefficient, $\Gamma$, through the relationship $\Gamma = {\left(\text{VSWR} - 1\right)}/{\left(\text{VSWR} + 1\right)}$. The reflection loss in decibels (dB) is subsequently determined as $L_{\text{reflect}} = -20 \log_{10} |\Gamma|$ \cite{Ludwig2000}. The reflection loss metric is imperative for ensuring efficient power transfer and minimizing signal degradation across the transmission, from the mixer to the power amplifier. However, total loss, denoted as $L_{\text{PS}}$ (dB), includes not only reflection loss, $L_{\text{reflect}}$ (dB), but also insertion loss, $L_{\text{ins}}$ (dB), which represents the intrinsic power loss through the phase shifter and transmission medium \cite{PSAccess}. Thus, the total loss (gain) of a phase shifter can be expressed as
\begin{equation}
    L_{\text{PS}}\, \text{(dB)} =-G_{\text{PS}}\,\text{(dB)}= L_{\text{reflect}}\,\text{(dB)} + L_{\text{ins}}\,\text{(dB)}.
    \label{eq:PS_total_loss}
\end{equation}

Based on the analysis of passive components, as derived from (\ref{WL}), we see that $W_{\text{PS}} = L_{\text{PS}}$.

\subsubsection{\textbf{Antenna}} 

The passive antenna is a critical component for the efficiency and performance of the communication system. 
The radiation efficiency, denoted as $\eta_{\text{rad}}$, quantifies the capability of the antenna to convert input RF power, \(P_{\text{in}}\), into radiated power, \(P_{\text{rad}}\). 
However, total antenna efficiency is a broader concept that encompasses radiation efficiency while also accounting for additional losses such as impedance mismatches and ${I^{2} R}$ losses (conduction and dielectric) within the structure of the antenna.

Impedance mismatch loss arises when the impedance of the antenna \(Z_{\text{Ant}}\) fails to align with the output impedance of the transmitter or receiver \(Z_{\text{out}}\), leading to power reflection. The mismatch can be quantified by the \ac{VSWR}, leading to a reflection coefficient defined as
\begin{equation}
\Gamma = \frac{Z_{\text{Ant}} - Z_{\text{out}}}{Z_{\text{Ant}} + Z_{\text{out}}}= \frac{\mathrm{VSWR}-1}{\mathrm{VSWR}+1}.
\end{equation}
The \ac{VSWR}, a significant metric found in antenna datasheets, reflects the degree of impedance mismatch, influencing the efficiency of RF power transfer. In particular, the \ac{VSWR} is integral for evaluating antenna performance, facilitating the computation of mismatch loss 
\begin{equation}
L_{\text{mismatch}} =  1 - |\Gamma|^{2}.
\end{equation}

Combining antenna radiation efficiency and mismatch gives a comprehensive view of the antenna efficiency
\begin{equation}
\eta_{\text{Ant}} = \eta_{\text{rad}}  \cdot (1 - |\Gamma|^2).
\end{equation}

Finally, \textit{W} for the antenna can be derived as the reciprocal of the overall antenna efficiency
\begin{equation}
\label{anteta}
W_\text{Ant} = \frac{1}{\eta_{\text{Ant}}} = \frac{1}{\eta_{\text{rad}}  \cdot (1 - |\Gamma|^2)}.
\end{equation}

\subsection{Waste Factor of Active Components}
\label{sec:active_components}

This section analyzes \textit{W} associated with active components found in the \ac{RU} and \ac{UE} of Fig. \ref{fig:TXRXStructure}. Define $P_\text{DC}$ as the power consumed from the \ac{DC} source, $P_\text{in}$ as the input power to the active element, and $P_\text{out}$ as the output power from the active element. The increment in signal power facilitated by the active component, known as the added signal power, is thus calculated as $P_\text{added-signal}=P_\text{out}-P_\text{in}$.
Consequently, the power that does not contribute to signal amplification is wasted power (see eq.~(\ref{powercontotal}) and eq.~(\ref{powerwasted})), referred to as non-signal power, determined by $P_\text{non-signal}=P_\text{DC}-P_\text{added-signal}$.
\textit{W} of the active component is thus defined as \cite{Rappaport2024Microwave}
\begin{align}
W &= \frac{P_\text{out}+P_\text{non-signal}}{P_\text{out}}  \nonumber 
= \frac{P_\text{out}+(P_\text{DC}-P_\text{added-signal})}{P_\text{out}}  \nonumber \\
     &= \frac{P_\text{out}+P_\text{DC}-(P_\text{out}-P_\text{in})}{P_\text{out}} = \frac{P_\text{DC}+P_\text{in}}{P_\text{out}}.
     \label{ActiveW}
\end{align}

\subsubsection{\textbf{DAC}} 
\acp{DAC} are essential for transforming digital signals into analog signals. DACs take a digital signal as input that is registered internally \cite{DaiDAC}. Thus, it does not directly participate in the signal cascade (the digital data applied to the DAC may be viewed as an input source to a cascade, as shown in Fig. \ref{figa1}).
The power consumption for DACs can be expressed as \cite{CuiDAC}
\begin{equation}
    P_{\text{DAC}} = \beta (P_\text{s} + P_\text{d}),
\end{equation}
where \(P_\text{s}\) is static power, \(P_\text{d}\) is dynamic power, and \(\beta\) is a correcting factor to incorporate second-order effects \cite{CuiDAC}. Such power consumption can be easily found in the datasheet of any DAC.

Based on eq.~(\ref{ActiveW}), \textit{W} for a DAC, assuming linear operation, can be defined using \eqref{defW}
\begin{equation}
\label{WDAC1}
    W_{\text{DAC}} = \frac{P_{\text{DAC}}}{P_{\text{out}}},
\end{equation}
with \(P_{\text{out}}\) denoting the power of the analog output signal, which can be calculated based on the output voltage and current of the DAC.

It is crucial to note that DAC technical datasheets often additionally report the power efficiency metric, computed \ as \(\eta_{\text{DAC}} = {P_{\text{out}}}/{P_{\text{DAC}}}\). Consequently, $W_{\text{DAC}}$ can also be computed as the reciprocal of the power efficiency as given in the DAC datasheet
\begin{equation}
    W_{\text{DAC}} = \frac{1}{\eta_{\text{DAC}}} = \frac{P_{\text{DAC}}}{P_{\text{out}}}.
\end{equation}
In the analysis of DAC gain, we assume linearity, negating any signal distortions due to Integral Nonlinearity (INL) or Differential Nonlinearity (DNL) errors, as detailed in \cite{johns2008analog}. Consequently, we consider the gain of the DAC to be 1, indicating no discrepancies between the digital and analog domains. Time averaging of the various power levels over the range of digital inputs, or creating a common definition such as the maximum amplitude output \cite{Rappaport2024Microwave} may be used to further define \textit{W} for DACs.

\subsubsection{\textbf{ADC}}

\acp{ADC} transforms analog signals into the digital domain and is usually at the last stage before the sink at a receiver. Normally, a device closest to the sink will have a great impact on receiver efficiency (see \eqref{Wcas}). The power consumption of ADCs, associated with their operational characteristics, is computed in \cite{SkriReceiver} as
\begin{equation}
    P_{\text{ADC}} = \text{FoM} \cdot f_{s} \cdot 2^n,
\end{equation}
where \(f_s\) is the sampling frequency, \(n\) is the bit resolution, and FoM represents the ADC Walden’s figure of merit \cite{LeeADC}, also referred to as the energy per conversion step. The power consumption of an ADC, $P_{\text{ADC}}$, is defined as the total power consumption in datasheets, and since the power consumption is not related in a linear way to the input power, we may treat $P_{\text{ADC}}$ as part of the non-path power, $P_{\text{ADC}} \subseteq P_{\text{non-path}}$. Here, we assume the ADC operates in its linear range with a linear transfer function~\cite{johns2008analog}, and this assumption is crucial as it ensures that there are no errors in the conversion from analog to digital signals that could affect the power consumption estimation of ADC. 

\begin{figure}[!t]
    \centering
    \includegraphics[width=2.8in]{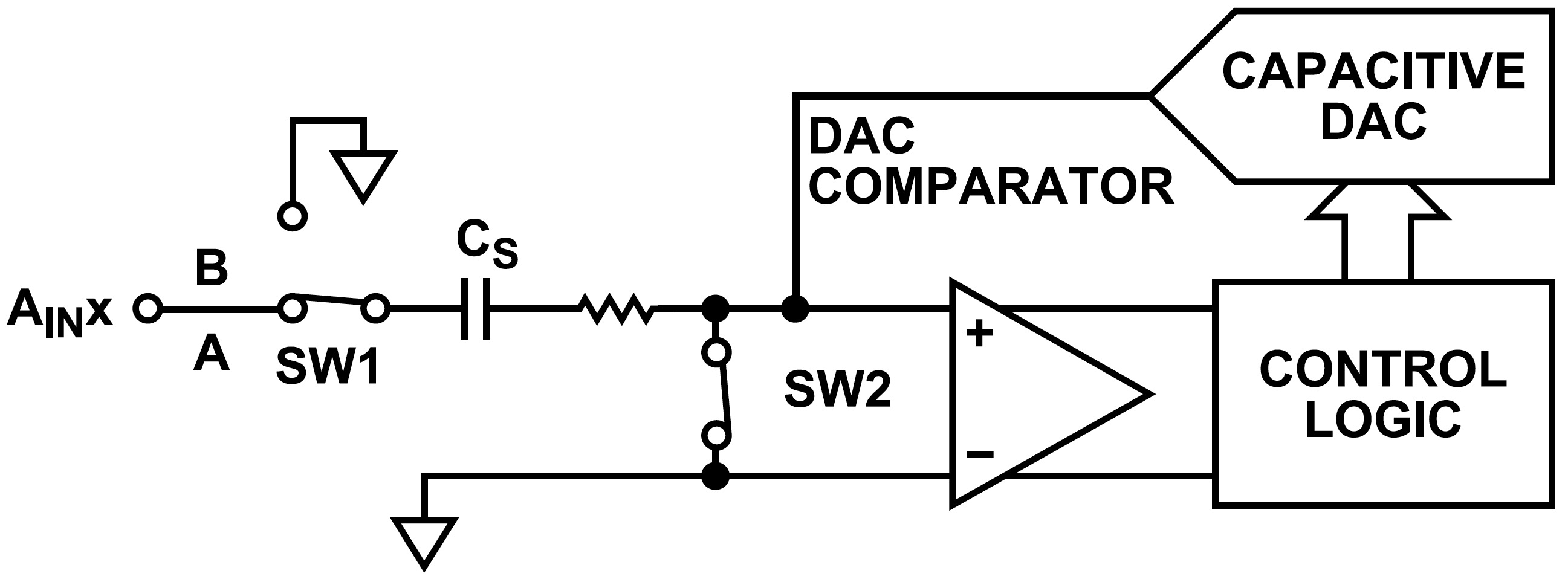}
    \caption{Schematic of the ADC acquisition phase, illustrating the assumption of impedance matching in the final stage \cite{AD_datasheet}.}
    \label{fig:ADCAcq}
\end{figure}

In Fig.~\ref{fig:ADCAcq}, the acquisition phase of a sample-and-hold or a charge redistribution Successive Approximation Register (SAR) ADC is depicted. The input to the ADC sees a DC block capacitor--electrical short at high frequencies--and a matched resistor, underscoring the premise that the ADC presents a perfectly matched impedance at the final stage and thereby validating its categorization as a non-path power component.
For analysis of signal-path power consumption using \textit{W}, here we assume an ADC may be replaced by an ideal wire (e.g. it is ignored for the value of \textit{W} for the cascade in eq.~\ref{Wcas}).

\subsubsection{\textbf{Power Amplifier}} 
Power Amplifiers enhance signal power, which is crucial for applications requiring signal transmission over lossy channels.
 
\textit{W} for a PA is given by
\begin{equation}
\label{eq:PAeff}
    W_{\text{PA}} = \frac{P_{\text{DC}} + P_{\text{in}}}{P_{\text{out}}},
\end{equation}
where \(P_{\text{DC}}\) is the power consumed by the PA, 
\(P_{\text{in}}\) represents the input power to the PA, and \(P_{\text{out}}\) is the output power delivered by the PA to the load. 

The PA efficiency, often referred to as the total power-added efficiency (PAE), is a key parameter providing a comprehensive measure of how a PA converts supplied power into usable output power. The PAE is traditionally~\cite{steer2019microwave} defined as
\begin{equation}
\label{PAE1}
    \text{PAE} = \eta_{\text{PA}} = \frac{P_{\text{out}}}{P_{\text{DC}} + P_{\text{in}}}.
\end{equation}

Note from \eqref{defW} the inverse relationship between \textit{W} and \eqref{PAE1}. A superior PAE correlates with a diminished \textit{W}, highlighting the operational energy efficiency of the amplifier.

An alternate definition of PAE was introduced by Walker in 1993 \cite{walker1993high} and is presented as
\begin{equation}
\text{PAE}^{\#2} = \frac{P_{\text{out}} - P_{\text{in}}}{P_{\text{DC}}}.
\label{PAE2}
\end{equation}
 
This alternative definition of PAE$^{\#2}$ is shown to be related to $W$ in \cite{Rappaport2024Microwave} as
\begin{equation}
W = \frac{1}{\text{PAE}^{\#2}}  \left[ \left( 1 + \frac{P_{\text{in}}}{P_{\text{DC}}} \right)  \left( 1 - \frac{1}{G} \right) \right].
\label{W_PAE2}
\end{equation}
Importantly, the widespread use of PAE as a figure of merit, typically reported in datasheets, and its direct relation with $W$, simplify the process of evaluating \textit{W} of the amplifier.

\subsubsection{\textbf{Low Noise Amplifier}}
LNAs are pivotal in enhancing weak signals with minimal noise addition, playing a critical role in preserving the integrity of signals in sensitive receiving equipment. Distinct from PAs, LNAs exhibit unique power consumption characteristics at low input signal levels, as delineated in \cite{szczepkowski2014study}. Their power consumption does not exhibit a linear relationship with input power. Within certain power ranges, LNAs maintain consistent gain and efficiency, which translates into a stable quiescent power consumption that is vital for their operational performance \cite{skrimponis2020power}. The methodology for calculating LNA power consumption, $P_{\text{LNA}}$, is provided in \cite{Ickhyun2008LNAFoM} as
\begin{equation}
\label{PLNA}
    P_{\text{LNA}} = \frac{G}{\text{FoM}_{\text{LNA}} (F-1)},
\end{equation}
where $\text{FoM}_{\text{LNA}}$ signifies the Figure of Merit for LNAs \cite{Ahlgren1999LNAFoM}, and $F$ represents the Noise Factor. This eq.~(\ref{PLNA}) is commonly employed for evaluating the power consumption of LNAs.

Reference \cite{kanhere2022power} posits that $W_{\text{LNA}} = 1$, suggesting a constant DC power consumption by the LNA, leading to an approximation where $P_{\text{LNA}} \approx P_{\text{DC}}$ (e.g., modeled as a high gain ideally efficient amplifier). Such an assumption is valid for LNAs operating near saturation but not within their entire linear range. The calculation of $W_{\text{LNA}}$ integrates eqs. \eqref{addnoise}, \eqref{ActiveW}, and \eqref{PLNA}
\begin{align}
    \label{WLNA}
    \nonumber
    W_{\text{LNA}} &= \frac{P_{\text{in}} + P_{\text{DC}}}{P_{\text{out}}} = \frac{P_{\text{LNA}}}{P_{\text{out}}} = \frac{G}{\text{FoM}_{\text{LNA}} (\textit{F}-1)P_{\text{out}}}\\ \nonumber
    &= \frac{G}{\text{FoM}_{\text{LNA}} \text{SNR}_{\text{in}} (F-1)  G N_{\text{in}}}\\
    &= \frac{G}{\text{FoM}_{\text{LNA}}  \text{SNR}_{\text{in}}P_{\text{additive-noise}}},
\end{align}
where $\text{SNR}_{\text{in}} = P_{\text{in}}/N_{\text{in}} = P_{\text{out}}/(G N_{\text{in}})$ denotes the input Signal-to-Noise Ratio. Typically, $W_{\text{LNA}}$ approaches unity under standard operational conditions, as shown in eq.~(\ref{WLNA}). It is important to note that the power consumption of LNAs is generally minuscule when compared to other components, and the large gain of LNAs, can be shown in eq.~(\ref{Wcas}) to minimize the impact of overall \textit{W} of cascade components when moving closer to the source~\cite{Rappaport2024Microwave}.

\subsection{Analyzing Quiescent Power in Amplifiers}
Active components like PAs and LNAs, require a DC power supply for operation. Specifically, Class AB PAs, a prevalent choice in such systems, exhibit variable DC power consumption ($P_{\text{DC}}$), a detail typically specified in datasheets. The $P_{\text{DC}}$ of these amplifiers adjusts in response to the input power. Through the examination of DC voltage, current, and power dissipation metrics, the waste factor $W$ can be determined using either (\ref{eq:PAeff}) or \eqref{W_PAE2} or measured in situ using an agreed-upon standard or average value range (see Fig. \ref{fig:RUmeasnexample}).

When amplifiers operate in a quiescent or static state—be it standby or ``off" modes—they consume power without facilitating signal amplification or signal transfer. Such consumption, termed quiescent or static power, remains substantial for a PA in the absence of signal output. In the quiescent state, deeming $W = \infty$ from (\ref{defW}) as a representation for active components in standby mode is impractical. It becomes evident that the quiescent power for devices that are not actively amplifying should be categorized as non-path power, allowing for an accurate depiction of the system total power usage in (\ref{powercontotal}) that includes stand-by power. Thus, linear amplifiers have a designated $W$ during their operational state, with quiescent power consumption falling under the non-path category, denoted as $P_{\text{quiescent}} \subseteq P_{\text{non-path}}$.

Incorporating quiescent power into $P_{\text{non-signal}}$ would be misleading, as it would suggest null total power usage in the absence of an output signal, which contradicts the actual power dynamics. Consequently, it is imperative to treat the quiescent or static/stand-by power of a PA as non-path power to ensure an accurate assessment of a system overall power consumption along an active signal path.
Our analysis on LNA and PA further illustrates how \textit{W} is applicable to a wide range of transmission schemes, including single carrier, OFDM, and carrier aggregation.

\subsection{Waste Factor of the Wireless Channel}
In wireless communication systems, the concept of \textit{W} for a channel within a cascade is critical to comprehensively understanding the efficiency of data transmission processes (e.g., as in Section III). Let us consider a scenario where the power transmitted from the source is denoted as \(P_\text{TX}\), and the power received at the receiver is \(P_\text{RX}\). The PL of a lossy channel, defined as \(L_\text{C} = {P_\text{TX}}/{P_\text{RX}}\), inversely relates to the channel gain, expressed as \( G_\text{C}= {P_\text{RX}}/{P_\text{TX}} = L_\text{C}^{-1}\).

Assuming the \(i^\text{th}\) stage of a cascaded system represents a wireless channel, the signal power at this stage, \(P_{\text{signal}_i}\), is derived as follows
\begin{equation}
\label{Wchan1}
    P_{\text{signal}_i} = G_\text{C} P_{\text{signal}_{i-1}},
\end{equation}
where \(P_{\text{signal}_{i-1}}\) corresponds to the output power of the transmitter \(P_\text{TX}\), and \(P_{\text{signal}_i}\) equates to the input power at the receiver \(P_\text{RX}\). Consequently, the non-signal power \(P_{\text{non-signal}_i}\) at this stage, representing the fraction of transmitted power that is not successfully received, can be formulated as
\begin{equation}
\label{Wchan2}
    P_{\text{non-signal}_i} = (1-G_\text{C})P_{\text{signal}_{i-1}},
\end{equation}
where \(1-G_\text{C}\) quantifies the proportion of the signal power lost due to various factors such as attenuation, scattering, and absorption in any type of channel (e.g. optical, copper, wireless, sonar, etc.). Thus, multiplying \(1-G_\text{C}\) by \(P_{\text{signal}_{i-1}}\) yields the amount of power dissipated (e.g. wasted) during transmission.

Waste Factor for the lossy channel can be calculated using eq. \eqref{Wchan1} and \eqref{Wchan2}
\begin{align}
    W_\text{C} &= \frac{ P_{\text{signal}_{i}} +  P_{\text{non-signal}_i} }{P_{\text{signal}_{i}} }\nonumber\\
      &= \frac{ G_\text{C}P_{\text{signal}_{i-1}} +  (1-G_\text{C})P_{\text{signal}_{i-1}} }{G_\text{C}P_{\text{signal}_{i-1}} } \nonumber\\
      &= \frac{1}{G_\text{C}} = L_\text{C}.
      \label{WL}
\end{align}

Notably, unlike electrical devices, a lossy channel does not have a non-path power component (\(P_\text{non-path}\)), rendering it akin to a passive attenuator~\cite{Rappaport2024Microwave}. Therefore, \textit{W} for a lossy channel (\(W_{\text{C}}\)) is equivalent to that of a passive attenuator (\(W_\text{atten} = L_\text{atten}\)).
The above demonstrates that \(W_{\text{C}}\) for a wireless channel is inherently related to the PL of the channel  (\(L_{\text{C}}\)).

Considering the antenna gains at both the transmitter and receiver completes our understanding of \(W\) for a wireless channel. Let \(G_{\text{TX}}^{\text{ant}}\) and \(G_{\text{RX}}^{\text{ant}}\) represent the antenna gains at the transmitter and receiver in absolute (not dB) values, respectively. These gains improve the effective signal power and are typically greater than 1 (or 0 dB). The effective PL, \(L_{\text{C}}^{\text{eff}}\), considering the antenna gains, is defined as
\begin{align}
    \nonumber L_{\text{C}}^{\text{eff}} &= \frac{L_{\text{C}}}{G_{\text{TX}}^{\text{ant}} G_{\text{RX}}^{\text{ant}}}, \\ 
    L_{\text{C}}^{\text{eff}} \text{(dB)} &= L_{\text{C}} \text{(dB)} - G_{\text{TX}}^{\text{ant}} \text{(dB)} - G_{\text{RX}}^{\text{ant}} \text{(dB)},
\end{align}
where \(L_{\text{C}}\) is the PL of the wireless channel without considering the antenna gains. Consequently, \(W_{\text{C}}^{\text{eff}}\), which now includes the effect of antenna gains, is recalculated as
\begin{equation}
    W_{\text{C}}^{\text{eff}} = \frac{1}{G_{\text{C}}^{\text{eff}}} = L_{\text{C}}^{\text{eff}} = \frac{L_{\text{C}}}{G_{\text{TX}}^{\text{ant}} G_{\text{RX}}^{\text{ant}}}.
    \label{eq:Wchannel}
\end{equation}

The formulation of \(W_{\text{C}}^{\text{eff}}\) demonstrates that the waste factor for a wireless channel, when factoring in the antenna gains, corresponds to the effective path loss (PL). The definition of effective PL is crucial for modeling the channel efficiency of a source-to-sink cascade.

\subsection{Waste Factor of RAN}
Referring to Fig.~\ref{fig:TXRXStructure}, we now use the Waste Factor theory to model the energy efficiency of the Radio Unit (RU), which comprises a DAC, a mixer, \(N_{\text{TX}}\) phase shifters (PSs), \(N_{\text{TX}}\) PAs, and \(N_{\text{TX}}\) antennas. We model the RU setup as a SIMO system and calculate the overall Waste Factor using the method outlined in Section III. Each parallel cascade in the RU includes a PS, a PA, and an antenna. We assume that each parallel cascade in the RU is constructed with equivalent components, thus having the same Waste Factor. The power outputted by the $i^\text{th}$ antenna of the RU, \(P_{\text{t},i}\), is assumed to be the same for each antenna, i.e., \(P_{\text{t},i} = P_\text{t} \; \forall i\). For simplicity, we consider non-coherent combining here, but the approach for coherent combining discussed in Section III can also be applied.

Given the aforementioned assumptions, \textit{W} for the parallel structure of PSs, PAs, and antennas of the RU depicted in Fig.~\ref{fig:TXRXStructure}, simplifies to
\begin{align}
\label{W_par_RU}
    W_{\text{M}\parallel}^{\text{RU}} = \frac{\sum_{i=1}^{N_{\text{TX}}} {P_{\text{t},i} \times  \left( W_{\text{Ant}} + \frac{W_{\text{PA}} - 1}{G_{\text{Ant}}} + \frac{W_{\text{PS}} - 1}{G_{\text{PA}} G_{\text{Ant}}}   \right)}}{\sum_{i=1}^{N_{\text{TX}}} {P_{\text{t},i}}} \nonumber \\ 
    = W_{\text{Ant}} + \frac{W_{\text{PA}} - 1}{G_{\text{Ant}}} + \frac{W_{\text{PS}} - 1}{G_{\text{PA}} G_{\text{Ant}}}.
\end{align}

Next, we determine the waste factor of the cascade comprising the DAC and the mixer of the RU
\begin{equation}
    W_0^{\textrm{RU}} = W_{\textrm{Mix}}+\frac{W_{\textrm{DAC}}-1}{G_{\textrm{Mix}}}.
\end{equation}

Subsequently, the overall waste factor of the RU, comprising the DAC, the mixer, and the parallel structure of PSs, PAs and antennas, is calculated as
\begin{align}
W_{\text{RU}} &=  W_{\text{M}\parallel}^{\text{RU}}  + \frac{(W_{0}^{\textrm{RU}} - 1)}{G_{\text{M}\parallel}^{\text{RU}}} \nonumber \\
&= W_{\text{Ant}} + \frac{W_{\text{PA}} - 1}{G_{\text{Ant}}} + \frac{W_{\text{PS}} - 1}{G_{\text{PA}} G_{\text{Ant}}} \nonumber \\
&+ \frac{W_{\text{Mix}} - 1}{G_{\text{PS}} G_{\text{PA}} G_{\text{Ant}}} + \frac{W_{\text{DAC}} - 1}{G_{\text{PS}} G_{\text{PA}} G_{\text{Ant}} G_{\text{Mix}}}.
\label{eq:W_RU}
\end{align}

The $W_{\text{RU}}$ in eq.~\ref{eq:W_RU} succinctly quantifies the power efficiency of the transmitter cascade by accounting for \textit{W} of each component. This insight is pivotal for pinpointing inefficiencies and boosting the energy efficiency of RF transmission systems. Emphasizing efficient component selection and strategic design minimizes power waste, enhancing overall system performance. Following this approach, we extend our analysis to include Waste Factor for the receiver ($W_{\text{UE}}$) shown in Fig.~\ref{fig:TXRXStructure}, ensuring a comprehensive evaluation of both transmitting and receiving ends of communication systems for improved sustainability.

The UE comprises an ADC (treated as a non-path component as discussed previously), a mixer, $N_\mathrm{RX}$ phase shifters, $N_\mathrm{RX}$ LNAs, and $N_\mathrm{RX}$ antennas. 
We model the UE as a MISO system, where each parallel cascade consists of a phase shifter, an LNA, and an antenna. Similar to the RU case, we assume that each parallel cascade in the UE is constructed with equivalent components, thus having the same waste factor, and each antenna $i$ of the UE receives the same power, $P_{\text{r},i} = P_\text{r}\;\forall i$.
Consequently, \textit{W} for the parallel structure of antennas, LNAs, and PSs in the UE, defined in eq.~(\ref{W_par_inco_gamma}), is simplified as follows
\begin{align}
\label{W_par_UE}
    W_{\text{M}\parallel}^{\text{UE}} = \frac{\sum_{i=1}^{N_{\text{RX}}} {P_{\text{r},i} \times  \left( W_{\text{PS}} + \frac{W_{\text{LNA}} - 1}{G_{\text{PS}}} + \frac{W_{\text{Ant}} - 1}{G_{\text{PS}} G_{\text{LNA}}}   \right)}}{\sum_{i=1}^{N_{\text{RX}}} {P_{\text{r},i}}} \nonumber \\ 
    = W_{\text{PS}} + \frac{W_{\text{LNA}} - 1}{G_{\text{PS}}} + \frac{W_{\text{Ant}} - 1}{G_{\text{PS}} G_{\text{LNA}}}.
\end{align}
\textit{W} of the mixer in the UE is formulated as
\begin{equation}
    W_{N+1}^{\text{UE}} = W_\text{Mix}.
\end{equation}

Finally, the overall waste factor of the UE is calculated by combining the contributions from the mixer and parallel cascades, leading to
\begin{align}
    W_\text{UE} &= W_{N+1}^\text{UE} + \frac{W_{\text{M}\parallel}^{\text{UE}}-1}{G_{N+1}^\text{UE}} \nonumber \\
    &= W_{\text{Mix}} + \frac{W_{\text{PS}} - 1}{ G_{\text{Mix}}} + \frac{W_{\text{LNA}} - 1}{G_{\text{Mix}} G_{\text{PS}} } + \frac{W_{\text{Ant}} - 1}{ G_{\text{Mix}} G_{\text{PS}} G_{\text{LNA}}   }.
\label{eq:W_UE}
\end{align}

Considering the defined Waste Factor for the RU in eq.(\ref{eq:W_RU}), for the wireless channel in eq.(\ref{eq:Wchannel}), and for the UE in eq.(\ref{eq:W_UE}), 
we see that \textit{W} of the \textit{entire cascade} comprising the RU, the wireless channel, and the UE can be written as
\begin{equation}
    W= W_\textrm{UE}+\frac{W_{\text{C}}^{\text{eff}}-1}{G_\textrm{UE}} + \frac{W_\textrm{RU}-1}{G_{\text{C}}^{\text{eff}}G_\textrm{UE}}.
    \label{eq:overall_W}
\end{equation}
Eq. \eqref{eq:overall_W} demonstrates the power of the Waste Factor theory as it permits energy efficiency analysis of a complete source-to-sink path, including any type of parallel structure and the channel.

\subsection{Computing W through measurements}
\label{sec:WFmeasurements}

In certain scenarios, determining the \textit{W} of systems by analyzing their internal components on the cascade --as outlined in eq.~(\ref{eq:W_RU}) and eq.~(\ref{eq:W_UE}) for the RU and UE-- may not be feasible due to insufficient knowledge regarding the hardware components specifications. Nonetheless, \textit{W} is a versatile metric that can also be estimated using a measurement-based approach to solve \eqref{defW}.

Here we explore the computation of \textit{W} for an RU using power consumption measurements. It is important to note that these measurements can be obtained not only from laboratory equipment but also from cell-level data autonomously reported by BSs, as referenced in~\cite{piovesan2023power,lopez2023data}.

\begin{figure}
    \centering
    \includegraphics[scale=0.85]{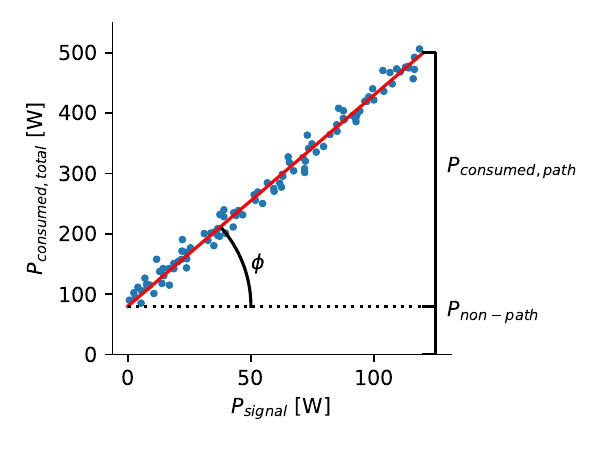}
    \caption{Numerical example from \cite{piovesan2022machine} showing the power consumed by an RU at different transmit power levels. Note that the Waste Factor of the RU corresponds to the constant slope $\phi$. From measurements, \textit{W} for the RU is equal to the slope of 3.5.}   \label{fig:RUmeasnexample}
\end{figure}

Fig.~\ref{fig:RUmeasnexample} illustrates a numerical example of energy consumption measurements collected by a RU, adapted from~\cite{piovesan2022machine}. In particular, the figure shows the relationship between the total power consumed, $P_\text{consumed,total}$, and the output signal power, $P_\text{signal}$. 
The dataset described in \cite{piovesan2022machine} includes hourly measurements of energy consumption by 7500 RUs over a period of 12 days from a real network. Notably, the study in~\cite{piovesan2022machine} reports only normalized values of energy consumption, hence the values in Fig.~\ref{fig:RUmeasnexample} have been rescaled to represent realistic energy consumption in Watts, following \cite{han2020energy}. 
A linear fitting technique is employed to mitigate the impact of noisy measurements, which can be influenced by operational temperature and tolerances of hardware components.
 
The non-path power consumption, which is independent of the transmitted signal power, is identified as the power consumed when the output signal power transmitted by the RU, $P_\text{signal}$, is zero.
Then, the power consumed along the RU signal path, $P_\text{consumed,path}$ is determined by subtracting $P_\text{non-path}$ from the RU measured $P_\text{consumed,total}$, considering eq.~(\ref{powercontotal}),
\begin{equation}
P_\text{consumed,path}=P_\text{consumed,total} - P_\text{non-path}.
\label{eq:conspath}
\end{equation}

Finally, the $W$ of the RU is calculated by applying the definition of W provided in eq.~(\ref{defW}),
\begin{equation}
    W_{\text{RU}}=\frac{P_\text{consumed,path}}{P_\text{signal}}.
    \label{eq:WRUmeas}
\end{equation}

Using eqs. (\ref{eq:conspath}) and (\ref{eq:WRUmeas}), measurements of the energy consumed by the RU in Fig.~\ref{fig:RUmeasnexample} show that the slope of the total consumed power to delivered output power, $\phi$, is equal to $W_\text{RU}=3.5$ at all operational regions.

Detailed analysis of the $W$ of each internal hardware component, or direct measurement of the energy consumption of the devices at varying output signal power levels or over an entire cascade is considered by Waste Factor theory (this is similar to Noise Factor). This versatility makes $W$ an invaluable tool in characterizing and optimizing signal path efficiency thereby reducing energy consumption in practical engineering applications.

\section{Analysis and Results for RAN Energy Efficiency}
\label{sec:Analysis_RAN}

\subsection{Impact of channel and components characteristics on RAN power waste}
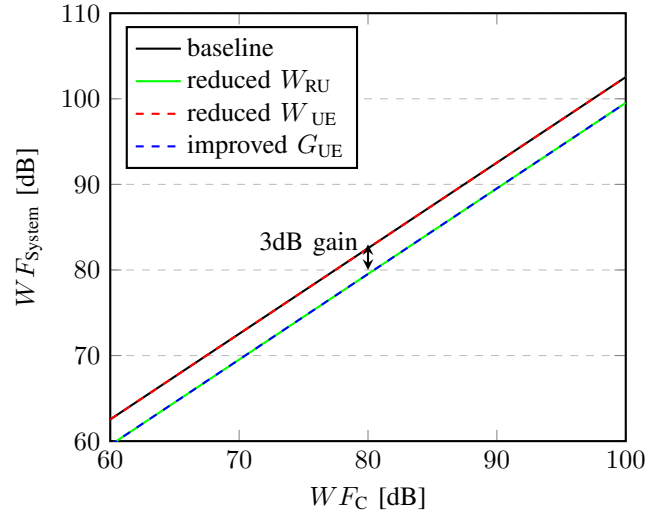
\begin{figure}[!t]
\input{Figures/fig_channel_W}
\caption{Comparative analysis of system waste figure reduction strategies.}
\label{fig:channel_W}
\end{figure}

\begin{table}[!t]
\caption{Parameters of the internal components of the RU and UE}
\centering
\begin{tabular}{|c|c|c|c|}
\hline
\textbf{Device} & \textbf{Component} & \textbf{Parameter} & \textbf{Value} \\ \hline \thickhline
RU& Antenna & Efficiency & 0.6          \\ \cline{3-4} 
  &                                     & VSWR                                & 1.5                             \\ \cline{2-4} 
& Power amplifier                     & Power added efficiency              & 0.48                            \\ \cline{3-4} 
&                                     & Gain                                & 50 dB                           \\ \cline{2-4} 
& Phase shifter                       & VSWR                                & 1.5                             \\ \cline{3-4} 
&                                     & Reflection loss                     & 14 dB                           \\ \cline{3-4} 
&                                     & Insertion loss                      & 3.5 dB                          \\ \cline{2-4} 
& Mixer                               & Conversion loss                     & 8.2 dB                          \\ \cline{2-4} 
& DAC                                 & Efficiency                          & 0.91                            \\ \hline \thickhline
UE                               & Antenna                             & Efficiency                          & 0.7                             \\ \cline{3-4} 
&                                     & VSWR                                & 1.5                             \\ \cline{2-4} 
& LNA                   & 
Gain                                & 20 dB                           \\ \cline{2-4} 
& Phase shifter                       & 
Insertion loss                      & 6 dB                            \\ \cline{2-4} 
 & Mixer                               & Conversion loss                     & 6.7 dB                          \\ \cline{1-4} 

\end{tabular}
\label{tab:RU_UE_params}
\end{table}

In the realm of system efficiency analysis, \textit{W} shows promise as a unifying metric due to its holistic nature of laying bare the wasted power along a source-to-sink path. \textit{W} facilitates the evaluation of both passive and active components within the signal path, including the channel, thus enabling a detailed evaluation of the power waste contribution of each component to the overall energy inefficiency of the system.

This section presents a comprehensive quantitative evaluation of \textit{W} of the system that encompasses the RU, the wireless communication channel, and the UE. 
Our methodology for evaluating the waste factor integrates a detailed analysis of the constituent components within both the RU and UE, building upon the foundational frameworks outlined in Section~\ref{sec:RAN_intro}-\ref{sec:passive_components} and Section~\ref{sec:RAN_intro}-\ref{sec:active_components} (e.g. the mathematical formulations presented in eq.~(\ref{eq:W_RU}) and eq.~(\ref{eq:W_UE}) to compute \textit{W} for the RU and UE, respectively).

The specific Waste Factors for both the RU and UE components are reported in Table~\ref{tab:RU_UE_params}. Through evaluation of \eqref{eq:W_RU}, the RU exhibits a waste factor of 3.5, whereas the UE is characterized by a significantly higher waste factor of 18.6. Casual survey of actual equipment and their power efficiency confirms that these values are reasonable.

Referencing Fig.~\ref{fig:channel_W}, the analysis using eqs.  (\ref{eq:Wchannel}),  (\ref{eq:W_RU}), (\ref{eq:W_UE}), (\ref{eq:overall_W}),  and Table~\ref{tab:RU_UE_params}, reveals the Waste Factor of the wireless channel is a dominant factor influencing the overall inefficiency of the system. The observed linear correlation between the \textit{W} of the channel and the overall system indicates the paramount importance of minimizing the channel losses between the RU and UE to obtain improved energy efficiency. This motivates small cells for a future direction of sustainable wireless networks.

Given the practical challenges associated with reducing channel losses, alternative approaches must be explored to reduce energy wastage across the signal path.
By examining the key variables in (\ref{eq:overall_W}), it is clear that strategic improvements on \textit{W} of the RU, $W_\textrm{RU}$, \textit{W} of the UE, $W_\textrm{UE}$, and the gain of the UE, $G_\textrm{UE}$, present the most viable avenues for enhancing the system efficiency if channel losses cannot be improved upon.
Accordingly, we now explore the impact of three distinct strategies:
\begin{itemize}
\item Halving the Waste Factor of RU, $W_\textrm{RU}$
\item Halving the Waste Factor of UE, $W_\textrm{UE}$
\item Doubling the gain of UE, $G_\textrm{UE}$
\end{itemize}

The outcomes of these strategies are visually depicted in Fig.~\ref{fig:channel_W}. Reducing by half the waste factor of the RU, $W_\textrm{RU}$, results in a significant decrease of 3 dB in \textit{W} for the system (e.g. doubling of the energy efficiency).
Similarly, replacing the baseline LNA with a gain of 20 dB with a more advanced LNA exhibiting a gain of 23 dB, thus doubling the UE gain, leads to an identical decrease of 3 dB in the overall \textit{W} of the system (for a fixed SNR at the UE).

In contrast, reducing the \textit{W} of the UE by 50\% achieves only minuscule improvements in overall system energy efficiency (a tiny fraction of a dB).
These results highlight that the most effective approach to enhance the overall source-to-sink \textit{W} involves acting on either the \textit{W} of the RU or the gain of the UE or increasing channel gain~\cite{Rappaport2024Microwave}.

\subsection{Comparison of RU power waste using W}

In this section, we assess \textit{W} of two distinct RUs through measurements, as shown in Section~\ref{sec:RAN_intro}-\ref{sec:WFmeasurements}.

Following the methodology depicted in Fig.~\ref{fig:RUmeasnexample}, we quantify the Waste Factor by measuring the power consumption, denoted as $P_\text{consumed,total}$, of each RU across a range of output power levels, $P_\text{signal,out}$. The non-path power consumption, $P_\text{non-path}$, which remains invariant with changes in the output signal power, is ascertained at a zero output signal condition ($P_\text{signal} = 0$). Conversely, the power consumed on the signal path, $P_\text{consumed,path}$, is derived from eq.~(\ref{powercontotal}), i.e., $P_\text{consumed,path}=P_\text{signal}+P_\text{non-signal}=P_\text{consumed,total}-P_\text{non-path}$.

Subsequent measurements indicate that both RUs deliver an output signal power of $P_\text{signal} = 120\,\text{W}$ and have a power consumption of $P_\text{consumed,total} = 500\,\text{W}$. However, distinctions emerge in their non-signal and non-path power consumption; RU-A consumes $P_\text{non-signal} = 240\,\text{W}$ and has non-path power of $P_\text{non-path} = 140\,\text{W}$, whereas RU-B exhibits values of $300\,\text{W}$ and $80\,W$ for $P_\text{non-signal}$ and $P_\text{non-path}$, respectively.

When adopting the energy efficiency metric for the RU defined by ITU in \cite{ituL13102020} and presented in Section~\ref{sec:RAN_intro}-\ref{sec:EEmetrics} (see eq. \eqref{eq:EE_RF}) we obtain the same energy efficiency values for both RUs
\begin{equation}
\mathrm{EE}_\mathrm{RU,A} = \mathrm{EE}_\mathrm{RU,B} = \frac{P_\text{signal}}{P_\text{consumed,total}} = \frac{120}{500} = 24\%. \\
\end{equation}
However, W, which provides insight into the efficiency of power conversion into useful signal transmission, differs between the two
\begin{align}
    W_A&=\frac{P_\text{non-signal}+P_\text{signal}}{P_\text{signal}}=\frac{240+120}{120}=3, \\
    W_B&=\frac{P_\text{non-signal}+P_\text{signal}}{P_\text{signal}}=\frac{300+120}{120}=3.5.
\end{align}

From this analysis, it becomes clear that although RU-A and RU-B display identical EE values using the ITU energy efficiency metric, RU-B wastes less power in signal transmission, due to its enhanced ability to convert power into useful data transmission. This difference highlights the limitations of relying solely on the EE metric defined by ITU~\cite{ituL13102020} to evaluate the energy performance of an RU. 
The EE metric in \cite{ituL13102020}, while commonly employed, only considers the total energy consumed by the RU, ignoring variations and dependencies inherent in the definition over operational conditions and differences in equipment types. Such oversight leads to the erroneous conclusion that RU-A and RU-B are equally energy efficient. It should be clear that the introduction and analysis of \textit{W} clearly demonstrate that the two RUs are not equivalent, underscoring the importance of incorporating \textit{W} for a more accurate assessment of the RU energy performance.

\subsection{Waste Factor: A Key Value Indicator for sustainability and optimization}

As demonstrated in Section~\ref{sec:RAN_intro}-\ref{sec:problemsEE}, \textit{W} offers a significant advantage over existing standard \ac{EE} metrics from ITU, ETSI, and 3GPP because \textit{W} is unaffected by traffic load variations at which it is quantified.  This unique characteristic of \textit{W} through the parsing of signal path and non-signal path energy consumption, and the partitioning of quiescent or static/stand-by power from energy consumed during information flow, enables a more accurate assessment of the efficiency of \acp{RU} for data communications.
Currently, \ac{EE} metrics need to be frequently recalculated to evaluate the efficiency of \acp{RU} under varying traffic conditions. In contrast, each \ac{RU} has a unique \textit{W} determining the intrinsic energy efficiency in data transmission. The nature of \textit{W} has a substantial impact on network optimization and provides mobile network operators with a means to identify less efficient \acp{RU} independent of their operating condition. For instance, these inefficient \acp{RU} can be replaced with more energy-efficient hardware or can be shut down during low-traffic periods, with users being transferred to more efficient \acp{RU} (e.g., with lower \textit{W}).
Moreover, \textit{W}, when coupled with other relevant network Key Performance Indicators (KPIs) or Key Value Indicators (KVIs), offers a powerful tool to drive decision-making strategies that improve the energy efficiency of any communication network. 

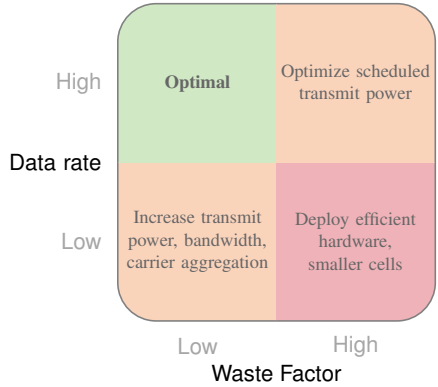
\begin{figure}[!t]
    \centering
\begin{tikzpicture}[scale=0.7, every node/.style={scale=0.7}]
\begin{scope}
\path[clip, preaction={draw, very thick, gray}] [rounded corners=5mm] (-3,-3) rectangle (3,3);
\foreach \anch [count=\xi] in {north east, north west, south east, south west}{%
    \ifcase\xi
    \or \def\filc{dorange}    
    \or \def\filc{dred}   
    \or \def\filc{dgreen}   
    \or \def\filc{dorange}      
    \fi
    \node[fill=\filc, anchor=\anch,%
        minimum size=4cm, outer sep=0, inner sep=0] (n\xi) at (0,0) {};
    \node[anchor=\anch, font=\fontsize{60}{22}\selectfont\sffamily, text=black!20, inner sep=0, outer sep=0] at (0,0) {};

}
\end{scope}

\begin{scope}[text=gray!70,font=\large\sffamily]
\foreach \lbla/\lblb [count=\fact starting from 0] in {
    Low/Low,
    \textcolor{black}{Data rate}/\textcolor{black}{Waste Factor},
    High/High
}{
    \pgfmathsetmacro\addendo{-1.5+(\fact*1.5)}
    \node[text width=2cm, align=right, anchor=east] at (-3.2,\addendo) {\lbla};
    \ifnum\fact=1
    \node at (\addendo,-4) {\lblb};
    \else
    \node at (\addendo,-3.5) {\lblb};
    \fi
}
\end{scope}

\begin{scope}[text width=3cm,text=black!60, align=center]
\node[font=\bfseries] (n1) at (-1.5,1.5) {Optimal};
\node (n2) at (1.5,1.5) {Optimize scheduled transmit power};
\node[] (n3) at (1.5,-1.5) {Deploy efficient hardware, smaller cells};
\node[] (n4) at (-1.5,-1.5) {Increase transmit power, bandwidth, carrier aggregation};

\end{scope}

\end{tikzpicture}
\caption{Optimal energy strategy based on Rate and Waste Factor}
\label{fig:strategy_R_W}
\end{figure}

Fig.~\ref{fig:strategy_R_W} shows the best decision to improve the energy efficiency (i.e., decrease the energy consumption or increase the rate) in different operational conditions in terms of carried traffic (e.g. data rate) and \textit{W}.
The operational conditions are divided into four regions, according to the low or high level of data rate and \textit{W}. Importantly, each region can be mapped to a general energy-optimized strategy.
\begin{itemize}
\item (High) Rate, (Low) \textit{W}: Ideal scenario because high transmission rate is combined with a low \textit{W}, indicating efficient energy use during signal transmission.
\item (High) Rate, (High) \textit{W}: While the transmission rate is high, \textit{W} is also elevated, indicating a significant energy consumption for signal transmission. A viable approach to reduce energy consumption is to reduce the transmitted power (for instance, by activating energy-saving methods that aim at reducing the transmit power scheduled to the UEs, thus reducing energy consumption).
\item (Low) Rate, (Low) \textit{W}:  The transmission rate is low but with a minimal \textit{W}, suggesting low energy consumption for signal transmission. The low waste factor presents an opportunity to improve the transmission rate by either increasing the transmit power or expanding the bandwidth of the cell since the hardware is energy efficient given a low W.
\item (Low) Rate, (High) \textit{W}: This scenario is the least desirable as it combines a low transmission rate with a high \textit{W}, leading to extremely inefficient energy use. The best approach to reduce RU energy consumption is to deploy more sustainable hardware --equipment having a smaller $W$-- or design smaller cell sizes to enhance channel conditions in order to reduce the Waste Factor of the channel, $W_\text{C}$, thereby reducing the need for high transmission power.
\end{itemize}

\begin{figure}[!t]
    \centering
\begin{tikzpicture}[scale=0.7, every node/.style={scale=0.7}]
\begin{scope}
\path[clip, preaction={draw, very thick, gray}] [rounded corners=5mm] (-3,-3) rectangle (3,3);
\foreach \anch [count=\xi] in {north east, north west, south east, south west}{%
    \ifcase\xi
    \or \def\filc{dgreen}    
    \or \def\filc{dorange}   
    \or \def\filc{dorange}   
    \or \def\filc{dred}      
    \fi
    \node[fill=\filc, anchor=\anch,
        minimum size=4cm, outer sep=0, inner sep=0] (n\xi) at (0,0) {};
    \node[anchor=\anch, font=\fontsize{60}{22}\selectfont\sffamily, text=black!20, inner sep=0, outer sep=0] at (0,0) {};

}
\end{scope}

\begin{scope}[text=gray!70,font=\large\sffamily]
\foreach \lbla/\lblb [count=\fact starting from 0] in {
    Low/Low,\textcolor{black}{Consumed\\power}/\textcolor{black}{Waste Factor},
    High/High
}{
    \pgfmathsetmacro\addendo{-1.5+(\fact*1.5)}
    \node[text width=2cm, align=right, anchor=east] at (-3.2,\addendo) {\lbla};
    \ifnum\fact=1
    \node at (\addendo,-4) {\lblb};
    \else
    \node at (\addendo,-3.5) {\lblb};
    \fi
}
\end{scope}

\begin{scope}[text width=3cm,text=black!60, align=center]
\node[] (n1) at (-1.5,1.5) {Shutdown methods, efficient cooling};
\node (n2) at (1.5,1.5) {Optimize scheduled transmit power};
\node[] (n3) at (1.5,-1.5) {Deploy efficient hardware};
\node[font=\bfseries] (n4) at (-1.5,-1.5) {Optimal};
\end{scope}
\end{tikzpicture}
\caption{Optimal energy strategy based on Consumed Power and Waste Factor}
    \label{fig:strategy_PC_W}
\end{figure}
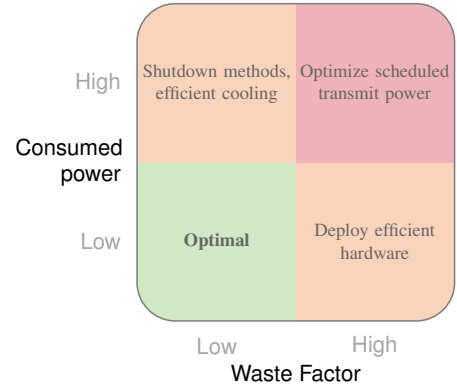

Similarly, Fig.~\ref{fig:strategy_PC_W} shows the best decisions to improve the energy efficiency (i.e., decrease the energy consumption and/or increase the rate) in different operational conditions in terms of consumed power and \textit{W}.
\begin{itemize}
    \item  (Low) Consumed Power, (Low) \textit{W}: Ideal situation, indicating efficient energy usage with minimal waste.
    \item (High) Consumed Power, (Low) \textit{W}: Despite high power consumption, the low \textit{W} indicates efficient energy conversion into a transmitted signal. Energy-saving efforts should, therefore, focus on reducing non-path power consumption, which can be achieved through implementing shutdown mechanisms\cite{lopez2022survey,piovesan2022machine} or using less quiescent power consumption, and more efficient cooling or power conversion within the infrastructure.
    \item (Low) Consumed Power, (High) \textit{W}: Although the overall power consumption is low, the high \textit{W} suggests inefficiency in energy used for signal transmission. Improvements can be made by employing equipment with both lower \textit{W} and reduced non path energy consumption, as well as adopting a small-cell approach to lower overall \textit{W} of each link.
    \item (High) Consumed Power, (High) \textit{W}:  This is the least favorable scenario, characterized by both high energy consumption and a high \textit{W}, indicating significant energy inefficiency. A viable approach to reduce energy consumption in this scenario is to decrease the transmit power or replace the RU with more sustainable hardware that has a smaller \textit{W}, while reducing cell size.
\end{itemize}

The discussions and findings presented in Fig. \ref{fig:strategy_R_W} and \ref{fig:strategy_PC_W} elucidate how the Waste Factor theory can be effectively integrated into both the planning and operational phases of network design. This integration is particularly useful for making informed decisions regarding capital expenditures and managing daily or real-time operations. Moreover, the real-time adaptability of \textit{W}, potentially enhanced through AI/ML technologies at the component, subsystem, or RAN level, further underscores its utility in optimizing network efficiency. The application of \textit{W} is thus not only a theoretical construct but a practical tool in the ongoing pursuit of network optimization.

\subsection{Assessing the energy efficiency of distributed MU-MIMO systems using W}

In this section, we use the Waste Factor to assess and compare the energy efficiency of simulated distributed multi-user MIMO (MU-MIMO) systems at 3.5, 17, and 28 GHz, where multiple BSs are assumed to transmit to multiple UEs in a coordinated manner. The concept of coordinated transmission first appeared in Coordinated Multi-Point (CoMP) communication, which was initially investigated during the LTE-Advanced study in 2008 \cite{3GPPRel9} to meet IMT-Advanced requirements \cite{ITURM2134}. CoMP was later included as a potential solution in 3GPP LTE Release 11 \cite{3GPPTR36824Rel11}.
The first BS diversity CoMP channel measurements at 73 GHz were conducted in 2018  \cite{maccartney2019millimeter}. In 5G, CoMP has been called \textit{distributed MU-MIMO}, where the coordinated transmissions to different users leverage beam-specific control signals from multiple BSs simultaneously.

For downlink (DL) transmission, coordinated transmissions from various BSs can significantly boost SNR performance at the UE, by limiting interference through joint beamforming. For uplink (UL), multiple BS reception points can greatly enhance the link performance through interference cancellation~\cite{you2021distributed, demir2021foundations}. In the following simulation study, we use \textit{W} to quantify the energy efficiency of the entire system considering the DL performance of a distributed MU-MIMO system with multiple randomly distributed BSs and UEs implemented in three different frequency bands (e.g., 3.5, 17, and 28 GHz). We use a novel modeling method that exploits MISO and MIMO results given in Section III.

\begin{table}[htbp]
\centering
\caption{Simulation parameters for distributed MU-MIMO using \textit{W}}
\begin{tabular}{|>{\raggedright\arraybackslash}p{0.6\columnwidth}|>{\centering\arraybackslash}p{0.2\columnwidth}|}
\thickhline
\textbf{Parameter} & \textbf{Value} \\ \thickhline
Number of UEs & 1024 \\ \hline
Number of BSs & 1 to 20 \\ \hline
Radius of the region under study & 1 km \\ \hline
BS height & 15 m \\ \hline
UE height & 1.5 m \\ \hline
Minimum BS separation & 200 m \\ \hline
rage cell radius per BS & 200 m \\ \hline
Transmission Bandwidth & 400 MHz\\ \hline
Target SNR at UEs & 10 dB \\ \hline
Noise Figure at UEs & 5 dB \\ \hline
Maximum transmit power per BS & 50 dBm \\ \hline
Waste Factor of each BS & 15 \\ \hline
Waste Factor of each UE & 33 \\ \hline
System gain of each UE & 11 dB \\ \hline
System gain of each BS & 30 dB \\ \hline
Non-path power of each BS & 140 W \\ \hline
Non-path power of each UE & 1 W \\  \thickhline
\end{tabular}
\label{tab:simulation}
\end{table}

\begin{figure}[!t]
    \centering
    \includegraphics[width=0.95\columnwidth]{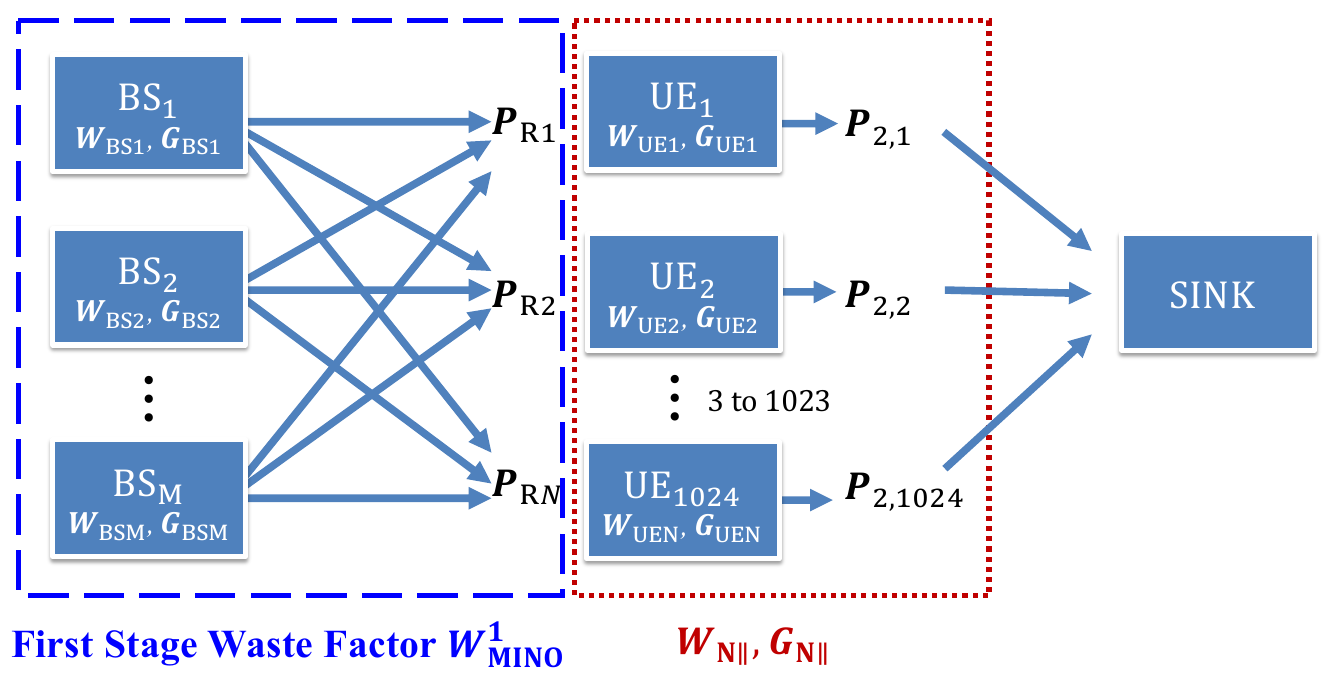}
    \caption{An M-input (where M ranges from 1 to 20) and 1024-output distributed MU-MIMO system (A single path termination model is used to analyze the energy efficiency of the entire multi-user system using a single metric \textit{W}).}
    \label{fig:MINO_UEBS}
\end{figure}

\begin{figure} [!t]
\centering
\includegraphics[width=0.9\columnwidth]{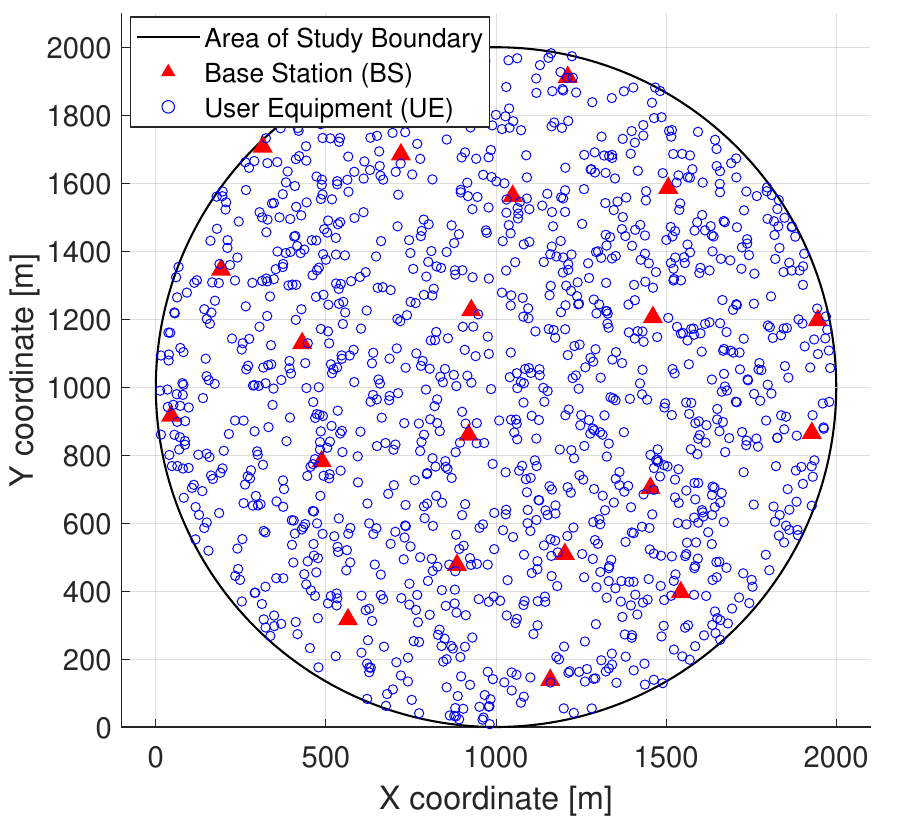}
\caption{Base station and user equipment layout with uniformed distribution in a distributed MU-MIMO system (20 BSs and 1024 UEs). }
\label{fig:UEBSLayout}
\end{figure}

\begin{table*}[htbp]
\centering
\caption{Path loss and antenna parameters at three different frequencies for distributed MU-MIMO system of Fig. \ref{fig:MINO_UEBS}.}
\begin{tabular}{|c|c|c||c|c!{\vrule width 1pt}c|c|}
\thickhline
\multirow{2}{*}{Frequency} & \multicolumn{2}{c||}{LOS Path Loss Parameters} & \multicolumn{4}{c|}{Antenna Parameters} \\ \cline{2-7}
 & Omni PLE $n$ & Omni $\sigma$ (dB) &  UE $A_{e}$ (m\(^2\)) & UE $G_\text{Ant}$ (dB) & BS $A_{e}$ (m\(^2\)) & BS $G_\text{Ant}$ (dB) \\ \thickhline
3.5 GHz & 1.82 & 4.89 & 0.00072  & 0.90  & 0.8 & 31.36 \\ \hline
17 GHz& 2.00 & 6.60 & 0.00072  & 14.63 & 0.8 & 45.09 \\ \hline
28 GHz& 2.02 & 8.98 & 0.00072  & 18.97 & 0.8 & 49.42 \\ \thickhline
\end{tabular}
\label{tab:combined-parameters}
\end{table*}

The analyzed distributed MU-MIMO system includes 1024 UEs, with the number of BSs varying from 1 to 20 over a circular area with a diameter of 2~km. Each UE has one single antenna receiver within it as shown in Fig. \ref{fig:MINO_UEBS}. The positions of UEs and BSs are randomly and uniformly distributed across the area as shown in Fig. \ref{fig:UEBSLayout}, and all the BSs are connected to a network controller for joint transmission synchronization \cite{nayebi2015cell}. Each BS is separated from the others by a minimum distance of 200 m, and each BS provides coverage within a radius of 200 m. This setup ensures overlapping coverage areas between adjacent BSs, enabling coordinated transmission across the network. Also, BSs and UEs heights are assumed to be 15 and 1.5 m above ground, respectively. This configuration represents a typical urban microcell (UMi) environment, with a bandwidth of 400 MHz and a \textit{NF} of 5 dB for each of the UEs. The BS dynamically adjusts its transmit power to ensure that the SNR at each UE is equal to 10 dB. The transmit power is capped at a maximum of 10 dBm for each BS-UE link, which restricts some UEs from achieving the desired 10~dB SNR due to propagation variations. All UEs are able to obtain 10~dB SNR with 20 BSs in all bands using directional antennas given in Table \ref{tab:combined-parameters}, and for the nonsensical case of omnidirectional antennas, simulation results with 20 BSs show that the transmit power limitation still enables more than 90\% of UEs to achieve an SNR greater than 8 dB at a carrier frequency of 28 GHz, and this percentage increases to 95\% for 17 GHz, and reaches up to 99\% at 3.5 GHz.  Simulation parameters are summarized in Table~\ref{tab:simulation}.

A single Waste Factor value for the entire distributed MU-MIMO system is calculated following the methodology outlined in \mbox{Section~III-\ref{sec:MIMO}}. In Fig. \ref{fig:MINO_UEBS}, the BSs shown within the leftmost dashed box represent the M (range from 1 to 20) BSs used in the simulation, while the UEs shown within the middle dotted box represent the N UEs (N = 1024), and each UE is equipped with a single antenna (note: using eqs. \eqref{2I2O_W_nonco_gamma} and \eqref{W2I2O_1st_coh} in Section III, we could easily consider each of the UE together with the BSs in Fig. \ref{fig:MINO_UEBS} being a MIMO system). 

The entire distributed MU-MIMO network can be conceptualized as a three-stage source-to-sink cascade as shown in Fig. \ref{fig:MINO_UEBS}, wherein a final-stage imaginary sink is used to non-coherently combine the output power from all of the distributed UEs to determine a single ``sum of powers" \textit{W} value for the entire distributed MU-MIMO system. Such an approach is novel, yet sensible and convenient,  as it permits a very complex system of many transmitters and receivers to be characterized by a single, unifying metric, whereby each element of the network  (including all of the devices and even all of the channels), are assumed to combine in an imaginary sink. This modeling method enables the total power consumption and power waste of each element, e.g., the additive wasted power of each BS, UE, and channel, to be conveniently accounted for in Waste Factor analysis with a single metric that represents the signal-path power waste of the entire system. In this manner, the absolute and relative energy waste of any complex communication system with many users and parallel paths may be carried out with ease by adopting the Waste Factor theory since each element in the distributed MU-MIMO system can easily be modeled for its own energy efficiency.

Now, we consider the channels between the BSs and UEs shown in Fig. \ref{fig:MINO_UEBS}. The path loss for each BS-UE link determines the Waste Factor of the channel. In the simulated network with parameters shown in Table \ref{tab:simulation} and Table \ref{tab:combined-parameters}, the PL between the BSs and UEs is quantified using the Close-In (CI) free space reference distance model, which is anchored at a 1~m reference distance. Omnidirectional path loss models for 3.5, 17, and 28 GHz wireless channels adopted in the simulation are derived from extensive urban propagation studies conducted by various research groups \cite{shakya2024radio, Samman2016Channel17GHz, adegoke2021channel}. The channel parameters, including the Path Loss Exponent (PLE, denoted as $n$) and large-scale shadow fading standard deviation ($\sigma$) \cite{rappaport2015wideband}, and directional antenna gain for these frequencies are compiled in Table~\ref{tab:combined-parameters}. 

Simulations were conducted of the distributed MU-MIMO network illustrated in Fig.~\ref{fig:UEBSLayout}, utilizing both omnidirectional and directional antennas. The omnidirectional scenario assumes isotropic radiation for both BS and UE antennas, with a nominal gain of 0 dB, emitting signals uniformly in all directions. It is noteworthy that 0 dB gain antennas are not deployed in practice. However, they persist in theoretical system analysis results from researchers due to a naive, erroneous, and persistent lack of understanding regarding the practical physical implementation of antennas and the fact that practical modern wireless antennas always have directivity and gain, which both increase by the square of increasing frequency.

We also considered the more reasonable and practical scenario where directional antennas at both the BS and UE are used, and where we assumed the best and proper pointing directions are found from beam steering protocols and the LOS Omni PL mode was used. We further assumed that for each three frequency bands, the antennas maintained a physical size that was static, such that a constant physical area across all studied frequencies was assumed, as is done in practice. Our simulations assumed physical dimensions of 1~m $\times$ 1~m for the BS antennas and 3~cm $\times$ 3~cm for the UE antennas, with an assumed antenna efficiency of 80\% at all three bands of study \cite{balanis2016antenna, Rappaport2024wireless}.

The antenna gains $G_\text{Ant}$ for the directional and omnidirectional transmission across various frequencies are summarized in Table~\ref{tab:combined-parameters}. Antenna gains for each BS and UE are calculated using
\begin{equation}
G_\text{Ant}(\text{dB}) = 10 \log_{10}\left(\frac{4 \pi \eta_\text{Ant} A_p}{\lambda^2}\right),
\end{equation}
where $\eta_\text{Ant}$ is the antenna efficiency, $A_p$ is the physical area of the antenna, and $A_{e} = \eta_\text{Ant} A_p$ is the effective antenna aperture. The term $\lambda = c / f$ represents the wavelength, which varies inversely with the carrier frequency $f$, and where $c$ is the speed of light $c= 3\times10^{8}\, \text{m/s}$\cite{rappaport2015millimeter}.

To determine the \textit{W} and total power consumption of the entire distributed MU-MIMO network composed of M BSs and 1024 UEs, we first determine the \textit{W} for each BS and UE separately using eqs. \eqref{eq:W_RU} and \eqref{eq:W_UE}, respectively. 
Based on the detailed gain and efficiency and DC power consumption from data sheets of real-world components, we apply equations \eqref{eq:passive} and  \eqref{ActiveW} for passive and active components in the RAN system given in Section~\ref{sec:RAN_intro}, and \eqref{Wcas} to compute the Waste Factor for a UE and BS, and we assume all BSs have identical $W_{\text{BS}}=15$ and $G_\text{BS}=30\,\text{dB}$, and that all 1024 UEs have the same $W_\text{UE}=33$ and $G_\text{UE}=11\,\text{dB}$. Also, we assume that the non-path power for each BS is $P_{\text{non-path}}^{\text{BS}}= 140\,\text{W}$ (consistent with the values used in Section~\ref{sec:Analysis_RAN}-B), and the non-path power for each UE is $P_{\text{non-path}}^{\text{UE}}= 1\,\text{W}$, which is a typical value since the display and the processor of the UE will averagely consume around 1 W \cite{carroll2010analysis}. All channels shown in Fig. \ref{fig:MINO_UEBS} are different and yield different values of $W_{\text{C}}$ based on the simulation of the propagation, but we dynamically adjust the transmit power in the simulation (to model distributed MU-MIMO) for each of the parallel BSs to achieve a 10 dB SNR at the input of all UEs, however, in some cases, unrealistic propagation conditions for omnidirectional antennas do not allow all UE to achieve a 10 dB SNR as the BSs transmit power is limited to 10 dBm for each BS-UE link. 

Considering the distributed MU-MIMO system shown in Fig. \ref{fig:MINO_UEBS}, a sub-system comprising of M BSs from the leftmost box and one UE in the middle box can be considered a MISO system. The entire distributed MU-MIMO system in Fig. \ref{fig:MINO_UEBS} can thereby be considered a parallel system of several MISO networks where each UE is the single output terminus of all BS transmitters that are simulated to be within and serving the 200~m range of the UE, as shown in the MISO architecture of Fig. \ref{fig:MISO_M}. This same approach is applied over all 1024 UEs for the particular serving BSs within 200 m range (note that here we have assumed each UE has a single antenna, but we can also assume each UE could be equipped with multiple antennas, then each system involving multiple BSs and one UE can be considered as a MIMO system as shown in Section III. The input power at each UE can be calculated using coherent or non-coherent combining, as described by equations \eqref{PRX_coh} and \eqref{PRX_noncoh}
). For the $n^{th}$ UE in Fig.~\ref{fig:MINO_UEBS}, we compute the $W_{\text{M}||}^{n}$ for the MISO system that consists of parallel BSs and channels (see the leftmost dashed box of Fig. \ref{fig:MINO_UEBS}) that transmit to the input of each of the UEs using \eqref{W_par_inco_gamma}. By repeating the above step for every UE, a value of  $W_{\text{M}||}^{n}$ is computed for each UE using eq. \eqref{W_par_inco_gamma}, thereby providing values of $W_{\text{M}||}^{n}$ for all of the 1024 MISO systems in the leftmost dashed box of Fig. \ref{fig:MINO_UEBS}. 

The signal path power consumption of the leftmost dashed box of Fig. \ref{fig:MINO_UEBS}, which includes the channels, is then obtained by multiplying each of the 1024 $W_{\text{M}||}^{n}$ of the MISO system composed of multiple BSs and wireless channels, with the power received at the $n^{th}$ UE, $P_{\text{R}n}$ (it would be easy for the reader to apply Section III Fig. \ref{fig:MINO}, eqs. \eqref{PRX_noncoh} and \eqref{PRX_coh} to handle the input power for each UE that was itself a MIMO receiver), where we assume non-coherent combining at the input of each UE (we could also just as easily have assumed coherent combining at the input of each UE, as described in Section~III). The total signal path power consumption for the leftmost dashed box in Fig.~\ref{fig:MINO_UEBS} is then calculated using eq.~\eqref{defW}, where \( P_{\text{consumed,path}}^{1} \) is the sum of the product of received power at each UE (\( P_{\text{R}n} \)) and the corresponding waste factor (\( W_{\text{M}||}^{n} \)) for all UEs: $P_{\text{consumed,path}}^{1} = \sum_{n=1}^{1024} (P_{\text{R}n} W_{\text{M}||}^{n})$

Referring to Fig.~\ref{fig:MINO_UEBS}, the distributed MU-MIMO system is represented as a three-stage cascade. In the initial stage (leftmost, including the channels), the signal path power consumption, denoted as $P_{\text{consumed,path}}^{1}$, is established as discussed in the above paragraph. We can now compute the Waste Factor for this leftmost first stage dashed box in Fig.~\ref{fig:MINO_UEBS}, which involves dividing $P_{\text{consumed,path}}^{1}$ by the total power received by all parallel UEs (the output power of the first stage), expressed as $P_{\text{signal}}^{1}=\sum_{n=1}^{1024} P_{\text{R}n}$. The formula is detailed in eq.~\eqref{WMIMO_1st_UEBS},
\begin{equation}
\label{WMIMO_1st_UEBS}
    W_{\text{MINO}}^{1} = \frac{P_{\text{consumed,path}}^{1}}{P_{\text{signal}}^{1}} = \frac{\sum_{n=1}^{1024} \left(P_{\text{R}n} W_{\text{M}||}^{n} \right)}{\sum_{n=1}^{1024} P_{\text{R}n}}.
\end{equation}

 \begin{figure} [!t]
\centering
\includegraphics[width=0.95\columnwidth]{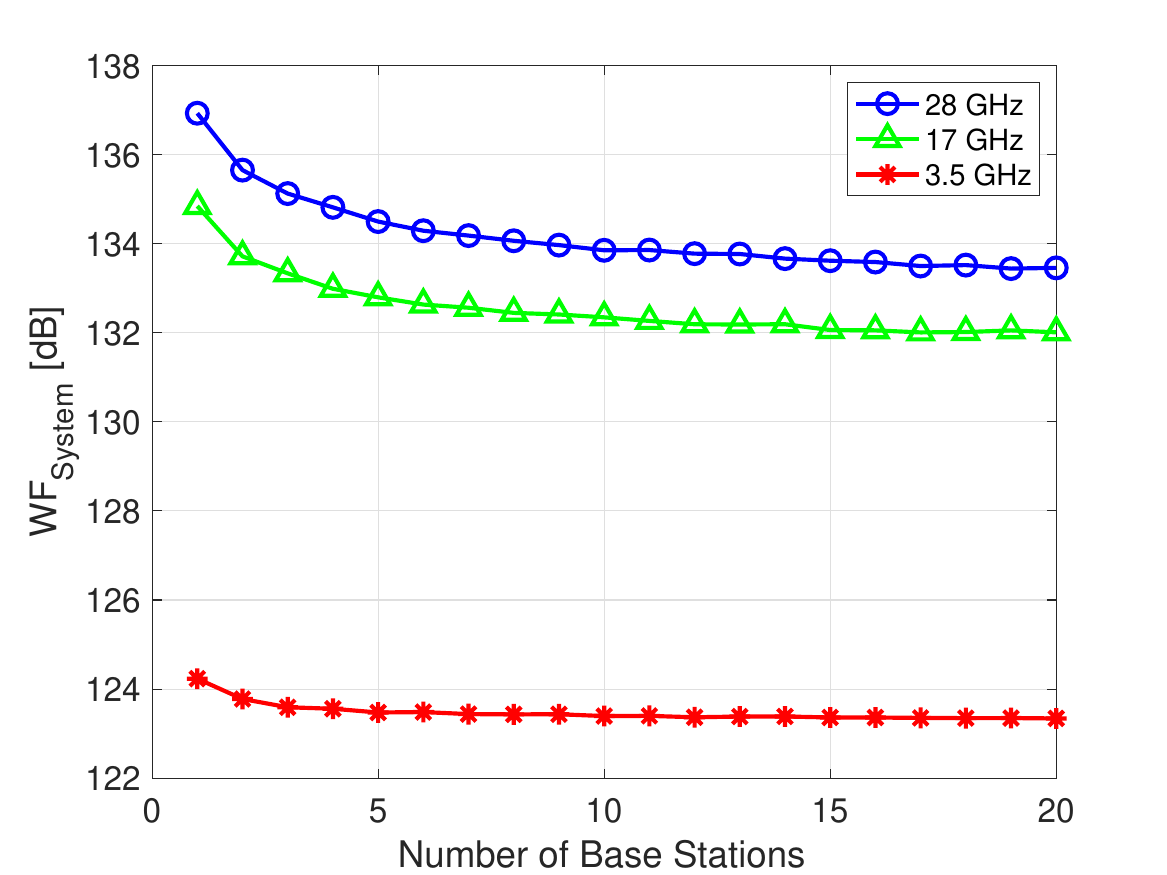}
\caption{Waste Figure of the entire distributed MU-MIMO system with different BS numbers per area of study in 3.5, 17, and 28 GHz DL distributed MU-MIMO system with (unrealistic) omnidirectional antennas for both BSs and UEs. }
\label{fig:CoMPLOS}
\end{figure}

\begin{figure} [!t]
\centering
\includegraphics[width=0.95\columnwidth]{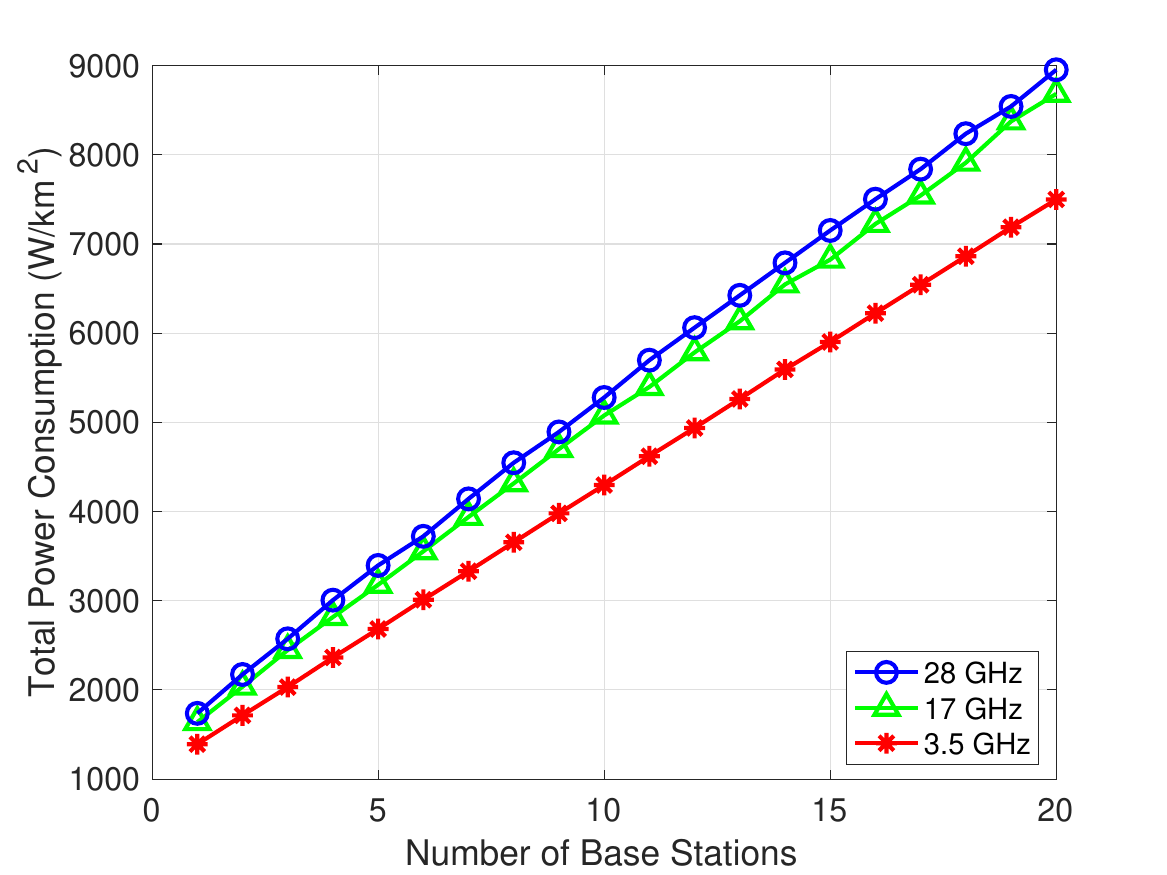}
\caption{Total power consumption of the entire distributed MU-MIMO system with different BS numbers per square kilometer area in 3.5, 17, and 28 GHz DL distributed MU-MIMO system with (unrealistic) omnidirectional antennas for both BSs and UEs. }
\label{fig:totpc_omni}
\end{figure}

For 1024 spatially distributed receivers represented in the middle dotted box in Fig.~\ref{fig:MINO_UEBS}, we assume the parallel 1024 UEs and the sink form a MISO system using non-coherent combining at the input of the sink, as taught in \mbox{Section III-\ref{sec:MISO}}. The Waste Factor for the MISO system of 1024 UEs, $W_{\text{N}||}$, is obtained by using eq.~\eqref{W_par_inco_gamma},
\begin{equation}
    \label{W_N_UE}
    W_{\text{N}||} = \frac{\sum_{n=1}^{1024} ( P_{2,n} W_{\text{UE}} )}{\sum_{n=1}^{1024} P_{2,n}} = W_{\text{UE}},
\end{equation}
and $G_{\text{N}||}$ can be computed using eq. \eqref{gain_co_nonco} for non-coherent combining, which is
\begin{equation}
    \label{G_N_UE}
    G_{\text{N}||} = \frac{\sum_{n=1}^{1024} \left(P_{\text{R}n} G_{\text{UE}}\right)}{\sum_{n=1}^{1024} P_{\text{R}n}} = G_{\text{UE}}.
\end{equation}

Given the first stage Waste Factor $W_{\text{MINO}}^{1}$ (e.g., the leftmost box in Fig. \ref{fig:MINO_UEBS}, which includes all BSs and all channels), and given $W_{\text{N}||}$ and $G_{\text{N}||}$ for the MISO system that consists of the parallel 1024 UEs in the dotted box (e.g., second stage) and the (imaginary) sink of Fig. \ref{fig:MINO_UEBS}, we are able to compute the Waste Factor for the entire distributed MU-MIMO system using eq. \eqref{W_MIMO_general}
\begin{equation}
    \label{W_MIMO_UEBS}
    W_{\text{System}} =  W_{\text{N}\mathbin{\parallel}}  + \frac{(W_{\text{MINO}}^{1} - 1)}{G_{\text{N}\mathbin{\parallel}}}.
\end{equation}

When the overall PL values between the BSs and UEs decrease, the \( W_{\text{cascade},i} \) in \eqref{W_par_inco_gamma} also decreases, and we observe a reduction in each of the \( W_{\text{M}\parallel}^{i}\).
As evident from equations \eqref{WMIMO_1st_UEBS} and \eqref{W_MIMO_UEBS}, the decrease in \( W_{\text{M}\parallel} \) leads to a decreasing \textit{W} (and increasing energy efficiency) of the entire distributed MU-MIMO system.
Note that we could have combined all UEs using MISO coherent combining at the imaginary sink, but non-coherent seems more practical in energy analysis as a ``sum of powers" for a widely distributed user base. 

\begin{figure} [!t]
\centering
\includegraphics[width=0.95\columnwidth]{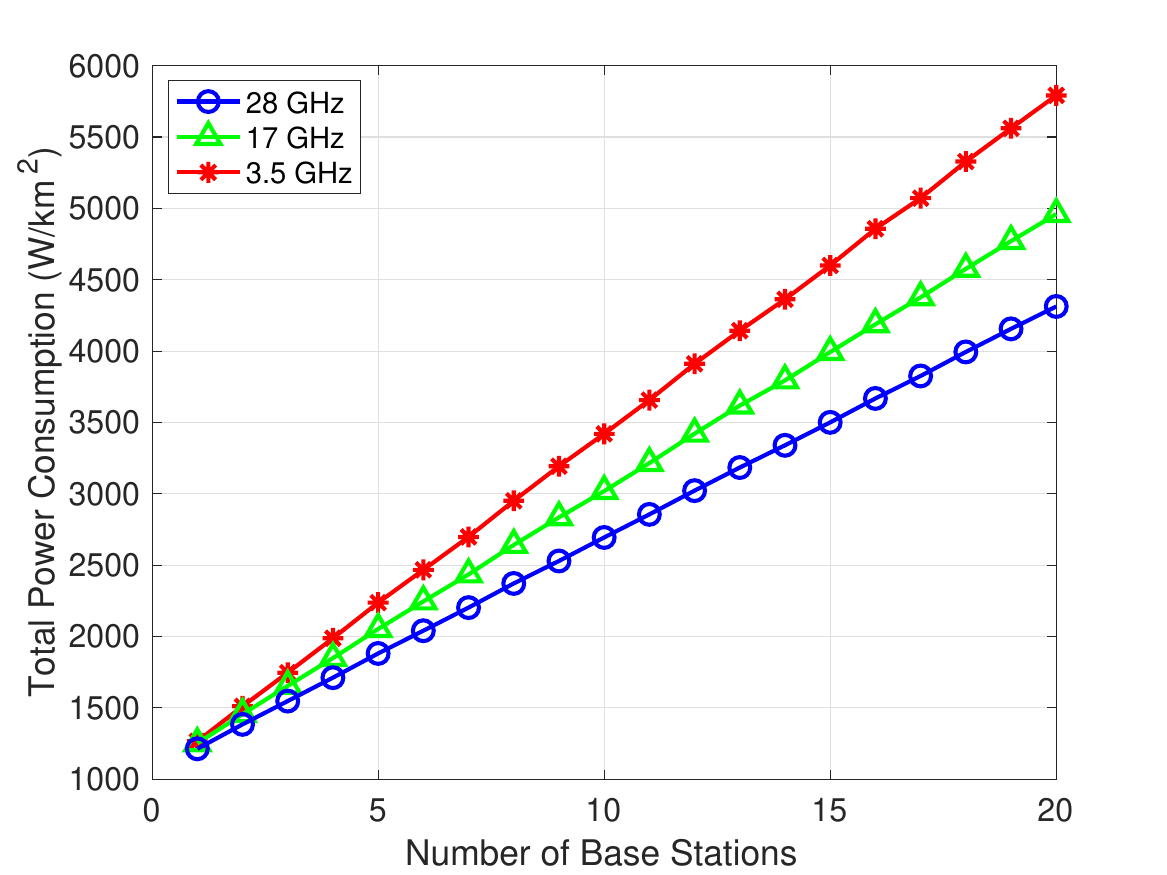}
\caption{Total power consumption of the entire distributed MU-MIMO system with different BS numbers per area of study in 3.5, 17, and 28 GHz DL distributed MU-MIMO system with (realistic) directional antennas listed in Table \ref{tab:combined-parameters} for both BSs and UEs. }
\label{fig:totpc_dir}
\end{figure}

Using \eqref{W_MIMO_UEBS}, we can compute the total power consumption of the system using \eqref{powercontotal}, which is given by
\begin{equation}
\label{MIMO_total}
       P_{\text{consumed,total}} = W_{\text{System}} P_{\text{System,out}} + P_{\text{non-path}}^{\text{System}},
\end{equation}
where $P_{\text{System,out}}= G_{\text{UE}} P_{\text{signal}}^{1}  = G_{\text{UE}} \sum_{n=1}^{1024} P_{\text{R}n}$ is the output power from the parallel UEs applied to the imaginary sink in Fig. \ref{fig:MINO_UEBS}, and $P_{\text{non-path}}^{\text{System}} = \text{M} \times P_{\text{non-path}}^{\text{BS}} + 1024 \times P_{\text{non-path}}^{\text{UE}}$ is the non-path power consumption of the distributed MU-MIMO system. We scale our simulation results from a $\pi\, \text{km}^{2}$ system of Fig. \ref{fig:UEBSLayout} to a standard area of 1~km$^2$, and the signal-path power consumption is divided by $\pi$ (the area of the circle).
The total power consumption per square kilometer is depicted in Fig.~\ref{fig:totpc_omni} with omnidirectional antennas and in Fig.~\ref{fig:totpc_dir} with directional antennas shown in Table \ref{tab:simulation}.

\begin{figure} [!t]
\centering
\includegraphics[width=0.95\columnwidth]{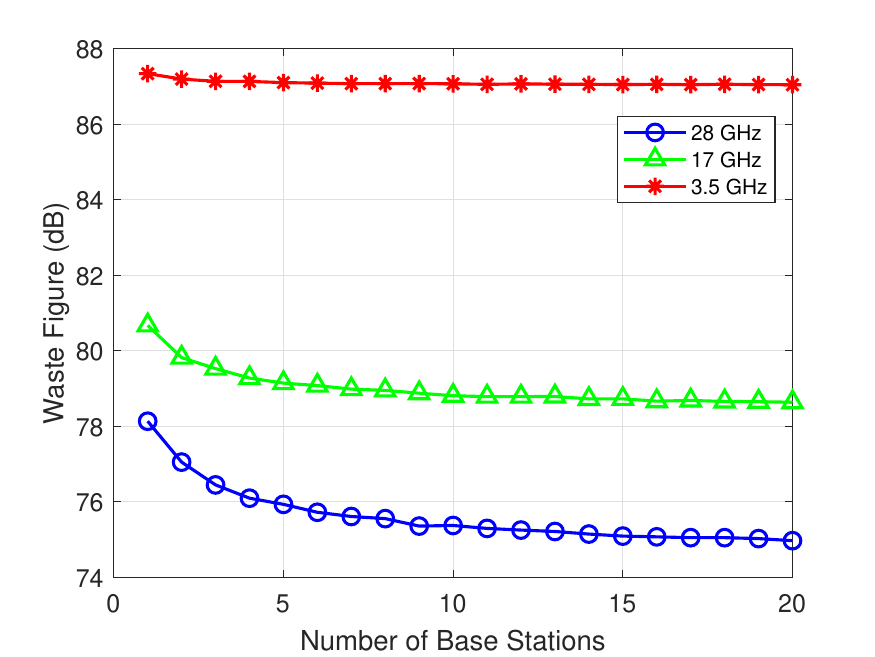}
\caption{Waste Figure of the entire distributed MU-MIMO system with different BS numbers per square kilometer in 3.5, 17, and 28 GHz DL distributed MU-MIMO system with (realistic) directional antennas listed in Table \ref{tab:combined-parameters} for both BSs and UEs. }
\label{fig:CoMPLOS_dir}
\end{figure}

Fig.~\ref{fig:CoMPLOS} provides incorrect insights because of the use of 0~dB gain antennas at all three simulated carrier frequencies. At 3.5 GHz, the \textit{W} of the entire distributed MU-MIMO network shows a slight decline with an increasing number of BSs, dropping from around 118 dB to 116 dB. At 17 GHz, the $\textit{WF}_{\text{System}}$ is higher than at 3.5 GHz for the same number of TXs, due to more pronounced propagation and shadowing losses. The rate of decrease in the $\textit{WF}_{\text{System}}$ initially exhibits a sharp decline with the addition of BSs; however, this rate of reduction slows significantly after surpassing a certain number of BSs. At 28 GHz, $\textit{WF}_{\text{System}}$ for small numbers of TXs is around 132 dB, decreasing only slightly with increased BS density to 128 dB with 20 BSs. These results indicate that even with a higher density of BSs, omnidirectional antennas incur substantial channel losses, leading to considerable power wastage. It is noteworthy to see how a single metric, \textit{W}, is able to determine the aggregate power waste for delivering signals through a very complex multi-user network that includes multiple parallel channels. It is also clear using eq. \eqref{MIMO_total} that total consumed power is easily computed using the Waste Factor theory by summing up the non-path powers of all BSs and UEs to determine the total power consumption of the entire distributed MU-MIMO network shown in Fig.~\ref{fig:MINO_UEBS}.

Fig.~\ref{fig:totpc_omni} illustrates the total power consumption of a distributed MU-MIMO system utilizing omnidirectional antennas across three carrier frequencies. By aggregating signal power from 1024 UEs at an imagined sink, and employing eq. \eqref{MIMO_total}, the simulation calculates the total power consumption of the simulated system.\footnote{
In the assessment of energy efficiency using \textit{W}, it is essential to clarify the relationship between power (P) and energy (E). The relationship between these quantities over a unit of time T in seconds is given by E (Joules) = P (Watts) \(\times\) T (s). In the context of our study, this implies that power efficiency and energy efficiency can be considered synonymous over a fixed time period. This clarification ensures that our analysis of true power levels over space (W/km\(^2\)) is well-grounded and accurately interpreted.
} The total power consumption at 3.5 GHz is considerably lower compared to higher frequencies due to the nonsensical assumption that omnidirectional antennas are used in all three bands such that channel loss increases by the square of frequency without any benefit of channel gain that comes from directional antennas. Here, we see clearly how the spherical radiation spreading for omnidirectional antennas \cite{Rappaport2024wireless, xing2018propagation, rangan2014millimeter, rappaport2015millimeter} include more propagation loss at 17 GHz, where the total power consumption is slightly lower than at 28 GHz, but higher than 3.5 GHz. At 28 GHz, the total power consumption is the highest among all three frequencies and increases with the number of BSs and is 30\% higher than at 3.5 GHz with 20 BSs.

Fig.~\ref{fig:CoMPLOS_dir} illustrates a more realistic mobile scenario where $\textit{WF}_{\text{System}}$ is computed for a distributed MU-MIMO system using directional antennas.
Here, the $\textit{WF}_{\text{System}}$ is remarkably reduced for all three bands as
compared to Fig. \ref{fig:CoMPLOS}, demonstrating how antenna gains yield less lossy channels that offer dramatic power savings and improved energy efficiency as compared to the omnidirectional systems. $\textit{WF}_{\text{System}}$ at 3.5 GHz, though reduced compared to the omnidirectional scenario of Fig.~\ref{fig:CoMPLOS}, is larger than the other higher frequencies due to the limited antenna gains at this band (see Table \ref{tab:combined-parameters}). At 17 GHz, the $\textit{WF}_{\text{System}}$ decreases rapidly with up to 5 BSs from 81 dB with 1 BS to 79 dB with 5 BSs, but the reduction rate stabilizes as the BS count rises further, highlighting the efficiency gains from directional antennas initially. For the 28 GHz band, the $\textit{WF}_{\text{System}}$ drops significantly with increasing BS density, dropping from 78 dB for 1 BS to 76 dB for 5 BSs and 75 dB for 15 BSs, showing a clear advantage due to directional antennas which counter the propagation and shadowing loss effects. The $\textit{WF}_{\text{System}}$ with 20 BSs is 12~dB lower than at 3.5~GHz with 20 BSs, highlighting the substantial power efficiency gains (more than an order of magnitude of power savings) at higher frequencies. Even greater efficiencies occur due to the fact that 5G mmWave technology is inherently more power efficient (by 2$\times$) than legacy 4G technology (see Section~\ref{sec:RAN_intro}), thereby showing the move to mmWave small cells offers much greater power efficiency over current cellular networks.

Fig.~\ref{fig:totpc_dir} shows the total power consumption of a distributed MU-MIMO system employing directional antennas across three different carrier frequencies. By collating signal path power outputs from 1024 parallel UEs at an imagined sink, we utilize eq. \eqref{MIMO_total} to compute the total power consumption. Notably, the 3.5 GHz band shows the highest total power consumption, which increases significantly with the number of BSs, being 34\% higher than at 28 GHz with 20 BSs. Meanwhile, power consumption at 17 GHz is less than at 3.5 GHz but higher than at 28 GHz, which is the lowest across the frequencies studied.

These findings show how \textit{W} may be used in an easy, simplified but unified manner to characterize the energy efficiency and total power consumption of very complex multi-user systems.  The results here show improved power efficiency and reduction in \textit{W} when moving higher in frequency and using directional antennas of constant physical area, challenging the traditional understanding that lower frequencies are inherently more energy-efficient~\cite{auer2010enablers}.

The above results highlight four key phenomena:
\begin{enumerate}
    \item Increasing the density of BSs at higher frequencies significantly enhances energy efficiency by improving the communication channel between the UE and BS. However, there is a limit to these benefits; beyond a certain point, further increases in BS density yield diminishing returns as the channel quality impact reaches a saturation point. This limitation might be influenced by specific constraints in our simulation settings, such as the minimum distance between BSs and the maximum BS coverage area.

    \item The use of directional antennas reveals an inverse relationship between $\textit{WF}_\text{System}$ and frequency. This advantage arises because directional antennas effectively counteract the increased free space losses typically seen in the first meter for higher-frequency wireless networks \cite{sun2016investigation, Rappaport2019Beyond6G, xing2018propagation}.

    \item An increase in the number of BSs naturally leads to a rise in total power consumption, observed with both omnidirectional and directional antennas. Specifically, at 28 GHz, the simulation using directional antennas demonstrates the lowest total power consumption required to maintain an SNR of 10 dB at the UE side, consistent across all three simulated frequencies. Consequently, to enhance the network energy efficiency while maintaining equivalent throughput, it is advisable to operate at higher carrier frequencies with high-gain antennas. Waste Factor is shown to enable the power consumption of complex systems to be analyzed easily.
    
    \item Observing the total power consumption results in Fig.~\ref{fig:totpc_dir}, it is crucial to note that the slope at 28 GHz is notably minimal when employing directional antennas. This observation underscores the potential for reduced electricity costs and more economically sustainable system growth as more BSs are added at higher frequencies. Moreover, if these BSs exhibit lower non-path power consumption (i.e., are more energy-efficient), there will be additional energy savings, reducing power consumed shown in Fig \ref{fig:totpc_dir}, which underscores the importance of utilizing higher frequencies.
    
    \item In analyzing the power consumption across different antenna configurations, our findings in Figure~\ref{fig:totpc_omni} and Figure~\ref{fig:totpc_dir} distinctly highlight the energy efficiency gains of directional antennas. Specifically, for 20 BSs operating at 28 GHz with directional antennas, the power consumption is markedly lower, at only 4.4 kW/km$^2$, compared to 9 kW/km$^2$ with omnidirectional antennas. This trend is consistent across different densities of BSs. Utilizing $W$ simplifies the calculation of total power and energy consumption in real networks, providing a clear and measurable insight into actual power and energy levels.

\end{enumerate}

\section{Limitations and Challenges of Waste Factor Adoption}

While \textit{W} presents a novel approach to evaluate energy efficiency in communication systems, some limitations and challenges need to be addressed for its broader adoption and application in the future:
\begin{itemize}
    \item \textbf{Linearity Requirement}: One of the primary limitations of \textit{W} is that it requires the components of the system to exhibit linear behavior. Real-world communication systems include non-linear components, which complicates the accurate application of \textit{W}. The applicability to linearized models, and the sensitivity of Waste Factor in the face of real-world non-linearities are open areas of work.
    
    \item \textbf{Novelty and Lack of Comprehensive Application}: As \textit{W} is a relatively new concept, there is limited practical experience in its application. Researchers and engineers are still exploring how to fully utilize \textit{W} in various contexts. This lack of comprehensive application knowledge means that further studies and practical implementations are necessary to understand its full potential and limitations.
    
    \item \textbf{Modeling Various Quiescent States}: Accurate modeling of different quiescent or ``stand-by'' states within communication systems may be challenging. The ability to effectively model these states and to assign the power consumption as being on the signal path or off the signal path is crucial for the successful usage of \textit{W}, and methodologies must be developed and refined through experience, learning, and trial-and-error.

    \item \textbf{Unexplored Network Architectures}: Certain network architectures, such as ring networking structures with multiple switches and routers, have not yet been thoroughly analyzed using \textit{W}. These architectures present unique challenges and opportunities for applying \textit{W}, and further research is needed to explore these areas.

    \item \textbf{Combining Method Variability}: The method of combining signals in the system plays a crucial role in determining the appropriate Waste Factor formula. In some network scenarios, parts of the devices may use non-coherent combining while others use coherent combining. This variability poses an open challenge that needs to be addressed further to ensure accurate and effective application of \textit{W}.
    
\end{itemize}

By acknowledging these limitations and challenges, we aim to provide a balanced perspective on the potential of the Waste Factor and highlight the need for ongoing research and practical experimentation to fully realize its benefits in improving energy efficiency in communication systems.

\section{Conclusion}
In this paper, we introduced \textit{W}, or \textit{WF} in dB, as a new figure of merit for power efficiency evaluation in circuits, cascaded, and parallel systems for components along the signal path in a device or cascade. By providing a systematic comprehensive analysis of power utilization and additive power waste across the system components on the signal paths in realistic RAN systems, we showed that \textit{W} overcomes the limitations of conventional energy efficiency metrics. The granular insights offered by \textit{W} enable a deeper understanding of the complex energy dynamics within network systems including various network configurations, such as MISO, SIMO, and MIMO systems. Simulation results for a distributed MU-MIMO system at 3.5, 17, and 28 GHz highlight the significance of \textit{W} in developing energy-efficient infrastructures, which reveals that both increasing BS density and using higher carrier frequencies improve system power utilization by reducing the Waste Factor, thereby enhancing network energy efficiency. Waste Factor equips network operators with valuable insights for optimizing next-generation wireless communications systems. The potential of \textit{W} to inform and drive energy optimization strategies makes it a promising tool in advancing the sustainability goals of 5G and 6G network technologies.

As a foundational metric, \textit{W} opens up new avenues for future research and practical applications in all facets of communications, including wireless communications. Future research could focus on refining non-path power to more accurately represent real-world total power consumption and incorporating the Waste Factor into various scenarios, such as different computing platforms. By leveraging this metric, the industry can make tangible progress toward designing energy-efficient network infrastructures, leading to a greener future in wireless communications.

\bibliographystyle{IEEEtran}
\bibliography{reference}
 
\input{acronyms}

\begin{IEEEbiography}[{\includegraphics[width=1in,height=1.25in,clip,keepaspectratio]{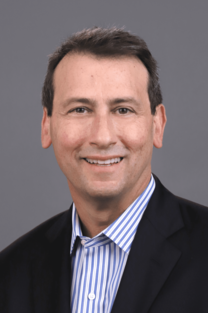}}]{Theodore S. Rappaport } (Fellow, IEEE)
is the David Lee/Ernst Weber Professor with New York University (NYU) and holds faculty appointments in the Electrical and Computer Engineering Department of the NYU Tandon School of Engineering, the Courant Computer Science Department, and the NYU Langone School of Medicine. He is the Founding Director of NYU WIRELESS, a multidisciplinary research center focused on the future of wireless communications and applications. His research has led the way for modern wireless communication systems. In 1987, his Ph.D. at Purdue University provided fundamental knowledge of indoor wireless channels used to create the first Wi-Fi standard (IEEE 802.11), and he conducted fundamental work that led to the first US Digital cellphone standards, TDMA IS-54/IS-136, and CDMA IS-95. He pioneered the use of site-specific design and optimization of wireless networks, and he and his students engineered the world’s first public Wi-Fi hotspots. His work proved the viability of millimeter waves for mobile communications and the global wireless industry adopted his vision for 5th generation (5G) millimeter wave cellphone networks. His most recent research has proven the viability of sub-terahertz wireless communications and position location for 6G, 7G and beyond. 

He founded three academic wireless research centers at Virginia Tech, The University of Texas, and NYU that have produced thousands of engineers and educators since 1990, and he has coauthored over 300 papers and twenty books, including the most cited books on wireless communications, adaptive antennas, wireless simulation, and millimeter wave communications. He co-founded two wireless companies, TSR Technologies and Wireless Valley Communication, which were sold to publicly traded companies, and he has advised many others. He co-founded the Virginia Tech Symposium on Wireless Communications in 1991, the Texas Wireless Summit in 2003, and the Brooklyn 5G Summit (B5GS) in 2014. He has more than 100 patents issued and pending, served on the Technological Advisory Council of the Federal Communications Commission (FCC), is a member of the National Academy of Engineering, a member of the Wireless History Foundation Wireless Hall of Fame, a member of the IEEE Vehicular Technology Society Hall of Fame, a Fellow of the Radio Club of America and the National Academy of Inventors, a life member of the American Radio Relay League, a licensed professional engineer in Texas and Virginia, and an amateur radio operator (N9NB). 

He has received IEEE’s Eric Sumner Award, ASEE’s Terman Award, The Sir Monty Finniston Medal from the Institution of Engineering and Technology (IET), the IEEE Vehicular Technology Society’s James R. Evans Avant Garde and Stu Meyer Awards, the IEEE Education Society William E. Sayle Award for achievement in education, the IEEE Communications Society Armstrong Award, and the Armstrong Medal and Sarnoff Citation from the Radio Club of America.
\end{IEEEbiography}

\begin{IEEEbiography}[{\includegraphics[width=1in,height=1.25in,clip,keepaspectratio]{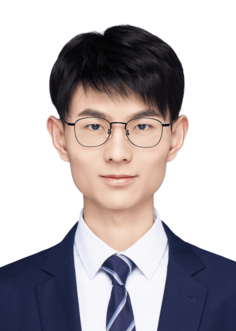}}]{Mingjun Ying } (Graduate Student Member, IEEE) received the B.E. degree in communication engineering from the Chongqing University of Posts and Telecommunications, Chongqing, China, in 2023. He is currently pursuing a Ph.D. degree in electrical engineering with the NYU WIRELESS Research Center, New York University Tandon School of Engineering, Brooklyn, NY, USA, under the supervision of Prof. Theodore S. Rappaport. His research interests include green communication, channel modeling, adversarial communication, satellite communications, and integrated sensing and communication.
\end{IEEEbiography}

\begin{IEEEbiography}[{\includegraphics[width=1in,height=1.25in,clip,keepaspectratio]{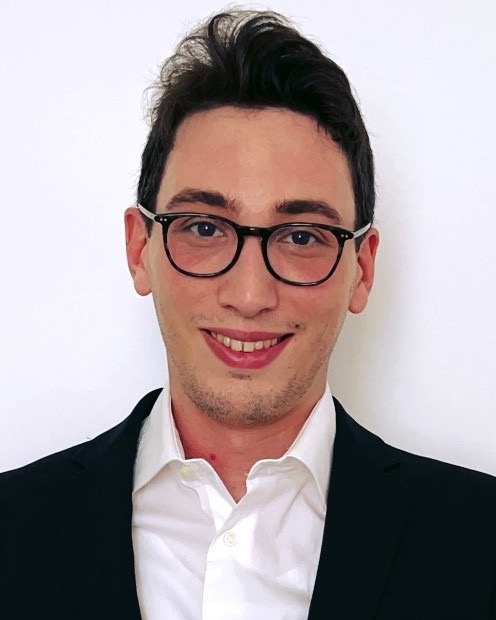}}]{Nicola Piovesan } (Member, IEEE)
is a Senior Researcher at Huawei Technologies in Paris, France. His work focuses on large-scale network modeling, data-driven network optimization, green networking, and the integration of artificial intelligence in the telecom domain. He earned his PhD in Network Engineering from the Polytechnic University of Catalonia (UPC) in Barcelona, Spain, in 2020. Awarded a Marie Skłodowska-Curie fellowship by the European Commission in 2016, he served as an Assistant Researcher at the Centre Tecnològic de Telecomunicacions de Catalunya (CTTC) in Barcelona, Spain, from 2016 to 2019. He has authored over 30 research articles and holds co-inventorship in over 10 patent applications. His achievements have been recognized with the Huawei GTS President Award in 2021 and the Huawei Quality Star Award in 2024, acknowledging his successful research application into product development.
\end{IEEEbiography}

\begin{IEEEbiography}[{\includegraphics[width=1in,height=1.25in,clip,keepaspectratio]{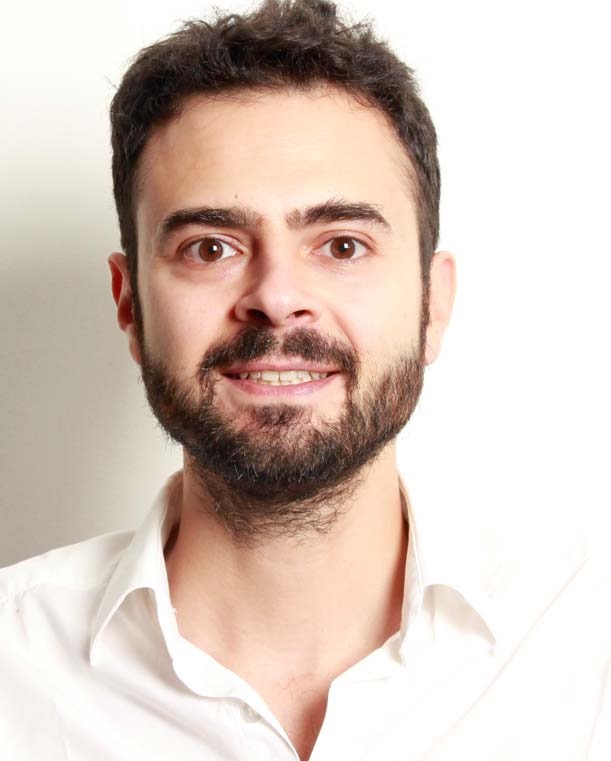}}]{Antonio De Domenico } (Member, IEEE)
received the M.Sc. degree in telecommunication engineering from the University of Rome La Sapienza in 2008 and the Ph.D. degree in telecommunication engineering from the University of Grenoble in 2012. From 2012 to 2019, he was a Research Engineer with CEA LETI MINATEC, Grenoble, France. In 2018, he was a Visiting Researcher with the University of Toronto, Canada. Since 2020, he has been a Senior Researcher with Huawei Technologies France SASU, Paris, France. He is the main inventor or a co-inventor of more than 25 patents. His research interests include heterogeneous wireless networks, machine learning, and green communications. Since 2023, Antonio is co-leading the network energy efficiency activities within the green future network project of the NGMN Alliance.
\end{IEEEbiography}

\begin{IEEEbiography}[{\includegraphics[width=1in,height=1.25in,clip,keepaspectratio]{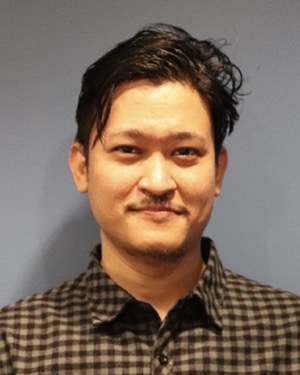}}]{Dipankar Shakya} (Graduate Student Member, IEEE)
received the B.E. degree in electronics and communications from Tribhuvan University, Nepal, in 2016, and the M.S. degree in electrical engineering from New York University in 2021. He is currently pursuing a Ph.D. degree in electrical engineering with the NYU WIRELESS Research Center, New York University Tandon School of Engineering, Brooklyn, NY, USA, under the supervision of Prof. Theodore S. Rappaport. In 2019, he joined the NYU WIRELESS Research Center, following three years of service as an engineer for flood early warning systems in South Asia. His research interests include mid-band, millimeter-wave, and Terahertz channel propagation measurements, modeling, RF circuit design, and metrology.
\end{IEEEbiography}

\end{document}

%% file: Figures/fig_channel_W.tex
\begin{tikzpicture}
\begin{axis}[
    xlabel={$WF_\textrm{C}$ [dB]},
    ylabel={$WF_\textrm{System}$ [dB]},
    xmin=60, xmax=100,
    ymin=60, ymax=110,
    legend pos=north west,
    legend cell align={left},
    ymajorgrids=true,
    grid style=dashed,
    thick,
]

\addplot[
    color=black,
    solid,
] coordinates {
    (60, 62.5376024)
    (120, 122.53755128)
};
\addlegendentry{baseline}

\addplot[
    color=green,
    solid,
] coordinates {
    (60, 59.52735356)
    (120, 119.52725133)
};
\addlegendentry{reduced $W_\textrm{RU}$}

\addplot[
    color=red,
    dashed,
] coordinates {
    (60, 62.53757683)
    (120, 122.53755128)
};
\addlegendentry{reduced $W_\textrm{ UE}$}

\addplot[
    color=blue,
    dashed,
] coordinates {
    (60, 59.52735356)
    (120, 119.52725133)
};
\addlegendentry{improved $G_\textrm{UE}$} 

\draw[<->, >=stealth, thick] (axis cs:80,80) -- (axis cs:80,83)
    node[pos=1,anchor=east] {3dB gain};

    
\end{axis}
\end{tikzpicture}

%% file: acronyms.tex
\begin{acronym}[AAAAAAAAA]
\acro{3GPP}{3rd Generation Partnership Project}
\acro{AAU}{active antenna unit}
\acro{ADC}{analog-to-digital converter}
\acro{BS}{base station}
\acro{DAC}{digital-to-analog converter}
\acro{DC}{direct current}
\acro{EE}{energy efficiency}
\acro{ETSI}{European Telecommunications Standards Institute}
\acro{IF}{intermediate frequency}
\acro{ITU}{International Telecommunication Union}
\acro{LNA}{low noise amplifier}
\acro{LTE}{long term evolution}
\acro{LO}{local oscillator }
\acro{NR}{new radio}
\acro{NGMN}{Next Generation Mobile Networks}
\acro{KPI}{key performance indicator}
\acro{OPEX}{operational expenditure}
\acro{PA}{power amplifier}
\acro{PS}{phase shifter}
\acro{RAN}{radio access network}
\acro{RF}{radio frequency}
\acro{RRU}{remote radio unit}
\acro{RU}{radio unit}
\acro{UE}{user equipment}
\acro{VSWR}{voltage standing wave ratio}
\end{acronym}

%% file: ojcoms.bbl
\begin{thebibliography}{10}
\providecommand{\url}[1]{#1}
\csname url@samestyle\endcsname
\providecommand{\newblock}{\relax}
\providecommand{\bibinfo}[2]{#2}
\providecommand{\BIBentrySTDinterwordspacing}{\spaceskip=0pt\relax}
\providecommand{\BIBentryALTinterwordstretchfactor}{4}
\providecommand{\BIBentryALTinterwordspacing}{\spaceskip=\fontdimen2\font plus
\BIBentryALTinterwordstretchfactor\fontdimen3\font minus \fontdimen4\font\relax}
\providecommand{\BIBforeignlanguage}[2]{{%
\expandafter\ifx\csname l@#1\endcsname\relax
\typeout{** WARNING: IEEEtran.bst: No hyphenation pattern has been}%
\typeout{** loaded for the language `#1'. Using the pattern for}%
\typeout{** the default language instead.}%
\else
\language=\csname l@#1\endcsname
\fi
#2}}
\providecommand{\BIBdecl}{\relax}
\BIBdecl

\bibitem{WorldBank2023}
\BIBentryALTinterwordspacing
{World Bank}, ``Green digital transformation: How to sustainably close the digital divide and harness digital tools for climate action,'' Climate Change and Development, Washington, D.C.: World Bank Group, 2023. [Online]. Available: \url{http://documents.worldbank.org/curated/en/099114411282332013/IDU0c5a6be52060f404fb10b6d501e477b3a9a16}
\BIBentrySTDinterwordspacing

\bibitem{itu2020greenhouse}
L.~Itu-T, ``Greenhouse gas emissions trajectories for the information and communication technology sector compatible with the unfccc paris agreement,'' \emph{document itu-t 1. 1470 (01/20)}, 2020.

\bibitem{silver2019mobile}
L.~Silver, A.~Smith, C.~Johnson, K.~Taylor, J.~Jiang, M.~Anderson, and L.~Rainie, ``Mobile connectivity in emerging economies,'' \emph{Pew Research Center}, vol.~7, 2019.

\bibitem{bieser2023review}
J.~C. Bieser, R.~Hintemann, L.~M. Hilty, and S.~Beucker, ``A review of assessments of the greenhouse gas footprint and abatement potential of information and communication technology,'' \emph{Environmental Impact Assessment Review}, vol.~99, p. 107033, 2023.

\bibitem{ayers2023measuring}
S.~Ayers, S.~Ballan, V.~Gray, and R.~McDonald, ``Measuring the emissions and energy footprint of the {ICT} sector: Implications for climate action,'' World Bank Group, Tech. Rep., 2023.

\bibitem{ying2023waste}
M.~Ying, D.~Shakya, H.~Poddar, and T.~S. Rappaport, ``Waste factor: A new metric for evaluating power efficiency in any cascade,'' in \emph{2023 IEEE Global Communications Conference (GLOBECOM)}, 2023, pp. 6735--6740.

\bibitem{Rappaport2013Itwillwork}
T.~S. Rappaport, S.~Sun, R.~Mayzus, H.~Zhao, Y.~Azar, K.~Wang, G.~N. Wong, J.~K. Schulz, M.~Samimi, and F.~Gutierrez, ``Millimeter wave mobile communications for {5G} cellular: It will work!'' \emph{IEEE Access}, vol.~1, pp. 335--349, 2013.

\bibitem{Rappaport2019Beyond6G}
T.~S. Rappaport, Y.~Xing, O.~Kanhere, S.~Ju, A.~Madanayake, S.~Mandal, A.~Alkhateeb, and G.~C. Trichopoulos, ``Wireless communications and applications above 100 {GHz}: Opportunities and challenges for {6G} and beyond,'' \emph{IEEE Access}, vol.~7, pp. 78\,729--78\,757, 2019.

\bibitem{Huawei2020}
\BIBentryALTinterwordspacing
C.~Gabriel, H.~He, and A.~Chern, ``{Green 5G: Building a sustainable world},'' {Huawei Technologies Co., Ltd.}, Tech. Rep., Aug. 2020. [Online]. Available: \url{https:// www.huawei.com/en/public-policy/green-5g-building-a-sustainable-world}
\BIBentrySTDinterwordspacing

\bibitem{Shakya2024Indoor}
D.~Shakya, M.~Ying, T.~S. Rappaport, P.~Ma, Y.~Wang, I.~Al-Wazani, and H.~Poddar, ``{Propagation measurements in Indoor Environment at 6.75 GHz FR1(C) and 16.95 GHz FR3 Upper mid-band Spectrum for 5G and 6G},'' in \emph{(submitted) 2024 IEEE Global Communications Conference (GLOBECOM)}, 2024, pp. 1--6.

\bibitem{Kang2024FR3}
S.~Kang, M.~Mezzavilla, S.~Rangan, A.~Madanayake, S.~B. Venkatakrishnan, G.~Hellbourg, M.~Ghosh, H.~Rahmani, and A.~Dhananjay, ``Cellular wireless networks in the upper mid-band,'' \emph{IEEE Open Journal of the Communications Society}, vol.~5, pp. 2058--2075, 2024.

\bibitem{Shakya2024Pen}
D.~Shakya, M.~Ying, T.~S. Rappaport, P.~Ma, Y.~Wang, I.~Al-Wazani, and H.~Poddar, ``{Wideband Penetration Loss through Building Materials and Partitions at 6.75 GHz in FR1(C) and 16.95 GHz in the FR3 Upper Mid-band spectrum},'' in \emph{(submitted) 2024 IEEE Global Communications Conference (GLOBECOM)}, 2024, pp. 1--6.

\bibitem{NGMN2021NetworkEnergy}
\BIBentryALTinterwordspacing
J.~von Perner, V.~Friderikos, J.~Erfanian, J.~Liu, S.~Ansari, D.~Dianat \emph{et~al.}, ``Network energy efficiency,'' Green Future Networks, NGMN Alliance, Frankfurt, Germany, Technical Report, v 1.1, Dec 2021. [Online]. Available: \url{https://www.ngmn.org/wp-content/uploads/211009-GFN-Network-Energy-Efficiency-1.0.pdf}
\BIBentrySTDinterwordspacing

\bibitem{gsma2024going}
\BIBentryALTinterwordspacing
E.~Kolta and T.~Hatt, ``Going green: Measuring the energy efficiency of mobile networks,'' GSMA Intelligence, Tech. Rep., 2024. [Online]. Available: \url{https://data.gsmaintelligence.com/api-web/v2/research-file-download?id=79791160&file=270224-Measuring-energy-efficiency-of-mobile-networks.pdf}
\BIBentrySTDinterwordspacing

\bibitem{GSMA20205Genergy}
\BIBentryALTinterwordspacing
T.~Hatt and E.~Kolta, ``{5G Energy efficiencis: Green is the new black},'' {GSMA}, Tech. Rep., Nov. 2020. [Online]. Available: \url{https://data.gsmaintelligence.com/api-web/v2/research-file-download?id=54165956&file=241120-5G-energy.pdf}
\BIBentrySTDinterwordspacing

\bibitem{ericsson2023}
\BIBentryALTinterwordspacing
P.~Jonsson, P.~Cerwall, A.~Lundvall, D.~von Koch, S.~Davis \emph{et~al.}, ``Ericsson mobility report november 2023,'' Ericsson, Tech. Rep., Nov 2023. [Online]. Available: \url{https://www.ericsson.com/en/reports-and-papers/mobility-report/reports/november-2023}
\BIBentrySTDinterwordspacing

\bibitem{chen2010energy}
T.~Chen, H.~Kim, and Y.~Yang, ``Energy efficiency metrics for green wireless communications,'' in \emph{2010 International Conference on Wireless Communications \& Signal Processing (WCSP)}.\hskip 1em plus 0.5em minus 0.4em\relax IEEE, 2010, pp. 1--6.

\bibitem{boyle2017energy}
D.~Boyle, R.~Kolcun, and E.~Yeatman, ``Energy-efficient communication in wireless networks,'' \emph{ICT-Energy Concepts for Energy Efficiency and Sustainability}, 2017.

\bibitem{hossfeld2022analytical}
T.~Ho{\ss}feld, S.~Raffeck, F.~Loh, and S.~Gei{\ss}ler, ``Analytical model for the energy efficiency in low power iot deployments,'' in \emph{2022 IEEE 8th International Conference on Network Softwarization (NetSoft)}.\hskip 1em plus 0.5em minus 0.4em\relax IEEE, 2022, pp. 19--24.

\bibitem{Mccune2019globecom}
E.~McCune, ``A comprehensive view of wireless link energy efficiency,'' in \emph{2019 IEEE Global Communications Conference (GLOBECOM)}, 2019, pp. 1--5.

\bibitem{murdock2013consumption}
J.~N. Murdock and T.~S. Rappaport, ``Consumption factor and power-efficiency factor: A theory for evaluating the energy efficiency of cascaded communication systems,'' \emph{IEEE Journal on Selected Areas in Communications}, vol.~32, no.~2, pp. 221--236, 2013.

\bibitem{kanhere2022power}
O.~Kanhere, H.~Poddar, Y.~Xing, D.~Shakya, S.~Ju, and T.~S. Rappaport, ``A power efficiency metric for comparing energy consumption in future wireless networks in the millimeter-wave and terahertz bands,'' \emph{IEEE Wireless Communications}, vol.~29, no.~6, pp. 56--63, 2022.

\bibitem{etsiES203228}
\BIBentryALTinterwordspacing
{ETSI Environmental Engineering}, ``Es2023228: Assessment of mobile network energy efficiency,'' October 2020. [Online]. Available: \url{https://www.etsi.org/deliver/etsi_es/203200_203299/203228/01.03.01_60/es_203228v010301p.pdf}
\BIBentrySTDinterwordspacing

\bibitem{etsiTS103786}
\BIBentryALTinterwordspacing
------, ``Ts103786: Measurement method for energy efficiency of wireless access network equipment; dynamic energy efficiency measurement method of {5G} base station (bs) dynamic energy efficiency measurement method of {5G} bs,'' December 2020. [Online]. Available: \url{https://www.etsi.org/deliver/etsi_ts/103700_103799/103786/01.01.01_60/ts_103786v010101p.pdf}
\BIBentrySTDinterwordspacing

\bibitem{ituL13502016}
\BIBentryALTinterwordspacing
ITU, ``Energy efficiency measurement for telecommunication equipment,'' 2016. [Online]. Available: \url{https://www.itu.int/rec/T-REC-L.1350-201610-I/en}
\BIBentrySTDinterwordspacing

\bibitem{ituL13102020}
\BIBentryALTinterwordspacing
------, ``L.1310: Energy efficiency metrics and measurement methods for telecommunication equipment,'' September 2020. [Online]. Available: \url{https://www.itu.int/rec/T-REC-L.1310-202009-I}
\BIBentrySTDinterwordspacing

\bibitem{3gpp28813}
3GPP, ``3rd generation partnership project; technical specification group services and system aspects; management and orchestration; study on new aspects of energy efficiency {(EE) for 5G (Release 17)},'' 3GPP, Tech. Rep. TR 28.813 V17.0.0 (2021-12), 2021.

\bibitem{Rappaport2024Microwave}
T.~Rappaport, M.~Ying, and D.~Shakya, ``Waste figure and waste factor: New metrics for evaluating power efficiency in any circuit or cascade,'' \emph{Microwave Journal}, vol.~67, no.~5, pp. 54--84, May 2024.

\bibitem{ying2024MIMOWF}
M.~Ying, D.~Shakya, and T.~S. Rappaport, ``Using waste factor to optimize energy efficiency in multiple-input single-output ({MISO}) and multiple-input multiple-output ({MIMO}) systems,'' in \emph{(submitted) 2024 IEEE Global Communications Conference (GLOBECOM)}, 2024, pp. 1--6.

\bibitem{a12}
H.~T. Friis, ``Noise figures of radio receivers,'' \emph{Proc. IRE}, vol.~32, no.~7, pp. 419--422, 1944.

\bibitem{SNROptTAP}
R.~Kumar and S.~Arnon, ``Snr optimization for leo satellite at sub-thz frequencies,'' \emph{IEEE Trans. Antennas Propag.}, vol.~70, no.~6, pp. 4449--4458, 2022.

\bibitem{LowNoiseAmp}
L.~Belostotski and S.~Jagtap, ``Down with noise: An introduction to a low-noise amplifier survey,'' \emph{IEEE J. Solid-State Circuits}, vol.~12, no.~2, pp. 23--29, 2020.

\bibitem{hemmati2006deep}
H.~Hemmati, ``Deep-space communications and navigation series,'' \emph{JPL, Wiley}, 2006.

\bibitem{tian2018ber}
R.~Tian, K.~Senda, and H.~Otsuka, ``{BER performance of OFDM-based 4096-QAM using soft decision Viterbi decoding in multipath fading},'' in \emph{2018 IEEE World Symposium on Communication Engineering (WSCE)}.\hskip 1em plus 0.5em minus 0.4em\relax IEEE, 2018, pp. 1--4.

\bibitem{a8}
J.~N. Murdock and T.~S. Rappaport, ``Consumption factor: A figure of merit for power consumption and energy efficiency in broadband wireless communications,'' in \emph{2011 IEEE GLOBECOM Workshops (GC Wkshps)}, 2011, pp. 1393--1398.

\bibitem{x1}
T.~S. Rappaport, J.~N. Murdock, and F.~Gutierrez, ``State of the art in {60-GHz} integrated circuits and systems for wireless communications,'' \emph{Proc. IEEE}, vol.~99, no.~8, pp. 1390--1436, 2011.

\bibitem{Rappaport2024wireless}
T.~S. Rappaport, \emph{Wireless communications: principles and practice}.\hskip 1em plus 0.5em minus 0.4em\relax Cambridge University Press, updated 2nd edition, 2024.

\bibitem{Shu2014Beamcomb}
S.~Sun, G.~R. MacCartney, M.~K. Samimi, S.~Nie, and T.~S. Rappaport, ``Millimeter wave multi-beam antenna combining for {5G} cellular link improvement in new york city,'' in \emph{2014 IEEE International Conference on Communications (ICC)}, 2014, pp. 5468--5473.

\bibitem{Habibi2019RAN}
M.~A. Habibi, M.~Nasimi, B.~Han, and H.~D. Schotten, ``A comprehensive survey of ran architectures toward {5G} mobile communication system,'' \emph{IEEE Access}, vol.~7, pp. 70\,371--70\,421, 2019.

\bibitem{3gpp38840}
3GPP, ``{3rd Generation Partnership Project; Technical Specification Group Radio Access Network; NR; Study on User Equipment (UE) power saving in NR (Release 16)},'' 3GPP, Tech. Rep. TR 38.840 V16.0.0 (2019-06), 2019.

\bibitem{dedomenico2023modeling}
A.~De~Domenico, D.~López-Pérez, W.~Li, N.~Piovesan, H.~Bao, and X.~Geng, ``Modeling user transfer during dynamic carrier shutdown in green {5G} networks,'' \emph{IEEE Transactions on Wireless Communications}, vol.~22, no.~8, pp. 5536--5549, 2023.

\bibitem{held1978conversion}
D.~N. Held and A.~R. Kerr, ``Conversion loss and noise of microwave and millimeterwave mixers: Part 1-theory,'' \emph{IEEE Transactions on Microwave Theory and Techniques}, vol.~26, no.~2, pp. 49--55, 1978.

\bibitem{Ludwig2000}
R.~Ludwig, \emph{RF Circuit Design: Theory \& Applications}.\hskip 1em plus 0.5em minus 0.4em\relax Pearson Education India, 2000.

\bibitem{PSAccess}
A.~Medina-Rull, F.~Pasadas, E.~G. Marin, A.~Toral-Lopez, J.~Cuesta, A.~Godoy, D.~Jimélnez, and F.~G. Ruiz, ``A graphene field-effect transistor based analogue phase shifter for high-frequency applications,'' \emph{IEEE Access}, vol.~8, pp. 209\,055--209\,063, 2020.

\bibitem{DaiDAC}
J.~Zhang, L.~Dai, Z.~He, B.~Ai, and O.~A. Dobre, ``Mixed-{ADC/DAC} multipair massive {MIMO} relaying systems: Performance analysis and power optimization,'' \emph{IEEE Transactions on Communications}, vol.~67, no.~1, pp. 140--153, 2019.

\bibitem{CuiDAC}
S.~Cui, A.~Goldsmith, and A.~Bahai, ``Energy-constrained modulation optimization,'' \emph{IEEE Transactions on Wireless Communications}, vol.~4, no.~5, pp. 2349--2360, 2005.

\bibitem{johns2008analog}
D.~A. Johns and K.~Martin, \emph{Analog integrated circuit design}.\hskip 1em plus 0.5em minus 0.4em\relax John Wiley \& Sons, 2008.

\bibitem{SkriReceiver}
P.~Skrimponis, N.~Hosseinzadeh, A.~Khalili, E.~Erkip, M.~J.~W. Rodwell, J.~F. Buckwalter, and S.~Rangan, ``Towards energy efficient mobile wireless receivers above 100 {GHz},'' \emph{IEEE Access}, vol.~9, pp. 20\,704--20\,716, 2021.

\bibitem{LeeADC}
H.-S. Lee and C.~G. Sodini, ``Analog-to-digital converters: Digitizing the analog world,'' \emph{Proceedings of the IEEE}, vol.~96, no.~2, pp. 323--334, 2008.

\bibitem{AD_datasheet}
\BIBentryALTinterwordspacing
\emph{{AD7386/AD7387/AD7388: 16-Bit/14-Bit/12-Bit, SAR ADC Data Sheet}}, Analog Devices, 2024, {Single-Ended Input, Quad, Simultaneous Sampling, 16-Bit/14-Bit/12-Bit, SAR ADC}. [Online]. Available: \url{https://www.analog.com/AD7388-4/datasheet}
\BIBentrySTDinterwordspacing

\bibitem{steer2019microwave}
M.~Steer, \emph{Microwave and RF Design: Amplifiers and Oscillators}, 3rd~ed.\hskip 1em plus 0.5em minus 0.4em\relax NC State University, 2019, vol.~5.

\bibitem{walker1993high}
L.~B.~J. Walker, \emph{High-Power GaAs FET Amplifiers}.\hskip 1em plus 0.5em minus 0.4em\relax Norwood, MA, USA: Artech House, 1993.

\bibitem{szczepkowski2014study}
G.~Szczepkowski and R.~Farrell, ``Study of linearity and power consumption requirements of cmos low noise amplifiers in context of lte systems and beyond,'' \emph{International Scholarly Research Notices}, vol. 2014, 2014.

\bibitem{skrimponis2020power}
P.~Skrimponis, S.~Dutta, M.~Mezzavilla, S.~Rangan, S.~H. Mirfarshbafan, C.~Studer, J.~Buckwalter, and M.~Rodwell, ``Power consumption analysis for mobile {mmWave and sub-THz} receivers,'' in \emph{2020 2nd 6G Wireless Summit (6G SUMMIT)}.\hskip 1em plus 0.5em minus 0.4em\relax IEEE, 2020, pp. 1--5.

\bibitem{Ickhyun2008LNAFoM}
I.~Song, J.~Jeon, H.-S. Jhon, J.~Kim, B.-G. Park, J.~D. Lee, and H.~Shin, ``A simple figure of merit of rf mosfet for low-noise amplifier design,'' \emph{IEEE Electron Device Letters}, vol.~29, no.~12, pp. 1380--1382, 2008.

\bibitem{Ahlgren1999LNAFoM}
D.~Ahlgren, N.~King, G.~Freeman, R.~Groves, and S.~Subbanna, ``Sige bicmos technology for rf device and design applications,'' in \emph{Proc. IEEE Radio Wireless Conf.}, 1999, pp. 281--284.

\bibitem{piovesan2023power}
N.~Piovesan, D.~López-Pérez, A.~De~Domenico, X.~Geng, and H.~Bao, ``Power consumption modeling of {5G} multi-carrier base stations: A machine learning approach,'' in \emph{ICC 2023 - IEEE International Conference on Communications}, 2023, pp. 3633--3638.

\bibitem{lopez2023data}
D.~L{\'o}pez-P{\'e}rez, A.~De~Domenico, N.~Piovesan, and M.~Debbah, ``Data-driven energy efficiency modelling in large-scale networks: An expert knowledge and ml-based approach,'' \emph{arXiv preprint arXiv:2401.00443}, 2023.

\bibitem{piovesan2022machine}
N.~Piovesan, D.~López-Pérez, A.~De~Domenico, X.~Geng, H.~Bao, and M.~Debbah, ``Machine learning and analytical power consumption models for {5G} base stations,'' \emph{IEEE Communications Magazine}, vol.~60, no.~10, pp. 56--62, 2022.

\bibitem{han2020energy}
S.~Han, S.~Bian \emph{et~al.}, ``Energy-efficient {5G} for a greener future,'' \emph{Nature Electronics}, vol.~3, no.~4, pp. 182--184, 2020.

\bibitem{lopez2022survey}
D.~López-Pérez, A.~De~Domenico, N.~Piovesan, G.~Xinli, H.~Bao, S.~Qitao, and M.~Debbah, ``A survey on {5G} radio access network energy efficiency: Massive {MIMO}, lean carrier design, sleep modes, and machine learning,'' \emph{IEEE Communications Surveys and Tutorials}, vol.~24, no.~1, pp. 653--697, 2022.

\bibitem{3GPPRel9}
{3GPP TSG RAN}, ``{Evolved Universal Terrestrial Radio Access (E-UTRA) and Evolved Universal Terrestrial Radio Access Network (E-UTRAN); Further Advancements for E-UTRA Physical Layer Aspects (Release 9)},'' 3GPP, Technical Report 36.814 v9.0.0, 2010.

\bibitem{ITURM2134}
{ITU-R}, ``{Requirements Related to Technical Performance for IMT-Advanced Radio Interface(s)},'' ITU-R, Report M.2134, 2008.

\bibitem{3GPPTR36824Rel11}
{3GPP Technical Specification Group Radio Access Network}, ``{Evolved Universal Terrestrial Radio Access (E-UTRA); LTE Coverage Enhancements (Release 11)},'' 3GPP, Technical Report 36.824 V12.0.0, June 2012.

\bibitem{maccartney2019millimeter}
G.~R. MacCartney and T.~S. Rappaport, ``Millimeter-wave base station diversity for {5G} coordinated multipoint ({CoMP}) applications,'' \emph{IEEE Transactions on Wireless Communications}, vol.~18, no.~7, pp. 3395--3410, 2019.

\bibitem{you2021distributed}
X.~You, D.~Wang, and J.~Wang, \emph{Distributed MIMO and cell-free mobile communication}.\hskip 1em plus 0.5em minus 0.4em\relax Springer, 2021.

\bibitem{demir2021foundations}
{\"O}.~T. Demir, E.~Bj{\"o}rnson, L.~Sanguinetti \emph{et~al.}, ``Foundations of user-centric cell-free massive {MIMO},'' \emph{Foundations and Trends{\textregistered} in Signal Processing}, vol.~14, no. 3-4, pp. 162--472, 2021.

\bibitem{nayebi2015cell}
E.~Nayebi, A.~Ashikhmin, T.~L. Marzetta, and H.~Yang, ``Cell-free {massive MIMO} systems,'' in \emph{2015 49th Asilomar Conference on Signals, Systems and Computers}.\hskip 1em plus 0.5em minus 0.4em\relax IEEE, 2015, pp. 695--699.

\bibitem{shakya2024radio}
D.~Shakya, S.~Ju, O.~Kanhere, H.~Poddar, Y.~Xing, and T.~S. Rappaport, ``Radio propagation measurements and statistical channel models for outdoor urban microcells in open squares and streets at 142, 73, and 28 {GHz},'' \emph{IEEE Transactions on Antennas and Propagation}, 2024.

\bibitem{Samman2016Channel17GHz}
A.~M. Al-Samman, T.~A. Rahman, M.~H. Azmi, N.~R. Zulkefly, and A.~M. Mataria, ``Path loss model for outdoor environment at 17 {GHz} mm-wave band,'' in \emph{2016 IEEE 12th International Colloquium on Signal Processing and Its Applications (CSPA)}, 2016, pp. 179--182.

\bibitem{adegoke2021channel}
E.~I. Adegoke, E.~Kampert, and M.~D. Higgins, ``Channel modeling and over-the-air signal quality at {3.5 GHz for 5G} new radio,'' \emph{IEEE Access}, vol.~9, pp. 11\,183--11\,193, 2021.

\bibitem{rappaport2015wideband}
T.~S. Rappaport, G.~R. MacCartney, M.~K. Samimi, and S.~Sun, ``Wideband millimeter-wave propagation measurements and channel models for future wireless communication system design,'' \emph{IEEE transactions on Communications}, vol.~63, no.~9, pp. 3029--3056, 2015.

\bibitem{balanis2016antenna}
C.~A. Balanis, \emph{Antenna theory: analysis and design}.\hskip 1em plus 0.5em minus 0.4em\relax John wiley \& sons, 2016.

\bibitem{rappaport2015millimeter}
T.~S. Rappaport, R.~W. Heath~Jr, R.~C. Daniels, and J.~N. Murdock, \emph{Millimeter wave wireless communications}.\hskip 1em plus 0.5em minus 0.4em\relax Pearson Education, 2015.

\bibitem{carroll2010analysis}
A.~Carroll and G.~Heiser, ``An analysis of power consumption in a smartphone,'' in \emph{2010 USENIX Annual Technical Conference (USENIX ATC 10)}, 2010.

\bibitem{xing2018propagation}
Y.~Xing and T.~S. Rappaport, ``Propagation measurement system and approach at 140 {GHz-moving to 6G and above 100 GHz},'' in \emph{2018 IEEE global communications Conference (GLOBECOM)}.\hskip 1em plus 0.5em minus 0.4em\relax IEEE, 2018, pp. 1--6.

\bibitem{rangan2014millimeter}
S.~Rangan, T.~S. Rappaport, and E.~Erkip, ``Millimeter-wave cellular wireless networks: Potentials and challenges,'' \emph{Proceedings of the IEEE}, vol. 102, no.~3, pp. 366--385, 2014.

\bibitem{auer2010enablers}
G.~Auer, I.~G{\'o}dor, L.~H{\'e}vizi, M.~A. Imran, J.~Malmodin, P.~Fazekas, G.~Bicz{\'o}k, H.~Holtkamp, D.~Zeller, O.~Blume \emph{et~al.}, ``Enablers for energy efficient wireless networks,'' in \emph{2010 IEEE 72nd Vehicular Technology Conference-Fall}.\hskip 1em plus 0.5em minus 0.4em\relax IEEE, 2010, pp. 1--5.

\bibitem{sun2016investigation}
S.~Sun, T.~S. Rappaport, T.~A. Thomas, A.~Ghosh, H.~C. Nguyen, I.~Z. Kovacs, I.~Rodriguez, O.~Koymen, and A.~Partyka, ``Investigation of prediction accuracy, sensitivity, and parameter stability of large-scale propagation path loss models for {5G} wireless communications,'' \emph{IEEE transactions on vehicular technology}, vol.~65, no.~5, pp. 2843--2860, 2016.

\end{thebibliography}
